\documentclass{article}

\usepackage{graphicx}

\usepackage[margin=1in]{geometry}
\usepackage{natbib}

\usepackage{amsmath} 
\usepackage{amssymb} 
\usepackage{amsthm}
\usepackage{tikz}
\usetikzlibrary{positioning,arrows.meta,,bending,arrows,automata, decorations.pathreplacing, calc}
\usepackage{xcolor}

\usepackage{cases}
\usepackage{subcaption}

\usepackage{authblk}

\usepackage{multirow}%
\usepackage{amsmath,amssymb,amsfonts}%
\usepackage{amsthm}%
\usepackage{mathrsfs}%
\usepackage[title]{appendix}%
\usepackage{xcolor}%
\usepackage{textcomp}%
\usepackage{manyfoot}%
\usepackage{booktabs}%
\usepackage{algorithm}%
\usepackage{algorithmicx}%
\usepackage{algpseudocode}%

\usepackage[normalem]{ulem}
\usepackage{subcaption}
\usepackage{tabularx}

\usepackage{setspace}
\definecolor{crimson}{rgb}{0.86, 0.08, 0.24}
\definecolor{royalblue}{rgb}{0.25, 0.41, 0.88}
\definecolor{lightcoral}{rgb}{0.94, 0.5, 0.5}
\definecolor{skyblue}{rgb}{0.53, 0.81, 0.92}
\definecolor{lightred}{rgb}{1.0, 0.75, 0.75}
\definecolor{lightblue}{rgb}{0.78, 0.88, 1.0}

\newcommand{\DRAFT}{} %

\ifdefined\DRAFT
\newcommand{\hl}[1]{{\color{red} [\textbf{HL:} #1]}}

\newcommand{\joj}[1]{{\color{purple}#1}}
\newcommand{\jb}[1]{{\color{orange} [\textbf{JB:} #1]}}

\newcommand{\ey}[1]{{\color{teal} [\textbf{EY:} #1]}}

\newcommand{\todo}[1]{{\color{red} [TODO: #1]}}

\else
\newcommand{\hl}[1]{ }
\newcommand{\hldelete}[1]{}
\newcommand{\deletehl}[1]{}
\newcommand{\joj}[1]{ }
\newcommand{\jb}[1]{ }
\newcommand{\ey}[1]{}

\newcommand{\todo}[1]{}

\fi

\newtheorem{theorem}{Theorem}
\newtheorem{lemma}{Lemma}
\newtheorem{corollary}{Corollary}
\newtheorem{proposition}{Proposition}

\newcommand{\Var}{\ensuremath{Var}}
\newcommand{\E}{\mathbb{E}}
\newcommand{\MDE}{\ensuremath{{MDE}}}

\tikzset{
  treenode/.style = {shape=rectangle,
                     draw, align=center},
  root/.style     = {treenode, font=\large},
  env/.style      = {treenode, font=\ttfamily\normalsize},
  dummy/.style    = {circle,draw}
}

\title{Operational Dosage: Implications of Capacity Constraints for the Design and Interpretation of Experiments}

\author[1]{Justin J. Boutilier}
\author[2]{Jónas Oddur Jónasson}
\author[3]{Hannah Li}
\author[4]{Erez Yoeli}

\affil[1]{\small Telfer School of Management, University of Ottawa. \text{boutilier@telfer.uottawa.ca}}
\affil[2]{\small Imperial College London. \text{joj@imperial.ac.uk}}
\affil[3]{\small Columbia Business School. \text{hannah.li@columbia.edu}}
\affil[4]{\small MIT Sloan School of Management. \text{eyoeli@mit.edu}}

\begin{document}
\date{}

\maketitle

\begin{abstract}
{We study RCTs that evaluate the impact of \textit{service interventions}, for example, teachers or advisors conducting proactive outreach to at-risk students, medical providers giving medication adherence support by calling or texting, or social service workers that conduct home visits. 
A defining feature of service interventions is that they are delivered by a capacity-constrained resource---teachers, healthcare providers, or social workers---whose limited availability creates causal inference complications. Because participants share a common and finite service capacity, adding more participants can reduce the timeliness or intensity of the service that others receive, introducing interference across participants. This generates hidden variation in the treatment itself, which we term \textit{operational dosage}.
We provide a mathematical model of service interventions using techniques from queueing theory and study the impact of capacity constraints on experimental outcomes. 
Our main insight is that treatment effects are both capacity- and sample-size-dependent, as well as decreasing in sample size once a critical threshold is exceeded. Interestingly, an implication is that statistical power of service intervention RCTs peaks at intermediate sample sizes---directly contradicting conventional power calculations that assume monotonically increasing power with sample size. 
We instantiate our insights using simulations calibrated to a real-world trial evaluating a behavioral health intervention aimed to increase medication adherence among tuberculosis patients in Kenya. 
Our simulation results suggest that a trial with high service capacity but limited sample size can obtain the same statistical power as a trial with lower service capacity but large sample size. 
Taken together, our results highlight the importance of capacity selection in experiment design and provide a mechanism for why experiments may fail to replicate or perform at scale.}
\end{abstract}

% \keywords{Randomized Controlled Trials, Experimentation, Interference, Spillovers, Capacity Constraints, Scaling, Service Interventions}

\section{Introduction}

\noindent Randomized Controlled Trials (RCTs) are a cornerstone of scientific progress across disciplines. When used to evaluate an intervention, RCTs are the ``gold standard'' because random assignment of the intervention facilitates causal inference---the attribution of differences in outcomes to the intervention \citep{pearl2009causal,imbens2015causal}. However, the use of randomization does not \emph{per se} ensure valid causal inference.  A growing literature identifies and addresses potential causal inference problems stemming from, e.g., spillovers \citep{miguel2004worms,hudgens2008toward}, network effects \citep{manski2013identification}, non-compliance \citep{imbens1994identification}, varying quality \citep{vanderweele2013causal,hernan2011compound}, unspecified operationalization \citep{hernan2008does}, and population heterogeneity \citep{heckman1997making}---calling for improved practices in both design and interpretation of RCTs \citep{athey2017econometrics,HernanRobins2020,leroy2022strengthening,brand2023recent,list2024optimally,imbens2024causal}.

In this paper we introduce and analyze a common source of causal inference interference which has hitherto received little attention: operational capacity constraints. In particular, we consider a prevalent class of interventions, which we label \emph{service interventions}, in which: (1) participants intermittently become eligible for a service that should be delivered `just-in-time' \citep{klasnja2015microrandomized, nahum2016just, smith2017design}, and (2) delivery of service involves a capacity-constrained resource. %
Service interventions span disciplines: healthcare providers conducting outreach to at-risk patients, teachers offering targeted learning support to students at risk of dropping out, or volunteers delivering social services to families experiencing financial stress.  Recent examples include: digital health interventions in which humans provide encouragement or counseling \citep{Yoeli_Rathauser_Bhanot_Kimenye_Mailu_Masini_Owiti_Rand_2019,volpp2017effect,gustafson2014smartphone,munoz2017impact,keum2021usefulness,cho2020effect,toro2020mobile,killian2023robust,dasgupta2024preliminary}; machine learning enabled Early Warning Systems, such as rapid response teams in clinical healthcare \cite{song2021strategies, chang2025rapid} and graduation support programs in education \cite{rossman2021maaps,schechtman2025discretion, perdomo2025difficult}; and public service interventions, such as emergency financial and public housing assistance, which have documented delays in service \cite{palmer2019does,painter1997does, simon2024target}.

Crucially, in a service intervention, capacity constraints can influence treatment effects.  
When the number of participants simultaneously requiring service exceeds the operational capacity, this reduces service quality, by, e.g., reducing service frequency and consistency, or increasing wait times. This creates what we term \textit{``operational dosage''}: systematic variation in treatment intensity determined by the participant-to-capacity ratio rather than by the intervention protocol. 
Evidence of operational dosage can be found in published trials. %
For instance, the `Heartstrong' intervention \cite{volpp2017effect}, which used electronic pill bottles to monitor heart attack patients' medication adherence, found ``considerable variation in delivery timing due to operational constraints such as [the] care team having limited capacity'' and greater effect on medication adherence and treatment outcomes when delivered ``quickly and consistently'' \citep{lekwijit2023evaluating}.  
Additionally, researchers have recognized capacity constraints as a practical implementation challenge in scaling interventions \citep{list2022voltage,list2024optimally} and operationalizing machine learning tools \citep{liu2025bridging}. %

As a service intervention's operational dosage can vary from one implementation to the next, this poses a challenge for causal inference.
For instance, when generalizing from a successful trial to an implementation at scale: if capacity is reduced, then the service intervention will be less effective at scale than during the trial. There are also important implications for the design of RCTs.  For instance, if a researcher seeking to increase the power of a trial increases its sample size, this will indeed reduce sampling variation and increase statistical precision. However, if capacity is not also sufficiently increased, operational dosage will fall, and the resulting effect on statistical power is ambiguous. Formally, a service intervention's capacity constraints generates violations of the Stable Unit Treatment Value Assumptions (SUTVA) \citep{rubin1980randomization,angrist1996identification,imbens2010rubin} through two mechanisms: `hidden variation' because quality of service varies with capacity, and `interference' \citep{cox1958planning} because, when a capacity constraint is binding, one participant's receiving of service necessarily comes at the expense of another's.

Despite the prevalence of service interventions, there has been little systematic study of how capacity selection and the resulting operational dosage affect experimental outcomes. %
Consequently, the typical experimental design, power analysis, and effect size interpretation usually implicitly assume that the treatment effects are invariant to these operational parameters.  
We will illustrate the consequences of this oversight, providing a theoretical framework for understanding capacity-induced interference, as well as practical guidelines for experimenters to properly design and interpret service intervention RCTs.

We begin by modeling the operational dosage effect of capacity constraints (see \S \ref{s:model}). Our model uses ideas from queueing theory, a form of mathematical modeling that is specifically designed to characterize congestion and wait times due to capacity constraints.  In the model, `participants' move stochastically between two states: a desirable state (representing, e.g., adherence to a medication, exercise, or diet regimen) and an undesirable one (representing failure to adhere). 
The experimenter introduces an intervention that is designed to increase the amount of time participants spend in the desired state.
The participants in the intervention group are eligible to receive a service that `boosts' the likelihood of switching from the undesirable state to the desirable one. However, they receive this boost only if they receive the intervention, which is provided by a limited number of servers---the intervention's constrained capacity.  If, at a given time, all servers are occupied, a queue forms, with service provided on a first come first serve basis. If the queue is short, most people who enter it receive service---and, thus, the `boost'---in a timely manner.
A longer queue makes it more likely that an individual in the queue will switch to the desirable state without receiving the service, thereby diluting the boost from the service intervention. In the limiting case as capacity diminishes, participants would have no practical chance of receiving the service and therefore experience the same transition probabilities as a control group which has no access to the intervention.

Analysis of the model (see \S \ref{s:mainresults}) yields the following four insights:
\begin{enumerate}
\item \textbf{Treatment effects depend on operational dosage and can vary from context to context.} Capacity constraints induce dependencies across experiment subjects, because one subject’s receipt of service can delay another’s receipt of service. Thus, treatment effects are conditional on available capacity and number of participants. 

\item \textbf{Queueing theory identifies a useful `critical threshold' $r$ delineating when limited capacity reduces treatment effects.}  When the participant-to-server ratio is below $r$, treatment effects change comparably little as more participants are added.  This is known as a ``quality-driven" (QD) queueing regime.  When the participant-to-server ratio exceeds $r$, queue lengths increase non-linearly, and, potentially quite rapidly, leading to corresponding reductions in treatment effects.  This is known as an ``efficiency-driven" (ED) queueing regime. 

\item \textbf{In service intervention trials, statistical power is not monotonically increasing in the number of participants---in contradiction with conventional wisdom.} Holding capacity constant, statistical power initially increases, but then decreases after the participant-to-server ratio exceeds $r$, and crosses into the ED regime.

\item \textbf{The critical threshold is a useful reference point for service intervention trial design.} A trial designer can use the critical threshold to improve trial planning, choosing a participant-to-server ratio that is appropriate for the trial's aims.  To maximize power, the participant-to-server ratio should be close to $r$ (e.g., using the queueing-theoretic ``square-root staffing'' heuristic we discuss in \S \ref{s:mainresults}). 
\end{enumerate}

After establishing these insights theoretically, we instantiate them 
using a calibrated simulation model that relaxes some of the assumptions we make in our technical analysis and allows us to simulate counterfactual RCTs (see \S \ref{s.keheala}). Our source of data is a completed, large-scale service intervention RCT designed to evaluate \textit{Keheala}---a service intervention that promotes tuberculosis (TB) medication adherence \citep{Yoeli_Rathauser_Bhanot_Kimenye_Mailu_Masini_Owiti_Rand_2019,Boutilier_Jónasson_Yoeli_2022}. Keheala provides digital services to support treatment adherence among TB patients in Kenya. Patients self-verify daily medication adherence via a mobile phone application and those who fail to do so become eligible for a support sponsor phone call, intended as a nudge towards adherence. While an RCT demonstrated significant reduction in adverse health outcomes among enrolled patients \citep{Yoeli_Rathauser_Bhanot_Kimenye_Mailu_Masini_Owiti_Rand_2019},
follow-up analysis has revealed that patients oftentimes experienced delays between becoming eligible for the service and receiving it \citep{Boutilier_Jónasson_Yoeli_2022}--- which raises the question of how well the intervention would have performed with more support sponsors dedicated to outreach. Focusing on this setting and these data, we examine how treatment effects and statistical power are affected by varying capacity levels and sample sizes, and use these to show how capacity of the trial might have \emph{ex-ante} been adjusted to maximize power, and to predict how future implementations of the intervention might perform under various assumptions on scale and capacity.

Our work is of both theoretical and practical significance. 
Theoretically, it provides a novel characterization of a common mechanism of SUTVA violations.  Practically, it suggests better practices for service intervention trial design and interpretation, thereby contributing to efforts to promote replication \citep{bryan2021behavioural, vivalt2020much, gelman2018dont, almaatouq2024beyond, almaatouq2023some} and improve scaling \citep{list2022voltage,list2024optimally}.  These contributions are explored in more detail in our discussion section (see \S \ref{s:discussion}), which also draws out the connection of our work to research on SUTVA violations, intervention targeting, and queueing.

\section{Model and Theoretical Analysis}
\label{s:model}

We first present a formal queueing model of subject behavior, both with and without the service intervention (\S\ref{ss:model_of_service_interventions}). This model captures both the temporal dynamics of individual behavior, e.g., a patient's outcomes and interactions at time $t$ affecting their future behavior, as well as the congestion that arises in the service intervention when multiple individuals need service simultaneously.

We then adapt this model to study experimental outcomes in a service intervention RCT (\S\ref{ss:model_rct}) and show how the congestion leads to operational dosage effects in the RCT.\footnote{A similar model can be found in \cite{simon2024target} which studies capacity constrained public housing assistance.}

\subsection{Queueing Model of Service Interventions} 
\label{ss:model_of_service_interventions}

We present a continuous time Markov chain model of participant behavior and their interaction with the service intervention (see SI \ref{si:sec:theory} for more  detail). %

\subsubsection{The Individual Participant}
Each subject alternates between a \textit{desirable} state and an \textit{undesirable} state. In the desired state, they are engaging in some desired behavior, such as adhering to a medication regimen, exercising, eating healthy, staying sober, etc. In the undesirable state, they have stopped this behavior. In the absence of any interaction with the intervention, they will alternate back and forth between the states. However, each time they are in the undesirable state, they become eligible for the \textit{service} of the service intervention. If they receive the service, their probability of transitioning back to the desirable state is increased.

Formally, we denote the state of subject $u$ at time $t$ as $x(u,t)\in\{0,1\}$, where $0$ represents the desirable state and $1$ the undesirable state, in which they are eligible for the intervention. Subjects transition from state $0$ to state $1$ with rate $\lambda$ and, without intervention, naturally return from state 1 to state 0 with rate $\tau$ (see Fig. \ref{fig:state_diagram_one_user}). 

\begin{figure}[t]
    \begin{subfigure}{\linewidth}
    \centering
    \begin{tikzpicture}[node distance=4cm,->,>=latex,auto,
      every edge/.append style={thick}]
      \node[state,fill=lightblue] (1) {\textcolor{black}{$x(u,t)=0$}};
      \node[state,fill=lightred] (2) [right of=1] {\textcolor{black}{$x(u,t)=1$}};  
      \path %
              (1)  edge[bend left]  node{$\lambda$}   (2)
              (2)  edge[bend left] node{$\tau$}     (1);
    \end{tikzpicture}
    \caption{Single user, in the absence of intervention.}
    \label{fig:state_diagram_one_user}
    \end{subfigure}
    
    \medskip
    
    \begin{subfigure}{\linewidth}
    \centering
    \begin{tikzpicture}[node distance=2cm,->,>=latex,auto,
      every edge/.append style={thick}]

    \node[state,fill=lightblue] (zero) {\textcolor{black}{$x(u,t)=0$}};
    \node[state,fill=lightred] (one) [right of=1, node distance=5cm] {\textcolor{black}{$x(u,t)=1$}};  

    \def\numlines{3}
    \def\queuesize{1.5cm}
    \node[draw=none] (queue_left) at (3.2, -2.0) {};
    \node[draw=none, right=\queuesize of queue_left.west] (queue_right) {};

    \draw[-] ([yshift=-0.5cm,xshift=\queuesize]queue_left.west) 
        -- ++(-\queuesize,0) 
        -- ++(0,1) 
        -- ++(\queuesize,0);
    \foreach \i in {1,...,\numlines} {
        \draw[-] ([yshift=0.5cm, xshift=(\queuesize/(\numlines+1))*\i]queue_left.west) -- ++(0,-1);
      }

    \node[anchor=center][left=1cm of queue_left] (dots) {\LARGE { $\vdots$}};
    \node[state,rectangle,text width=1.35cm,minimum height=0.5cm, align=center,] 
        (server1) at ([yshift=.1cm]dots.north) 
        {Server 1};
    \node[state,rectangle,text  
    width=1.35cm,minimum height=0.5cm,align=center] 
        (server3) [below=0.02cm of dots] {Server $M$};

    \path  (zero)  
        edge[bend left] node{$\lambda$}   
        (one);

    \path (one) edge[bend left,dashed] 
        ([xshift=-0.2cm,yshift=0.2cm]queue_right.north east);

      \draw[-] ([yshift=0.6cm, xshift=(1*\queuesize/(\numlines+1))]queue_left.west) 
        -- ++(0,0.1) 
        -- ++(.4,0) 
        -- ++(0,-0.1);
    \path ([xshift=(1.5*\queuesize/(\numlines+1)), yshift=0.7cm]queue_left.west) 
        edge[bend right = 30] 
        node[above]{$\tau$} 
        ([xshift=0.15cm,yshift=0.2cm]zero.south east);

    \path (queue_left) edge[thin]   
            (server1.east);
    \path (queue_left) edge[thin] 
            (server1.east |- dots.east);
    \path (queue_left) edge[thin] 
            (server3.east);
    
    \path (server1.west) 
        edge[bend left = 20] 
        node[pos=0.25,below,yshift=-0.1cm]{$\mu p$}
        (zero.south);
    \path (server3.west) 
        edge[bend left = 30] 
        node[pos=0.18,below,yshift=-0.2cm]{$\mu p$}
        (zero.south);

    \end{tikzpicture}
    \caption{Queueing system, in the presence of intervention.}\label{fig:state_diagram_treatment}
    \end{subfigure}
    \caption{A queuing model of participant behavior in a service intervention RCT. Participant $u$'s state at time $t$ is represented as a ball. Blue indicates the desirable behavior, $x(u,t) = 0$. Red indicates the undesirable behavior, $x(u,t) = 1$. %
    Participants transition from one behavior to another, as represented by an arrow, with the transition rate indicated alongside the arrow. (a) In the absence of a service intervention, participants in the control group are assumed to transition from the desirable state to the undesirable state with rate $\lambda$ and from the desirable state to the desirable behavior at rate $\tau$. (b) Participants in the treatment group become eligible for the service intervention upon transitioning from the desirable state to the undesirable state. At this point, they enter a queue with stochastic length, depending on the number of providers $M_1$ and the number of of participants $N_1$. Providers serve participants one at a time and finish with rate $mu$. At any time when participants are waiting in the queue, they transition to the desirable state with rate $\tau$, just as in the control.  However, if they are selected from the queue and `receive service,' they have a probability $p$ of transitioning to the desirable state. }
    \label{fig:state_diagram}
\end{figure}

\subsubsection{The Service Intervention}
The intervention involves $M$ providers (e.g., clinicians, teachers, or behavior coaches) who deliver the service to the subjects in the undesirable state to encourage them to return to the desirable state. For example, interventions can be personalized messages, voice calls, or medical intervention. Each provider can serve one subject at a time and completes each service with rate $\mu$. At the completion of service, the subject returns to the desirable state with probability $p\in[0,1]$, where $p$ denotes the service success probability. The combination of $\mu$ and $p$ together create an effective service rate of $\mu p$.

We assume that the provider observes the state of each subject and thus has an up-to-date list of subjects who are eligible for the service, at any point in time.\footnote{This is reasonable for any intervention where the subjects log their behavior (e.g., medication adherence, exercise, diet, sobriety) 
or where the provider directly monitors the subject's condition (e.g., school early warning systems observe student grades, hospital early warning systems observe patient vital signs).} For clarity, we assume that the service provider interacts with the subjects solely through the service component.\footnote{In most practical cases, the service provider would most likely interact with subjects in other ways (e.g., through online dashboards, automated messages, app-based games, etc.). Since those would not be capacity constrained, we do not model those interactions explicitly.} 

\subsubsection{The Queueing System}
Our starting insight is that in this type of system, the behavior of one subject can interfere with the trajectory of another subject. Specifically, when multiple subjects are simultaneously eligible for service, some must wait, potentially remaining in the undesirable state longer as a result. In line with the growing literature on experimental interference, we refer to this effect as \textit{congestion interference}.

To capture these dependencies, we analyze the system 
not as a set of independent subjects but as a closed queueing system with $N$ subjects sharing $M$ servers (see Fig. \ref{fig:state_diagram_treatment}). This approach makes the issue of congestion interference explicit through the language of queueing theory.

Mathematically, we describe the system as follows (we will use capital letters to describe attributes of the system, to contrast with the lower case letters that describe the individual subjects). The system state at time $t$ is the number of subjects in the undesirable state, denoted as $X_{M,N}(t)$. The system $\{X_{M,N}(t)\}$ for $t\geq 0$ is a Markov chain with transition rates induced by the dynamics described above and depending on both the number of subjects eligible for service and the available capacity (see SI \ref{si:ssec_model} for full description of dynamics). The special case of $M=0$ denotes a system where participants are not given access to the service intervention. 
This formulation will allow us to derive the long-run steady-state distribution $\pi_{M,N}$ and expected queue length $\overline{K}_{M,N}$, which corresponds to the expected number of subjects in the undesirable state.

\subsubsection{Model Applied to Service Intervention Examples}

The service intervention model encompasses a broad array of interventions with capacity constrained components.
In the Keheala and Heartstrong medication support interventions, the desirable and undesirable states refer to adhering to the daily medication and not, respectively. Providers conduct the service of outreach calls.

In the healthcare and education Early Warning Systems for proactive treatment, an algorithm tracks participant data over time and states correspond to algorithmic predictions. In the healthcare sepsis EWS \cite{chang2025rapid}, participants are hospital patients, the undesirable state corresponds to an ``at-risk'' prediction for sepsis, the provider is the rapid response team and the service is  administration of antibiotics and intravenous fluids. In the education advising EWS \cite{rossman2021maaps,schechtman2025discretion}, participants are students, the undesirable state corresponds to an ``at-risk'' prediction of dropping out, the provider is the academic advisor, and the service is a conversation with the student. 

In the emergency financial aid and public housing interventions \cite{palmer2019does,painter1997does}, participants are residents, the undesirable state corresponds to eligibility criteria such as financial need or eviction, 
the provider is a government or non-profit administration, and the service is the provision of financial aid or housing.

\subsection{Service Intervention Model Applied to Experimentation}
\label{ss:model_rct}
We adapt the service intervention queueing model to study RCT outcomes.
The objective of the RCT is to establish 
whether the intervention causes subjects to spend more time in the desirable state.

Our work focuses on the design choices of the number of subjects and providers to recruit. 
The experimenter runs a trial in which they recruit subjects and then randomize subjects into the treatment group, which receives access to the service intervention, and the control group, which does not. Let $N_1$ and $N_0$ denote the number of treatment and control subjects, respectively. For clarity, we fix $N_1=N_0$, although our analysis allows for different treatment proportions. The experimenter also recruits $M_1$ providers to conduct the service interventions for the treatment group. The experiment runs a time horizon of $T$.

We can model dynamics of the treatment and control groups as two separate Markov chains, since there are no interactions between the two groups. 
The treatment group evolves as a Markov chain $\{X_{M_1,N_1}(t)\}$ with $M_1$ servers. The control group operates according to the chain $\{X_{0,N_0}(t)\}$ with no servers. Note that subjects within the treatment group are dependent, whereas subjects in the control group evolve independently of each other.

We define the
treatment effect of the intervention $\theta^{ss}(M_1,N_1, N_0)$ %
to be the difference
between the average proportion of time subjects in the treatment group (with $M_1$ servers) spend in the desirable state versus the control group (with no servers):
\begin{align}
    \label{eq:treatment_effect}
    \theta^{ss}(M_1,N_1,N_0) 
    &= [1 - \E[x_{M_1,N_1}(u,t)]] - [1 - \E[x_{0,N_0}(u,t)]] \\
    &= [1 - \E[x_{M_1,N_1}(u,t)]] - [1 - \frac{\lambda}{\lambda+\tau}].
\end{align}
Note that this expression depends on the number of treatment group servers $M_1$ and treatment group subjects $N_1$, but not the number of control group subjects $N_0$. The capacity constraints in the treatment group induce dependencies between the treatment subjects, affecting the length of time an individual subject spends in the undesirable state, whereas subjects in the control group evolve independently. To emphasize the independence of $\theta^{ss}(M_1, N_1, N_0)$ and $N_0$, from here on out we write $\theta^{ss}(M_1, N_1)$.

Using the language of queueing theory, the treatment effect defined above can be equivalently expressed using the expected queue length, including subjects in service, $\overline{K}_{M_1,N_1}$ in the treatment arm of the trial (see SI \ref{ssec:methods_experiment_setup}, Lemma \ref{lemma:prop_verified_to_queue_length}). %
\begin{equation}
\label{eq:steady_state_queue_length}
    \theta^{ss}(M_1,N_1) = [1 - \frac{\overline{K}_{M_1,N_1}}{N_1}] - [1 - \frac{\lambda}{\lambda+\tau}]. 
\end{equation}
This characterization of  $\theta^{ss}(M_1,N_1)$ as a function of queue length will allow us to leverage results on queueing behavior to analyze the experiment outcomes.

A natural estimator for the treatment effect is the \textit{difference-in-means estimator} that compares the empirical mean of the proportion of time in the desirable state, between the treatment and control group. 
We define the difference-in-means estimator $\hat{\theta}(M_1, N_1, N_0, T)$ as:
\begin{align}\label{e:estimator}
    \hat{\theta}(M_1, N_1, N_0, T)  &=
    \left[\frac{\sum_{u: Z_u=1}1 - Y_{M_1,N_1}(u,T)}{N_1}\right]\\ 
    &- \left[ 
    \frac{\sum_{u: Z_u=0} 1- Y_{0, N_0}(u,T) }{N_0} \right], \nonumber
\end{align}
where $Y_{M,N}(u,T)$ represents the proportion of time subject $u$ spends in the undesirable state over the observation period $T$. We note that in general, as $T\to\infty$ this estimator converges to the true treatment effect $\theta^{ss}(M,N)$ only when $M=M_1$ and $N=N_1$ (SI \ref{si:ssec:clt}, Corollary \ref{cor:consistency}).

\subsection{Theoretical Analysis of Service Intervention RCT Model} \label{ss:analysis}

To study the estimator \eqref{e:estimator}, we need to characterize both its expected value and its variance. However, due to the complexity of service intervention systems, and the dependencies both within subjects over time and between subjects competing for limited capacity, exact analysis is difficult %
for systems with finite $N_1, N_0, M_1$, and $T$. 
Thus we use two asymptotic frameworks to provide tractable approximations of the finite system behavior.

The Central Limit Theorem provides asymptotic results for the estimator of an RCT with a fixed number of servers $M_1$ and subjects $N_1$ and $N_0$, as the observation time $T \to \infty$. This approach characterizes the sampling distribution of the treatment effect estimator and motivates variance approximations used for calculating the power of the test. 

The fluid model approximation analyzes the system behavior of the treatment group with $M_1$ servers and $N_1$ subjects as both $M_1, N_\to \infty$, while maintaining a fixed subject to server ratio. This analysis allows us to approximate the treatment effect $\theta^{ss}(M_1,N_1)$ and the mean of the estimator $\hat{\theta}(M_1, N_1, N_0, T)$ with interpretable, tractable expressions. %

\subsubsection{Central Limit Theorem}
\label{sss:clt}

In service intervention RCTs, the main challenges deriving a CLT are the dependence, at any given time, of a subject's current state with their future state in both treatment and control groups and the added complication of dependence across subjects in the treatment group. 
Thus, we do not have independent and identically distributed samples, a condition typically assumed to guarantee the existence of a CLT. However, leveraging queueing theory results from \cite{whitt1992asymptotic}, we can establish a CLT for service intervention RCTs. The proof is included in SI \ref{si:sec:theoryresults}.

\begin{theorem}[Central Limit Theorem]
\label{thm:clt}
Let $T\to\infty$ for $\hat{\theta}(M_1, N_1, N_0, T)$ as defined in [\ref{e:estimator}]. Then 
\begin{equation*}
T^{1/2} \cdot \left(\hat{\theta}(M_1, N_1, N_0, T)  - (\mu_1 - \mu_0) \right) \xrightarrow{d} N\left(0, \ 
\tilde{\sigma}_1^2 + \tilde{\sigma}_0^2\right). 
\end{equation*}

where $\xrightarrow{d}$ denotes convergence in distribution and 
\begin{align*}
\mu_1 &= 1 - \overline{K}_{M_1,N_1}/N_1, \\
\mu_0 &= \frac{\tau}{\lambda + \tau},\\
\tilde{\sigma}^2_1 &= \frac{2}{N_1^2} \sum_{j=0}^{N_1-1} \frac{\left[ \sum_{i=0}^j (i - \bar{K}_{M_1, N_1})\pi_{M_1,N_1}(i) \right]^2}{(N_1-j) \pi_{M_1,N_1}(j)}, \\
\tilde{\sigma}_0^2 &= \frac{1}{N_0} \cdot \frac{2 \lambda \tau}{(\lambda + \tau)^3}.
\end{align*}
\end{theorem}

Motivated by Theorem \ref{thm:clt}, for the remainder of this section, we use the following approximations for the mean and variance of the estimator: 
\begin{equation}
\label{eq:asymptotic_mean_approximation}
    \hat{\theta}(M_1, N_1, N_0, T) \approx  \left[ 1 - \overline{K}_{M_1,N_1}/N_1 \right] 
    - \frac{\tau}{\lambda + \tau}.
\end{equation}
\begin{equation}
\label{eq:asymptotic_var_approximation}
    \Var\left(\hat{\theta}(M_1, N_1, N_0, T)\right)
    \approx 
    (\tilde{\sigma}_1^2 + \tilde{\sigma}_0^2 )/ T.
\end{equation}
Simulations suggest that the above approximations are close to actual outcomes  (SI \ref{si:sec:theorynumerics}).

\subsubsection{Fluid Model Approximation}
\label{sss:fluid_model_main_text}

We construct a continuous time fluid approximation \cite{whitt2006fluid} of the system by replacing the (stochastic) evolution in the Markov chain of the number of participants $X_{M_1,N_1}(t)$ in the undesirable state with the (deterministic) system of a fluid mass of participants $z(t)$ in the undesirable state, governed by a system of differential equations. In SI \ref{si:sec:theoryresults}, we show that this fluid model has a unique equilibrium point, which we can characterize as an explicit function of the model parameters. 
This approach to analyzing the system results in close approximations of the finite, stochastic model steady state in large systems as well as simpler analytical expressions. 

We find, in the fluid model, that subject outcomes depend on the ratio $N_1/M_1$, allowing us to define a fluid limit treatment effect $\theta^*(N_1/M_1)$.
For systems with large enough $M_1$ and $N_1$, we can use the approximation $\theta^*(N_1/M_1)\approx \theta^{ss}(M_1,N_1)$. See SI \ref{si:sec:theorynumerics} for a numerical evaluation of the quality of the approximation.%

Importantly, the fluid model shows that the steady state equilibrium for the mass of participants in the undesirable state follows a threshold behavior, depending on whether the ratio of participant-to-servers is less than or greater than $r$. This will translate to a threshold behavior in the treatment effect, as we will show in \S\ref{ss:treatmenteffect}.

\section{Main Results}
\label{s:mainresults}
We utilize the model and theory developed in \S\ref{s:model} to highlight the main analytical insight of capacity-dependent treatment effects (\S\ref{ss:treatmenteffect}) and the resulting non-monotonic relationship between sample size and power (\S\ref{ss:power}). These two theoretical insights have numerous implications for trial design, interpretation, replication, and scaling (\S\ref{ss:discussion}). See SI \ref{si:sec:theorynumerics} for details about our numerical calibration.

\subsection{Capacity-Dependent Treatment Effects and the Threshold Relationship Between Sample Size and Treatment Effect}\label{ss:treatmenteffect}

\begin{figure}   
    
    \begin{subfigure}[b]{0.5\textwidth}
    \centering 
    \includegraphics[width=0.7\linewidth]{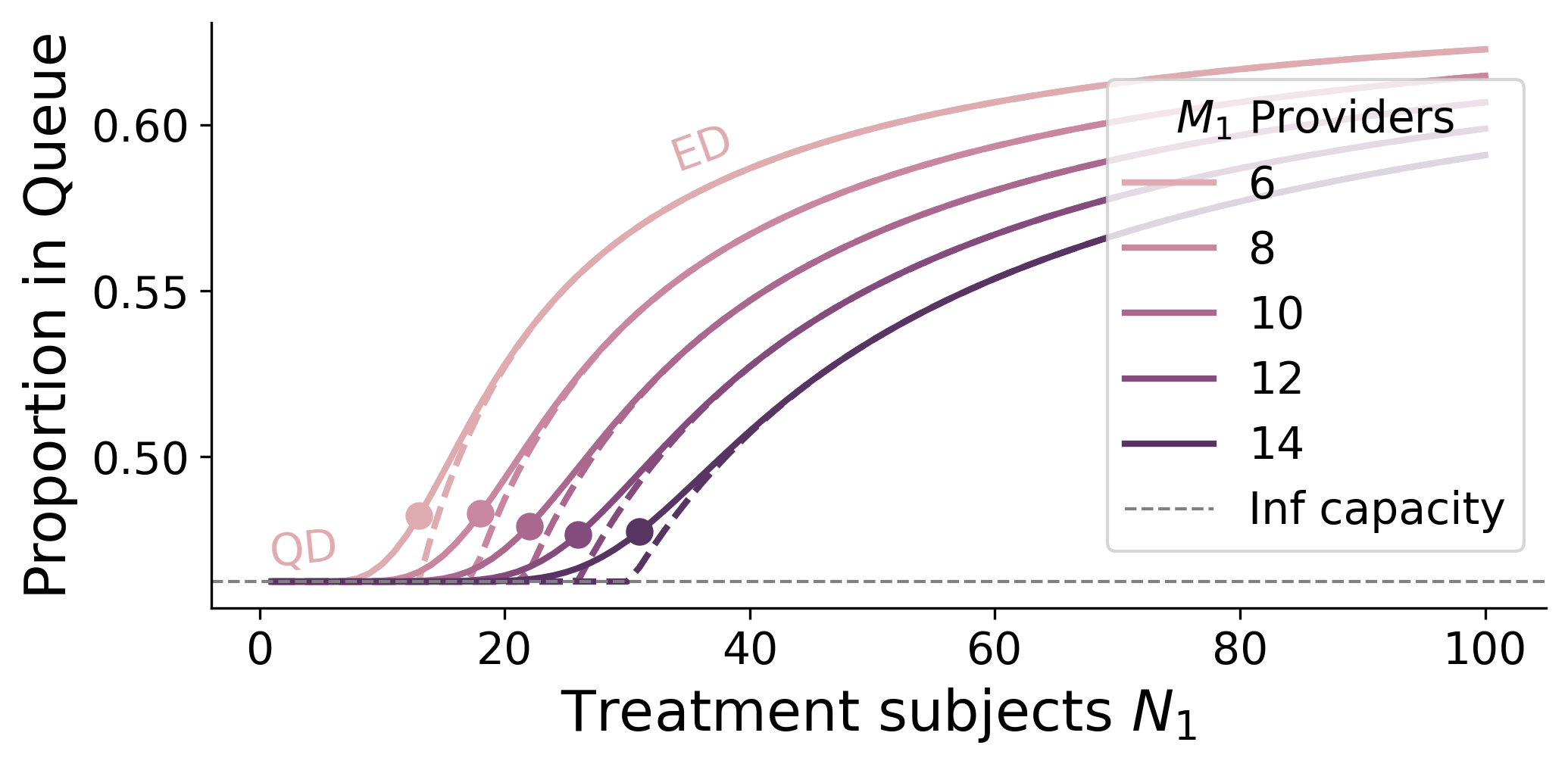}
    \caption{\small Proportion in queue and service}
    \label{fig:numerics_queue_prop_by_M_N}
    \end{subfigure}
    \begin{subfigure}[b]{0.5\textwidth}
    \centering 
    \includegraphics[width=0.7\linewidth]{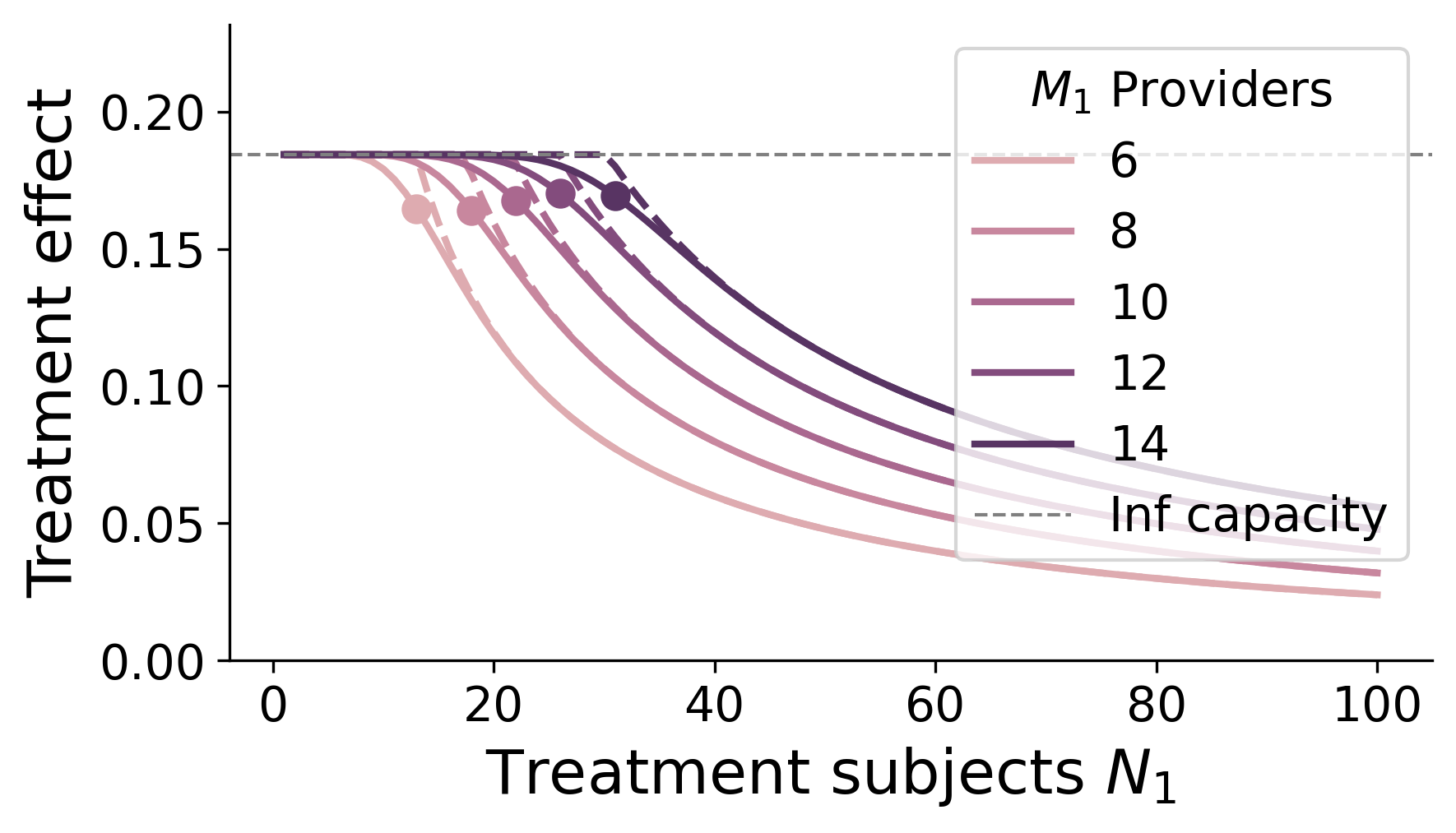}
    \caption{\small Treatment effect}
    \label{fig:numerics_effect_size_by_M_N}
    \end{subfigure}
    \begin{subfigure}[b]{0.5\textwidth}
    \centering
    \includegraphics[width=0.7\linewidth]{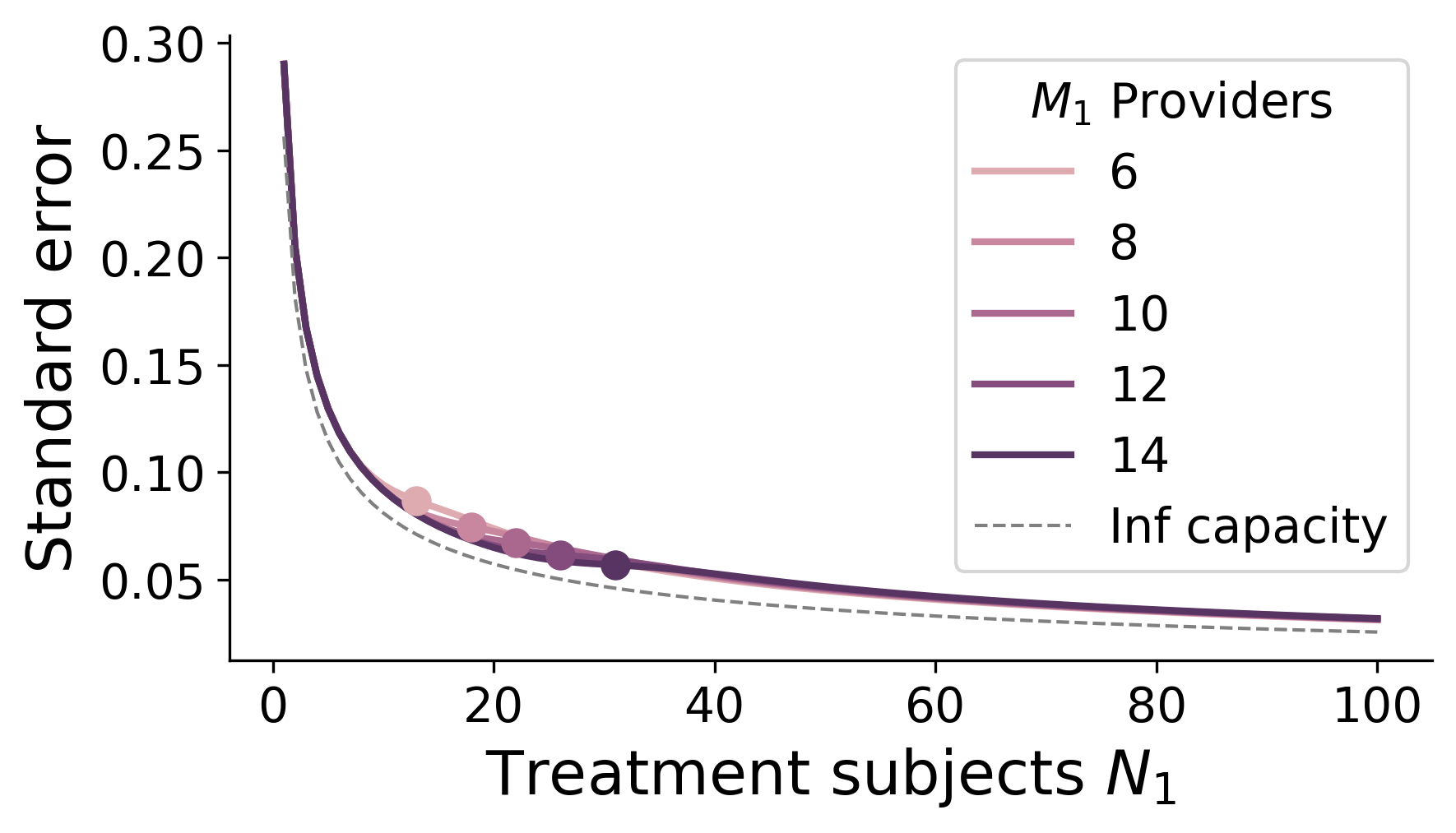} 
    \caption{\small Standard error}
    \label{fig:numerics_standard_error}
    \end{subfigure}
    \begin{subfigure}[b]{0.5\textwidth}
    \centering
    \includegraphics[width= 0.7\linewidth]{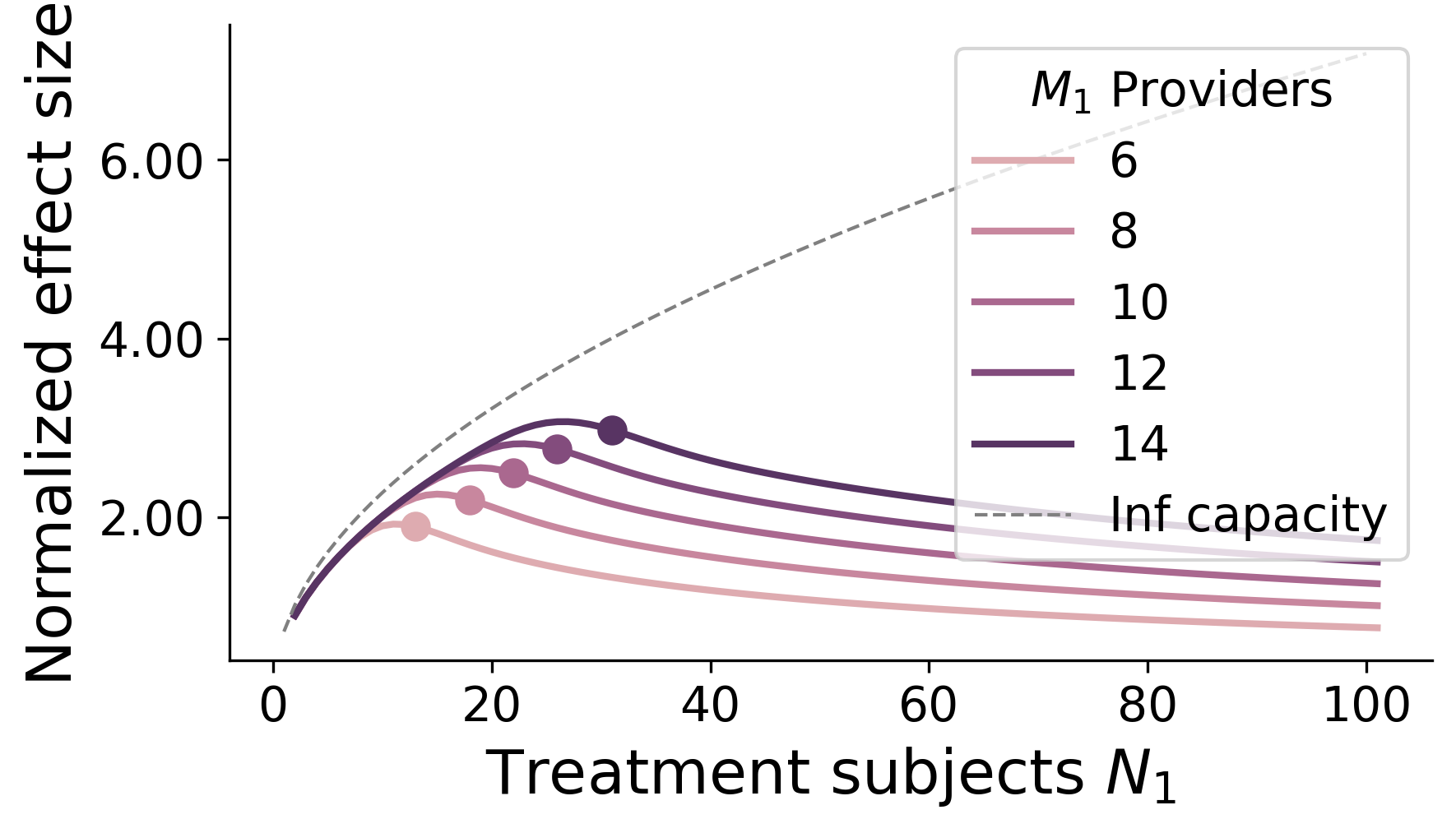} 
    \caption{\small Normalized effect size}\label{fig:numerics_normalizedeffectsize}
    \end{subfigure}
    \begin{center}
        
    \begin{subfigure}[b]{0.5\textwidth}
    \centering
    \includegraphics[width=0.7\linewidth]{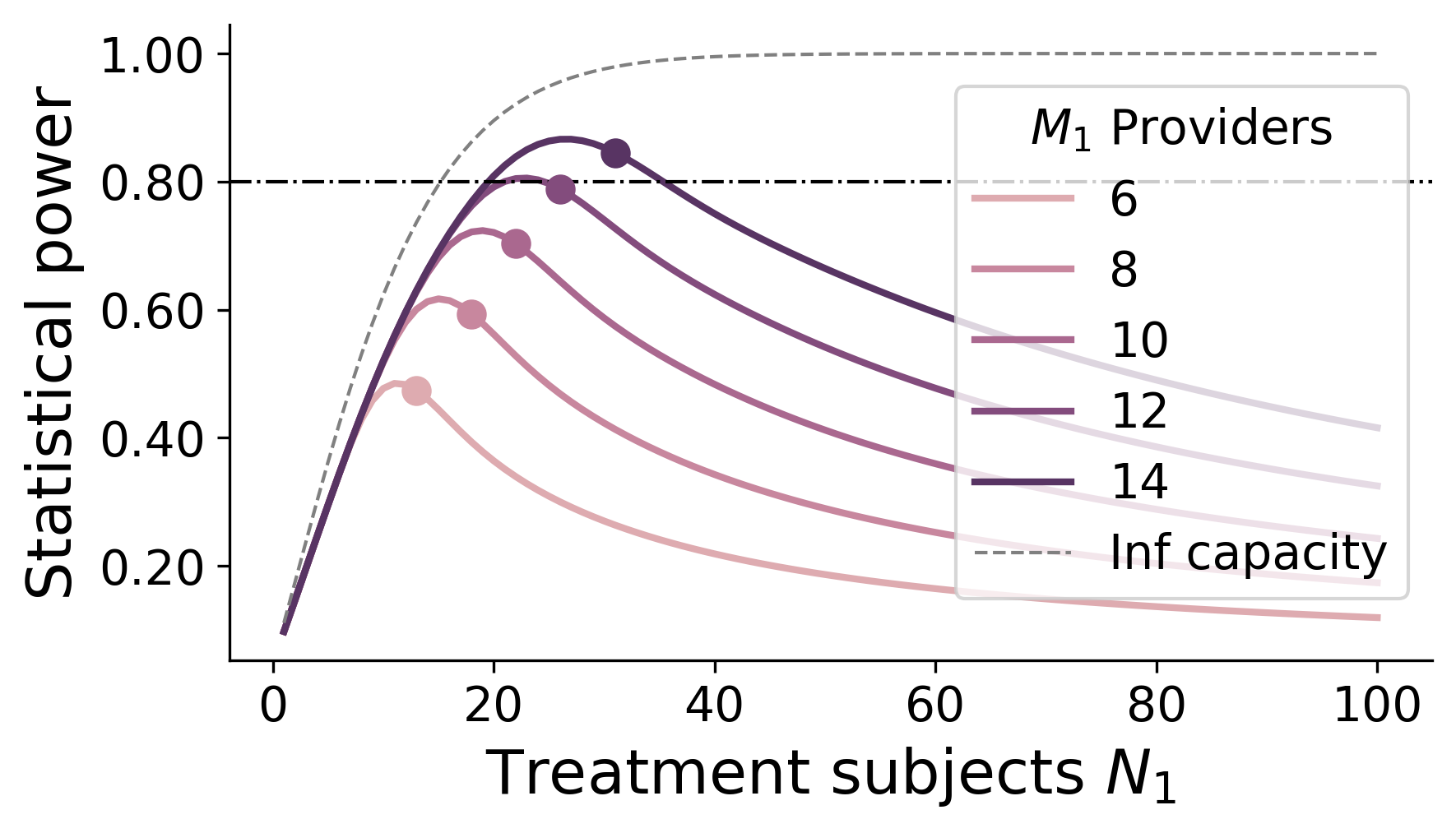}
    \caption{\small Statistical power}
    \label{fig:numerics_power}
    \end{subfigure}
    \end{center}

    \caption{Proportion in queue and service, treatment effect, standard error, normalized effect size, and statistical power for various combinations of capacity $M_1$ and sample size, fixing the size of the intervention group $N_1$ equal to the size of the control $N_0$. Dashed grey lines represent values with infinite capacity.
    For each $M_1$, the marker denotes the value of $N_1$ such that $N_1/M_1$ is equal to the critical threshold $r$, with the Quality-Driven regime to the left and the Efficiency-Driven regime to the right.  
    In (\ref{fig:numerics_queue_prop_by_M_N}) and (\ref{fig:numerics_effect_size_by_M_N}), solid lines show quantities from a stochastic model, while dashed lines correspond to deterministic approximations (SI \eqref{eq:ode_z_small}-\eqref{eq:ode_z_large} and  \eqref{eq:mean_field_treatment_effect}), respectively.
    The parameters employed to generate these figures are: 
    $\lambda=0.63, \tau=0.34, \mu=5.5, p=0.07$, leading to a critical threshold of $r=2.15$ (see SI \ref{si:sec:theorynumerics} for discussion of parameter calibration).
    }
    \label{fig:numerics}
\end{figure}

The approximation for the mean  (\eqref{eq:asymptotic_mean_approximation}) allows us to study how the estimated treatment effect size changes with respect to experiment design parameters. Fig.  \ref{fig:numerics_effect_size_by_M_N} plots in solid lines the effect size $\theta^{ss}(M_1, N_1)$ as a function of capacity levels $M_1$ and number of treatment subjects $N_1$, where we fix $N_1=N_0$. Recall that the effect size is independent of $N_0$.

Fig. \ref{fig:numerics_effect_size_by_M_N} shows that effect size decreases in the number of subjects $N_1$ and increases in the number of providers $M_1$. 
This observation is 
intuitive, as we expect the wait time for a call to increase in the number of subjects and decrease in the number of providers, thus changing the ``operational dosage'' of the intervention. Indeed, we see these effects reflected in the the average queue length, shown in Fig. \ref{fig:numerics_queue_prop_by_M_N}.  For each $M_1$, the effect size has an upper bound achieved at small $N_1$ and decreases to 0 as $N_1\to\infty$.

However, Fig.  \ref{fig:numerics_effect_size_by_M_N} also shows a phase transition as the number of enrolled subjects increases. For a fixed capacity, the treatment effect is constant in the number of subjects for small $N_1$, but decreases non-linearly when the number of subjects grows past a threshold, dependent on $M_1$. This pattern reflects the presence of an underlying queueing mechanism; such systems are well known to exhibit phase transitions as load increases. To formalize this connection with well-established queueing theory results, we use a fluid model approximation to identify a \textit{critical threshold} $r$ in the ratio of participants-to-servers:

\begin{equation}
    \label{eq:critical_ratio}
    r = (\lambda+\tau+\mu p) / \lambda.
\end{equation}
This threshold dictates the relationship between the effect size and the number of participants (see SI \ref{si:sec:theoryresults} for details) as follows:\footnote{This characterization is exact in the fluid model but also provides very similar predictions to \eqref{eq:asymptotic_mean_approximation}, illustrated in Fig. \ref{fig:numerics_effect_size_by_M_N}, where the fluid limit approximations are shown as dashed lines---closely tracking the finite system quantities shown as solid lines.}

\begin{theorem}[Fluid limit treatment effect]
\label{eq:fluid_limit_effect_size}
Fix a participant-to-server ratio $N_1/M_1$ in the treatment group.
The fluid model treatment effect is
\begin{equation*}
\theta^*(\kappa) = 
    \begin{cases}
    \frac{\lambda\mu p}{(\lambda+\tau)(\lambda+\tau+\mu p)} & \text{for } N_1/M_1 \leq r \ \text{(QD Regime)}\\
    \frac{\mu p}{\kappa(\lambda+\tau)} & \text{for } N_1/M_1 > r \ \text{(ED Regime)}
    \end{cases}
\end{equation*}
In particular, 
as the participant-to-server ratio $N_1/M_1$ increases, 
\[
\lim_{N_1/M_1\to\infty} \theta^{*}(N_1/M_1) 
= 0, 
\]
\end{theorem}

In particular, the above result shows two distinct regimes, which correspond to established notions of \textbf{Quality-Driven (QD)} and \textbf{Efficiency-Driven (ED)} regimes in queueing theory \citep{gans2003telephone}.

\begin{itemize}
    \item \textbf{Quality-Driven (QD) Regime ($N_1/M_1 < r$):} When the participant-to-server ratio lies below the critical ratio threshold, treatment effects remain relatively constant as sample size increases. 
    \item \textbf{Efficiency-Driven (ED) Regime ($N_1/M_1>r$):} When the participant-to-server ratio falls above this threshold, treatment effects decrease as sample size increases. 
\end{itemize}

The intuition behind these regimes is that $N_1/r$ is the \textit{offered load} on the system, or the expected demand for service (given the parameters $\lambda, \tau, \mu, \text{and } p$). Whether $M_1$ is smaller or larger than the offered load dictates the amount of extra capacity the system has and whether subjects receive prompt outreach. In the QD regime, there is excess capacity and additional subjects can be served without increasing wait times. In the ED regime, there is excess demand and additional subjects increase wait time, reduce the operational dosage, and reduce intervention effectiveness. 

Fig. \ref{fig:numerics}
demarcates the point where $N_1/M_1$ is equal to the critical threshold $r$. The QD regime lies to the left and the ED regime lies to the right.
As predicted by the fluid limit analysis, the effect size is constant in the QD regime and decreasing with $1/N_1$ in the ED regime.   

While the above results are derived using a deterministic model of large systems, numerics for finite systems (Fig. \ref{fig:numerics}) and simulations (see \S \ref{s.keheala}) confirm these insights for finite size stochastic systems.

\subsection{Statistical Power is Non-Monotonic in the Number of Subjects}\label{ss:power}
The dependence of the effect size on capacity and sample size---introduced in the previous section---leads to counterintuitive patterns in how readily a true effect can be distinguished from a null effect.

First, we note that the standard definition of power is not appropriate for a service intervention RCT.
The standard approach to power analysis for RCTs is to first assume an exogenous minimum detectable effect (MDE), usually based on some theory about the protocol describing the intervention---regardless of operational implementation. Subsequently, a minimum sample size is determined so that an effect of that predetermined size is detected with a high probability (often 80\%) and a low false positive rate (often 5\%), given the expected variability in the outcome variable. 
However, this framework is not appropriate for service interventions, since %
the effect size $\theta^{ss}(M_1,N_1)$ itself
changes with the sample size $N_1$. When effect sizes vary with sample size, applying the conventional definition of power at a fixed MDE can lead to misleading conclusions about experimental performance (as we will discuss in \S \ref{ss:discussion}).

To address this dependency, we define power as the probability of detecting an effect at least as large as the true effect size at the given $N_1$ and $M_1$. %
Mathematically, this corresponds to:
\begin{equation}
    Pr(\text{reject} \ H_0 | \text{true effect} = \theta^{ss}(M_1,N_1)).
\label{eq:power_def}
\end{equation} %

Under this definition, it is clear that statistical power depends on sample size through both the treatment effect and its variance. 
We evaluate the expression using the CLT approximations given in \eqref{eq:asymptotic_mean_approximation} and \eqref{eq:asymptotic_var_approximation}. (See SI \ref{si:sec:hyp_testing_power_analysis}, \eqref{eq:power_approximate_mde_true_effect_two_sided} for the full expression.) Notably, statistical power is a monotonically increasing function in the normalized effect size, defined below
\begin{equation}
\label{eq:normalized_effect_size}
    \frac{\theta^{ss}(M_1, N_1)}{\sqrt{\tilde{\sigma}_1^2  (M_1, N_1)/T
    +
    \tilde{\sigma}_0^2 (N_0)/T}}.
\end{equation}

Fig.  \ref{fig:numerics} shows the standard error of the estimator $\hat{\theta}^{ss}(M_1,N_1,N_0,T)$  (Fig. \ref{fig:numerics_standard_error}), the effect size normalized by its standard error  (Fig. \ref{fig:numerics_normalizedeffectsize}), and the resulting power (Fig. \ref{fig:numerics_power}). 
The main observation is that for any level of fixed capacity, statistical power is first increasing and then decreasing in sample size. 
This is in stark contrast with the insights of traditional power analysis, which assumes that statistical power increases with the number of subjects. For service interventions, increasing the number of subjects beyond a certain threshold will increase the cost of the experiment, but with no benefit to statistical power.

The mechanism behind this non-monotonicity is again related to the queueing behavior in the QD and ED regimes.
If the intervention had infinite capacity (dashed lines in the figures), then the effect size would be constant in $N_1$ (Fig. \ref{fig:numerics_effect_size_by_M_N}) and the standard error would be decreasing in $N_1$ (Fig. \ref{fig:numerics_standard_error}), resulting in increasing normalized effect size (Fig. \ref{fig:numerics_normalizedeffectsize}) and power (Fig. \ref{fig:numerics_power}). 
The above behavior holds true for a fixed, finite capacity as well in the QD regime, where the effect size is constant in $N_1$. However, in the ED regime, the effect size decreases in $N_1$ and can decrease faster than the standard error, leading to a decrease in the normalized effect size and overall power. Indeed Fig. \ref{fig:numerics_power} shows that statistical power is generally increasing to the left of the critical threshold (QD regime) and decreasing to the right of the 
critical threshold (ED regime). Importantly, for a fixed $M_1$, the $N_1$ that maximizes statistical power is near the critical threshold.

Furthermore, in \S\ref{sss:suboptimal_naive_power_analysis} we show how naive strategies for power analysis and experiment planning can lead to either under-powered experiments or excessive resource costs, due to this non-monotonic relationship between sample size and power. We show that established staffing heuristics from queueing theory (\S\ref{sss:sqrt_staffing}) can aid in designing high powered experiments with lower resource requirements.

\subsection{Implications for RCTs} \label{ss:discussion}
Our main findings (\S\ref{s:mainresults}) of capacity-dependent treatment effects and non-monotonicity of statistical power in sample size  have important implications for  interpretation (\S\ref{sss:multiple_estimands}), failure modes (\S\ref{sss:failure_modes}), and design (\S\ref{sss:suboptimal_naive_power_analysis}-\ref{sss:sqrt_staffing}) of service intervention RCTs.

\subsubsection{Multiple Estimands of Interest}\label{sss:multiple_estimands}

Our modeling framework allows us to distinguish between various potential estimands of interest, in the context of service intervention trials. This distinction helps to illustrate various challenges associated with the interpretation of outcomes, the replication of prior results, and the scaling up of service interventions. 

Consider an RCT with $M_1$ servers, $N_1$ treatment subjects, and $N_0$ control subjects. While researchers typically assume they are estimating ``the'' treatment effect of the intervention, our framework shows that there is a family of treatment effects. In fact, depending on the experimenter's objective, there may be multiple estimands of interest, including:
\begin{itemize}
    \item \textbf{The trial-implemented effect} $\theta^{ss}(M_1, N_1)$: The effect as implemented in the trial with $M_1$ servers treating $N_1$ subjects.
    \item \textbf{The capacity-unconstrained effect} $\theta^{ss}(\infty, N_1)$: The effect of the intervention if service capacity was not a limitation. This quantity is important since service intervention protocols are often conceived in terms of individual subject trajectories---where a given behavior immediately triggers the next step in service. %
    \item \textbf{The full-sample deployment effect} $\theta^{ss}(M_1, N_1+N_0)$: The effect of rolling out the intervention to all RCT participants, without changing the capacity. This is relevant if an experimenter splits their entire user base into a treatment and control group for the purposes of experimentation, but is ultimately interested in offering the intervention to everyone. 
    \item \textbf{The scaled deployment effect} $\theta^{ss}(M', N')$: The treatment effect if the intervention is scaled up so that the intervention is offered to more patients using more servers.
\end{itemize}

Clearly distinguishing the above estimands from each other is important when interpreting the outcomes of service intervention trial. The experimenter using standard RCT practices can only estimate the trial-implemented effect  $\theta^{ss}(M_1, N_1)$, which our results demonstrate can differ substantially from other treatment effects of interest. 

\subsubsection{Failure Modes for Experimentation}
\label{sss:failure_modes}
The ambiguities in the multiple estimands lead to three distinct failure modes for service intervention experimentation. 

\begin{itemize}
    \item \textbf{False negatives in identifying beneficial interventions.}
    If the experimenter observes a low or statistically insignificant treatment effect estimate $\hat{\theta}^{ss}(M_1, N_1, N_0, T)$, the correct conclusion is not that the intervention itself `has no impact' but rather that the specific implementation of the intervention (i.e., the capacity $M_1$ relative to the sample size $N_1$) has little or no impact. In other words, ${\theta}^{ss}(M_1, N_1)$ is small, but $\theta^{ss}(\infty, N_1)$ might still be large, indicating that the protocol (at the level of an individual subject) might be effective but sensitive to implementation.
    \item \textbf{Replication.} 
    As suggested in the `Heterogeneity Revolution'  \cite{bryan2021behavioural}, when multiple trials of a single intervention produce different effect size estimates, the issue may not be researcher misconduct, as previously hypothesized, but rather heterogeneity across experiments that is unaccounted for, such as changes in the study population or  intervention. Our work provides another such mechanism through which heterogeneity can arise in service intervention RCTs--when studies have different levels of capacity or ``operational dosage,'' the trials will predictably lead to different effect sizes. 
    \item \textbf{Scaling.} A recurring issue in policy-making is that many interventions are initially promising but fail to perform at scale \cite{list2022voltage,list2024optimally}. Capacity constraints again provide another mechanism that leads to this `voltage drop.' An experimenter may observe a large and statistically significant $\hat{\theta}^{ss}(M_1, N_1, N_0,T)$ in an RCT and wrongly assume that the same treatment effect holds for large-scale implementation (i.e., assumes that $\theta^{ss}(M_1, N_1)=\theta^{ss}(M', N')$, for some $M'\ne M_1$ and $N'\ne N_1$. Our work suggests predictable drops in the treatment effect as the number of subjects increases, which must be factored into policy-making decisions.
\end{itemize}

\subsubsection{Suboptimal Performance of Na\"ive Power Analysis Methods}
\label{sss:suboptimal_naive_power_analysis}
An implication of the non-monotonicity of power with respect to sample size (\S\ref{ss:power}) is that traditional power analysis methods
may, paradoxically, lead to lower power. 

Consider a standard procedure in which the experiment designer first runs a small scale \textit{pilot study}, conducts a \textit{power analysis} using the outcomes from the pilot study, and then uses the results to decide how many users to recruit in the full study. Let $M_1^p$, $N^p_1$, and $N^p_0$ denote the number of servers, treatment subjects, and control subjects in the pilot, respectively. For clarity in the following discussion, we fix $N^p_1=N^p_1$ and $N_1=N_0$, although the results hold for any fixed ratio of treatment to control subjects.

We consider two naive protocols for designing the large scale experiment. For more details see SI \ref{si:sec:hyp_testing_power_analysis}.
\begin{itemize}
    \item \textbf{Naive Policy 1--No $M_1$ Scale-up (Naive 1):} The experimenter is not aware of operational dosage effects. The experimenter estimates the variance across individuals in the pilot study and, under the assumption that all subjects are independent, calculates the minimum $N_1$ needed for the experiment. The experimenter does not increase capacity and sets $M_1 = M^p_1$. 
    \item \textbf{Naive Policy 2-- Proportional $M_1$ Scale-up (Naive 2).} The experimenter is aware that treatment effects depend on capacity. The experimenter chooses the same $N_1$ as Naive Policy 1, but proportionally increases $M_1$. 
\end{itemize}
Since power decreases in sample size once sample size is large enough, \textbf{Naive Policy 1--No $M_1$ Scale-up} can result in a large scale experiment that actually has \textit{less} power to detect an effect than the original pilot experiment, thus wasting experiment resources.

In numerics (SI \ref{si:sec:hyp_testing_power_analysis}) and RCT-calibrated simulations (\S\ref{sss:num:staffing}), we find that \textbf{Naive Policy 2--Proportional $M_1$ Scale-up} can alleviate this issue and achieve the desired power. 
However, this policy may require large $M_1$, $N_1$, and $N_0$ than necessary. In the next subsection, we present a heuristic that leverages the queueing structure to provide a high-powered design with fewer resources.

\subsubsection{Improved Power Analysis through Square Root Staffing}\label{sss:sqrt_staffing}

Motivated by the observation in \S\ref{ss:power} that, for a given $M_1$, power is often increasing in the small $N_1$ (QD) regime and decreasing in the large $N_1$ (ED) regime, we define a new power analysis leveraging an existing queueing-theory heuristic that targets the region between the QD and ED regimes.

The ``Square Root Staffing Rule''\citep{halfin1981heavy}
is a method to choose, for a given $N_1$, the number of providers $M_1$ in order to order to balance the benefits of the QD regime (high service quality) with those of the ED regime (low resource requirements). The Square Root Rule targets an intermediate region known as the \textbf{Quality-Efficiency-Driven Regime}.

For a given $N_1$, the existing ``Square Root Staffing Rule'' chooses $M_1$ that satisfies
\begin{equation}
    \label{eq:sqrt_staffing_rule}
    M_1^*(N_1) := \left\lceil  N_1/r + \gamma \sqrt{N_1}\right\rceil,
\end{equation}
where $r$ is the critical threshold defined earlier and $\gamma$ is a constant that depends on the desired probability of delay. 
The formula for the number of servers has two components; $M_1$ to match the offered load on the system $N_1/r$ (the amount of traffic the queue is expected to handle) and a safety buffer proportional to $\sqrt{N_1}$. %
For mathematical justification, see \cite{halfin1981heavy, shortle2018fundamentals}.

Building on the existing heuristic, we propose a \textbf{Square Root Staffing Informed Power Analysis} to support experimental design. We start by observing that the Square Root Staffing Rule gives a sequence of $(M_1, N_1)$ pairs that satisfy \eqref{eq:sqrt_staffing_rule}. Using the expressions in Theorem \ref{thm:clt}, we propose that the experimented estimates the treatment effect, variance, and statistical power for each of those $M_1$ and $N_1$ pairs and chooses the combination that entails the lowest number of servers and subjects.

We compare the outcomes of following this policy to those of the naive policies in calibrated simulations (\S\ref{sss:num.squareroot}) and numerics (SI \ref{si:sec:hyp_testing_power_analysis}). We find that this approach is likely to result in the maximum treatment effect, which, in turn, allows for meeting power requirements with minimal service capacity.

\section{Application to a TB Treatment Adherence Trial}\label{s.keheala}

To illustrate the implications of our insights for real-world service intervention trials, we leverage data from a completed RCT evaluating the impact of the digital health service \textit{Keheala}. Using data from this service intervention trial, we fit a double machine learning model to predict patient behavior, and simulate counterfactual trial outcomes at varying levels of capacity and sample size. 

Importantly, these simulations allow us to relax some of the assumptions of our theoretical models as well as to incorporate both patient-level heterogeneity and non-Markovian behavior.

\subsection{The Keheala Intervention}\label{ss:keheala}
Tuberculosis remains a global health emergency with over 10.8 million new infections estimated in 2023 \citep{who2024global}. Despite a known cure \citep{mrc1948streptomycin}, mortality rates remain high, particularly in resource constrained settings such as India and sub-Saharan Africa, partly driven by sub-optimal adherence to treatment plans \citep{who2024global}. 

Keheala addresses this adherence problem through a digital health mobile phone application that patients interact with during the course of their treatment, typically for six months. Patients self-verify their daily medication adherence and, if they fail to do so, receive a personalized message or phone call (the capacity-constrained service) from a \textit{support sponsor}---a service provider who gives advice for getting back on track. Patients become eligible for outreach after consecutive days of non-verification. %
In addition to the service component, patients also receive automated reminders and access to educational content.

This service intervention has been tested in two separate RCTs. The first trial was conducted in 17 clinics around Nairobi in 2016 and 2017, enrolling approximately 1190 patients (with an equal split between the treatment and control arms) and found that patients receiving the service had significantly better health outcomes, relative to the standard of care \citep{Yoeli_2019}. The second trial was more expansive in scope and geography, enrolling 15,500 patients in 4 arms \citep{nct04119375}. Our simulation model (described in the next section) is trained using data from the treatment arm (which included approximately 5000 patients) of the second trial, which offered the full Keheala service. 
While treatment success is the ultimate outcome of interest for Keheala, we will focus on the intermediate metric of the number of days in the self-verified adherence state.%

\subsection{Calibrating Simulations to the Keheala Trial} \label{ss:simsetup}
We calibrate our simulations using machine learning models trained on 416,071 patient-days 
from the service intervention arm of the second Keheala trial. Using double machine learning \citep{chernozhukov2018double} (a similar approach as \cite{baek2023policy}), we predict patient next-day adherence behavior as a function of their history and the platform's actions, isolating the causal impact of a single instance of support sponsor outreach. Using this framework, we predict adherence both with and without outreach, leading to AU-ROC scores of 0.92 (no outreach) and 0.79 (with outreach), and an estimated average treatment effect of a single instance of support sponsor outreach of 0.07 increase in next-day self-verification probability. 

For each simulated trial, we randomly draw patients from the real dataset, initialize their first ten days using observed behavior, and then simulate counterfactual trajectories for the remaining 170 days ($T=180$ total). Control arm patients receive no intervention; treatment arm patients 
who fail to adhere become eligible for outreach, delivered according to daily capacity $C$ drawn from a Poisson distribution. For each trial we estimate the treatment effect, standard error, t-statistic, and statistical power. 
We simulate %
25 replications for each combination of enrolled patients ($N_0$ and $N_1$, varying from 20 to 1300) and support sponsors ($M_1$, varying from 1 to 80) and present the averages  across simulation runs in Fig. \ref{fig:keheala_resultsplots}. %
Full calibration and simulation details are included in SI \ref{si:sec:Keheala}.

\subsection{Main Numerical Results}\label{ss:simresults}
In this section, we will present our main numerical results, which are depicted graphically in Fig. \ref{fig:keheala_resultsplots}. As in \S \ref{s:mainresults}, where we discussed the main theoretical results, we will first focus on how the combination of subjects and capacity affects the service rate (\ref{sss:num.treatmenteffect}). Second, we show how the treatment effect behavior impacts the statistical power of the RCT (\ref{sss:num.power}). These findings align with the insights from the theoretical model, despite the added heterogeneity and non-Markovian behavior.

\subsubsection{Threshold Relationship of Sample Size on Treatment Effect}\label{sss:num.treatmenteffect}

\begin{figure*}[h]
    \begin{subfigure}[b]{0.5\linewidth}
        \centering
        \includegraphics[width=.8\linewidth]{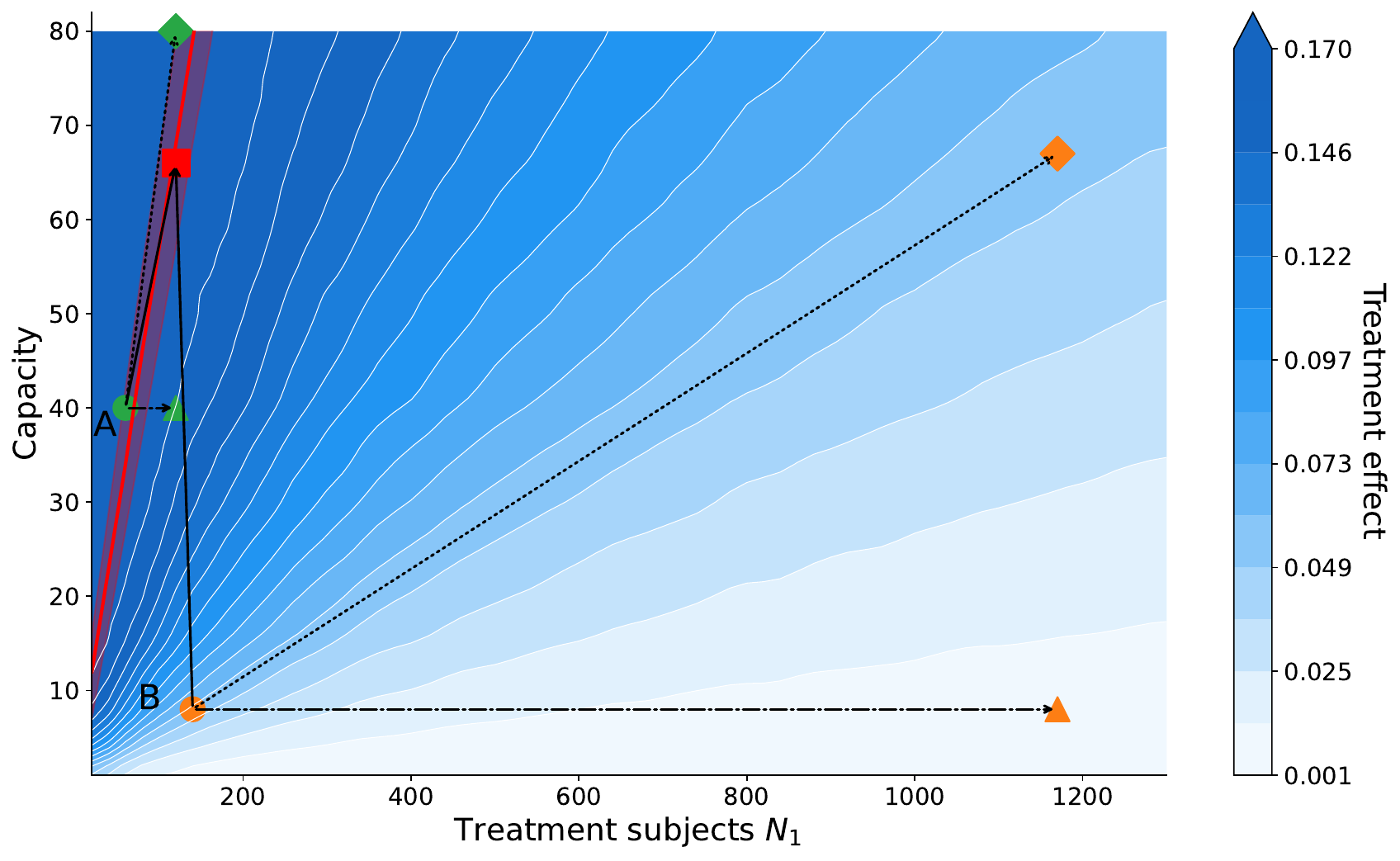} 
        \caption{Treatment Effect }
        \label{fig:keheala_resultsplots_effect_contour}
    \end{subfigure}
    \hfill
    \begin{subfigure}[b]{0.5\linewidth}
        \centering
        \includegraphics[width=.8\linewidth]{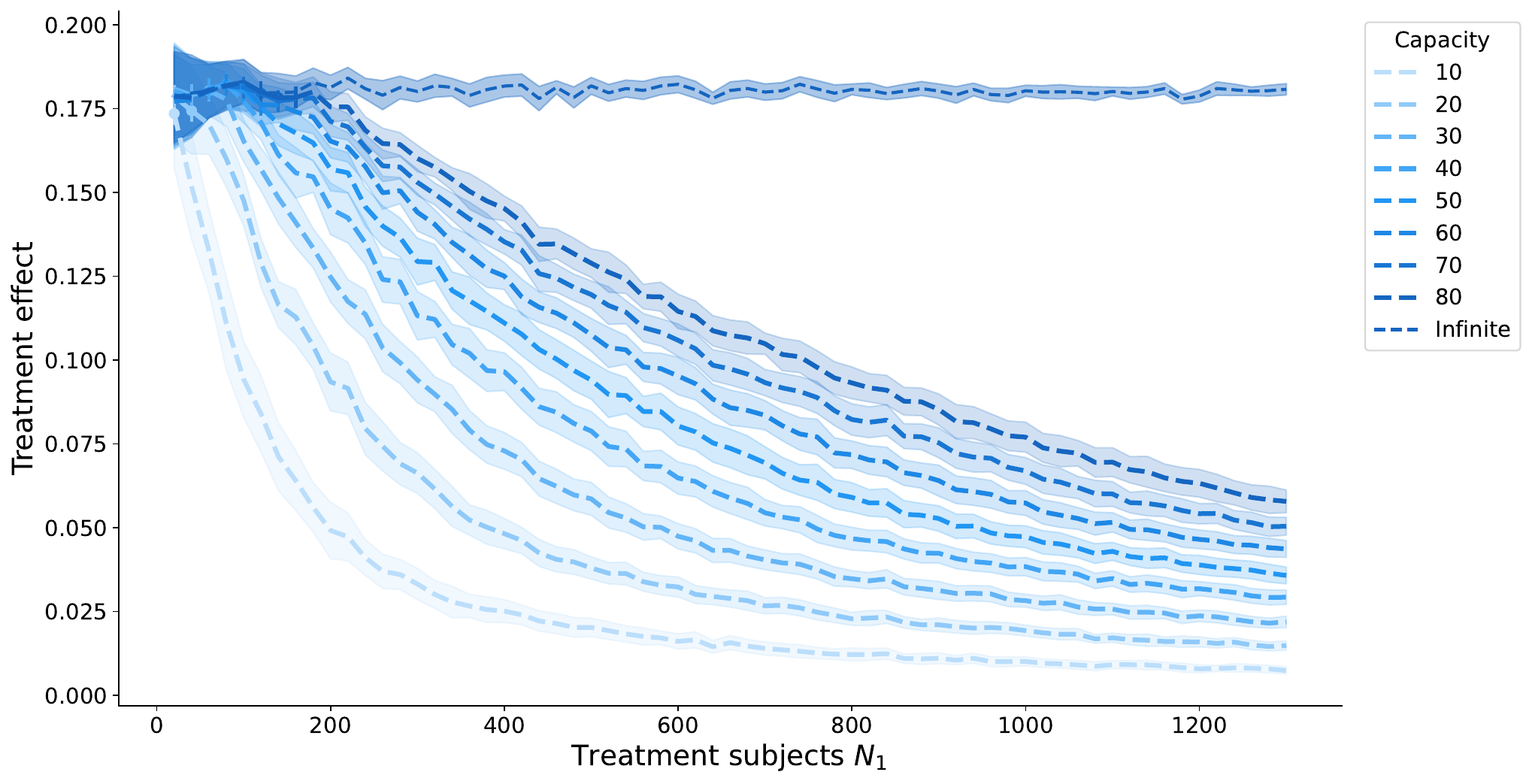}
        \caption{Treatment Effect}
        \label{fig:keheala_resultsplots_effect_line}
    \end{subfigure}

    \begin{subfigure}[b]{0.5\linewidth}
        \centering
        \includegraphics[width=.8\linewidth]{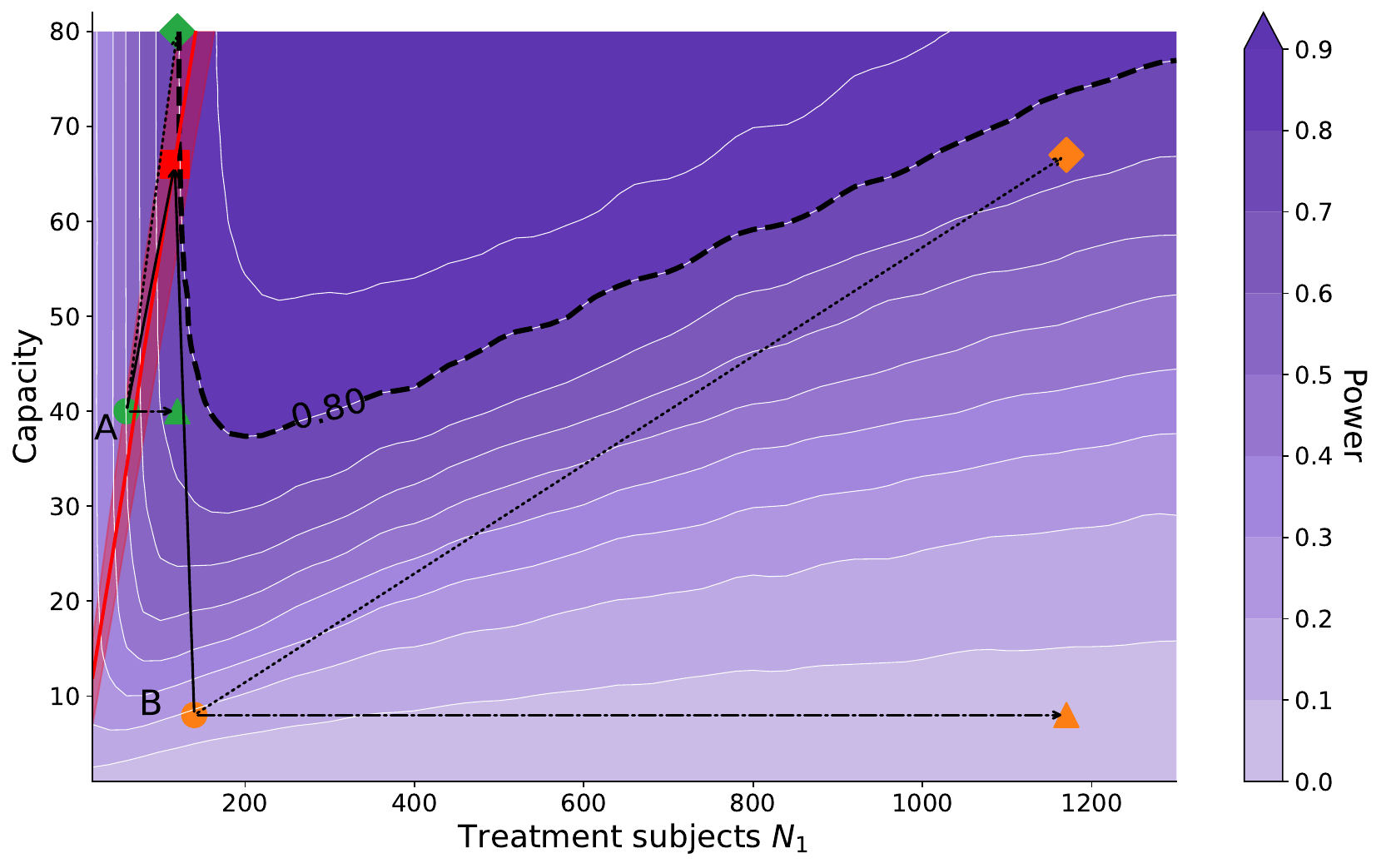}
        \caption{Statistical Power}
        \label{fig:keheala_resultsplots_power_contour}
    \end{subfigure}
    \hfill
    \begin{subfigure}[b]{0.5\linewidth}
        \centering
        \includegraphics[width=.8\linewidth]{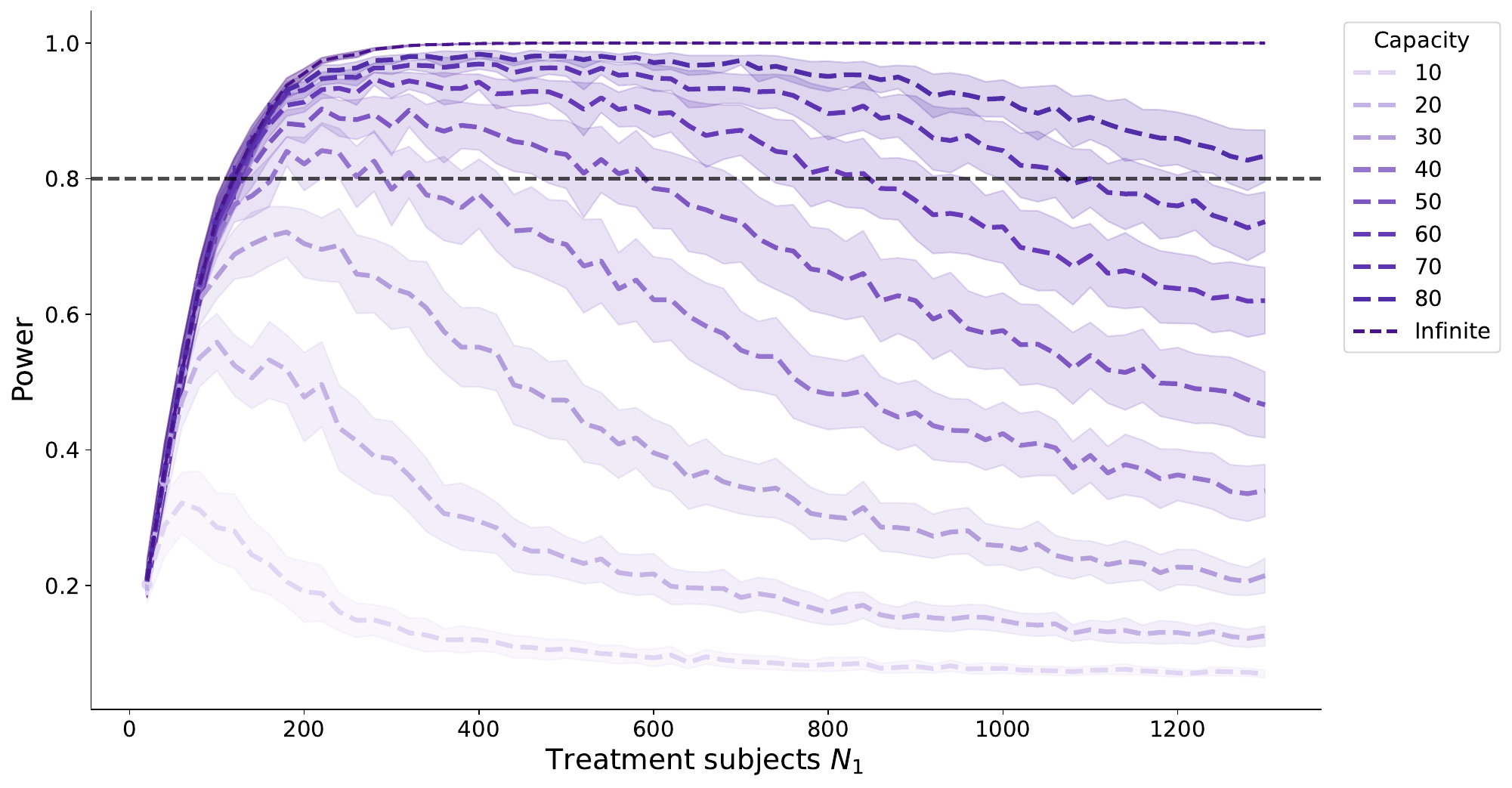}
        \caption{Statistical Power}
        \label{fig:keheala_resultsplots_power_line}
    \end{subfigure}

    \caption{
    Simulations of a service trial, calibrated to the Keheala RCT data. The figure shows the treatment effect (\ref{fig:keheala_resultsplots_effect_contour} and \ref{fig:keheala_resultsplots_effect_line}) and the statistical power  (\ref{fig:keheala_resultsplots_power_contour} and  \ref{fig:keheala_resultsplots_power_line}), for various combinations of patient enrollment and service capacity. The solid red lines in panels \ref{fig:keheala_resultsplots_effect_contour} and \ref{fig:keheala_resultsplots_power_contour} denote the threshold $N_1/M_1=r$, which represents the boundary between the quality-driven (above the boundary) and the efficiency-driven (below the boundary) regimes, with the shaded area denoting the square root rule, with $\gamma \in [-1,1]$. 
    In panels \ref{fig:keheala_resultsplots_effect_line} and \ref{fig:keheala_resultsplots_power_line}, the points labeled A and B correspond to pilot study scenarios discussed in \S \ref{sss:num:staffing}, with point A in the Quality-Driven regime and point B in the Efficiency-Driven regime. The  triangles correspond to Naive 1 Policy (No $M_1$ Scale-up), where the experimenter conducts a standard power analysis, assuming constant treatment effects, and scales up $N_1$ but not $N_1$. The diamonds correspond to Naive 2 Policy (Proportional $M_1$ Scale-up), which chooses the same $N_1$ as Naive 1 but scales up $M_1$ proportionally. The red square corresponds to queueing informed Square Root Staffing policy (\ref{sss:num.squareroot}). 
    }
    \label{fig:keheala_resultsplots}
\end{figure*}

Fig. \ref{fig:keheala_resultsplots_effect_contour} and \ref{fig:keheala_resultsplots_effect_line} show the estimated treatment effect for counterfactual versions of the \textit{Keheala} RCT, for various sample sizes and capacity levels. %

The treatment effect is sample-size dependent and 
has distinct behavior in the two regimes, as predicted in \S\ref{ss:treatmenteffect}. Fig. \ref{fig:keheala_resultsplots_effect_contour} shows the critical threshold ($r=1.77$ in our simulations) as a solid red line. To the top left of this threshold, the treatment group systems are in the Quality-Driven regime with ample capacity. In this regime, the treatment effect is stable at 0.18, regardless of the number of providers or subjects,
aligning with the upper bound $\theta^{ss}(\infty, N_1)$, the effect of the proposed intervention with infinite capacity. This region corresponds to the small $N_1$ portion of each line in Fig. \ref{fig:keheala_resultsplots_effect_line}. 
Fig. \ref{fig:keheala_resultsplots_effect_contour} shows that to the bottom right of the threshold, where the treatment group systems are in the Efficiency-Driven regime with low capacity, the treatment effect increases in the capacity and decreases in the number of subjects. This region corresponds to the large $N_1$ portion of each line in Fig. \ref{fig:keheala_resultsplots_effect_line}.

Importantly, as the number of subjects increases, the treatment effect approaches 0; thus, a promising intervention can look arbitrarily bad when there is too little capacity. For example, at capacity level $M_1=40$, increasing the number of treated participants from $100$ to $1000$ decreases the treatment effect by $78.3\%$, from $0.177$ to $0.038$.

\subsubsection{Statistical Power is Non-Monotone in the Number of Subjects}\label{sss:num.power}
Fig. \ref{fig:keheala_resultsplots_power_contour} and Fig. \ref{fig:keheala_resultsplots_power_line}
display the statistical power, for various sample sizes and capacity levels. %
Fig. \ref{fig:keheala_resultsplots_power_contour} shows that the power of the experiment is monotonically increasing in the amount of outreach capacity (i.e., the vertical dimension of the figure). However, in contrast to trials without a service component where power increases in sample size, the statistical power for our service trial is first increasing and then decreasing in the number of enrolled subjects. This is in line with our theory and demonstrates the trade-off between higher service levels (resulting in higher treatment effects) and higher number of enrolled subjects (resulting in a lower outcome variance).\footnote{Note that a service intervention with infinite capacity would have power increasing in sample size (see Fig. \ref{fig:keheala_resultsplots_power_line}), similar to an intervention without a service component.}

In other words, the same statistical power can be achieved at multiple levels of enrollment and service capacity.
For example, 80\% power can be achieved by running a trial with a high capacity and low enrollment (e.g., $M_1=57$ servers and $N_1=120$ treatment patients) or a trial with lower capacity and higher enrollment (e.g., $M_1=36$ servers and  $N_1=220$ patients). Which of those points is more suitable for a given trial then depends on the relative cost of hiring servers and recruiting patients, as well as the goal of the experiment. 

Fig. \ref{fig:keheala_resultsplots_power_line} tells a similar story, displaying a cross section of the surface of Fig.~\ref{fig:keheala_resultsplots_power_contour}. 
Observe that, for a given level of capacity, there is an upper bound to statistical power, regardless of sample size. As an extreme example, the lowest level of capacity (10 servers) has a maximum power of 32\% and never achieves 80\% statistical power, regardless of the number of participants. As we increase capacity, it becomes feasible to design a trial with high statistical power, but only for the right number of participants. 

Note that the $N_1/M_1=r$ boundary, plotted in Fig. \ref{fig:keheala_resultsplots_power_contour} as a solid red line, shows that the relationship between power and sample size depends on the queueing regime.
Broadly speaking, above the boundary (QD regime) power increases with sample size while below the boundary (ED regime) power increases with capacity. This illustrates that for trial planning purposes it is important for the experimenter to understand which region their planned trial belongs to.

\subsection{Implications of Numerical Results}\label{ss:simdiscussion}
In this subsection, we discuss the implications of the main numerical results, focusing on the same themes highlighted by our theoretical analysis in \S\ref{ss:discussion}. Specifically, the numerical simulations shed light on how multiple estimands of interest can complicate the interpretation of trial outcomes (\S\ref{sss:num.interpretation}), reveal potential failure modes for experimentation (\S\ref{sss:num.failure}), and illustrate the suboptimal performance of existing power analysis methods (\S\ref{sss:num:staffing}). Finally, they demonstrate how applying the Square Root Rule can improve power while reducing resource requirements (\S\ref{sss:num.squareroot}).

\subsubsection{Multiple Estimands of Interest}\label{sss:num.interpretation}
Our simulations show how much the estimands of interest (defined in \S\ref{sss:multiple_estimands}) can differ for real interventions. Consider the scenario where the experimenter runs a trial with $M_1=10$ and $N_1=N_0=100$. Suppose that, if the trial is successful, the experimenter plans to scale the intervention to a population of 500 participants, and recruit additional providers to reach 20 providers.

In this setting, the trial-implemented effect is $\theta^{ss}(10, 100)=0.094$. The full-sample deployment effect, or the effect if the intervention were made available to treatment and control subjects in the experiment, would be $\theta^{ss}(10, 200)=.049$, roughly half of the trial-implemented effect. At the planned rollout with a capacity of 20 and 250 subjects, the scaled deployment effect would be $\theta^{ss}(20, 250)=0.038$. Further, all of these effects are smaller than the capacity-unconstrained effect $\theta^{ss}(\infty, N_1)=0.184$ when there is ample capacity.

\subsubsection{Failure Modes for Experimentation}\label{sss:num.failure}

Our simulation results highlight the challenges in trial interpretation, replication, and scaling of service interventions. First, note that the operational dosage effects can lead the researcher to wrongly conclude that a promising result is ineffective.
Fig. \ref{fig:keheala_resultsplots_power_contour}  shows that any trial of the \textit{Keheala} intervention with a service capacity of 30 or less will have less than 80\% power, regardless of the number of participants. In other words, although 
the intervention can be effective under ample capacity, an experimenter is likely to conclude that treatment effects are insignificant, even under large sample size.

Second, the operational dosage effects can create challenges in trial replication, in line with the growing recognition that heterogeneity across trials leads to variation in effect size \cite{bryan2021behavioural, vivalt2020much, gelman2018dont, almaatouq2024beyond, almaatouq2023some}. If an experimenter attempts to replicate an existing trial but either is not aware of the operational dosage effects or the capacity levels were not documented in the existing trial, the experimenter may choose different levels of capacity or sample size, leading to seemingly incommensurate effect size estimates. Following the above example, a small trial with $N_1=N_0=100$ leads to a treatment effect of $\theta^{ss}(10, 100)=0.094$, whereas an experimenter that, perhaps, doubles the sample size in a replication trial to reduce variance would observe an expected effect size of $\theta^{ss}(10, 200)=0.049$.

\subsubsection{Suboptimal Performance of Existing Power Analysis Methods}
\label{sss:num:staffing}
We now consider the situation (introduced in \S\ref{sss:suboptimal_naive_power_analysis}) in which an experimenter runs a pilot experiment and conducts a power analysis to decide the required sample size for a full RCT. We assume a standard target of $0.80$ for statistical power. We consider two scenarios for the pilot: 
\begin{itemize}
    \item \textbf{Scenario A -- Large capacity (QD) pilot:} Initial pilot study with $M_1^p = 60$ servers and $N_1^p = N_0^p = 40$ subjects. This would result in a pilot effect size of $0.181$ and statistical power of $0.518$.
    \item \textbf{Scenario B -- Low capacity (ED) pilot:}  Initial pilot study with $M_1^p = 8$ servers and $N_1^p = N_0^p = 140$ subjects. This would result in a pilot effect size of $0.057$ and statistical power of $0.245$.
\end{itemize}
Since these initial pilot study parameters $M^1_p$ and $N_1^p$ are typically chosen for feasibility reasons and not optimized, a power analysis policy would ideally perform well given any instance of $M^1_p$ and $N_1^p$.
Fig. \ref{fig:keheala_resultsplots_power_contour} shows the outcome of each pilot study (the circles labeled A (green) and B (orange)) and the resulting RCTs informed by the Naive 1 (No $M_1$ Scale-up---represented by a triangle), Naive 2 (Proportional $M_1$ Scale-up---represented by a diamond), and Square Root Staffing (represented by a square) policies.

When the initial pilot is in the Quality Driven regime (Scenario A), Naive Policy 1 and Naive Policy 2 can lead to RCTs of reasonably high power: Naive Policy 1 (No $M_1$ Scale-up) suggests an RCT with $M_1=40, N_1=N_0=120$, resulting in effect size $0.171$ and power $0.762$. Since the initial pilot is in the QD regime, the effect size is large, there is excess capacity, and more subjects can be added without significantly diminishing the effect size. Naive Policy 2 (Proportional $M_1$ Scale-up) suggests an RCT with $M_1=80, N_1=N_0=120$, resulting in effect size $0.180$ and power $0.803$. The power of each trial is close to the target power $0.80$.

In contrast, when the initial pilot is in the Efficiency Driven regime (Scenario B), the Naive Policy 1 and Naive Policy 2 lead to severely underpowered designs or extremely high resource requirements, respectively. 
Naive Policy 1 (No $M_1$ Scale-up) suggests an RCT with $M_1=8, N_1=N_0=1170$, resulting in effect size $0.057$ and power of $.175$, substantially lower than the target of $0.80$. 
Because the pilot in the ED regime has a small effect size, the naive experimenter believes that many more subjects are required to detect the effect. However, as discussed in \S\ref{ss:power}, adding more subjects in the ED regime decreases the operation dosage, thus decreasing the effect size and power.
On the other hand, Naive Policy 2 (Proportional $M_1$ Scale-up) suggests an RCT with $M_1=67, N_1=N_0=1170$,  resulting in larger effect size $0.054$ and power $0.744$.
While this policy leads to high power, note from Fig. \ref{fig:keheala_resultsplots_power_contour} that the same power can be achieved with lower capacity and fewer subjects.

\subsubsection{Improved Power Analysis through Square Root Staffing}
\label{sss:num.squareroot}
Importantly, the recommendation of the Square Root Staffing Policy is the same for both Scenario A and B, regardless of the queueing regime of the pilot. The red square in Fig.  \ref{fig:keheala_resultsplots_power_contour} shows this recommendation, which suggests an RCT with $M_1=66, N_1=N_0=120$, resulting in effect size $0.182$ and $0.801\%$ power.

This policy obtains high power for both scenarios, but the benefits against the Naive policies is particularly evident in Scenario B. 
In Scenario B, Naive Policy 1 does not obtain sufficiently high power. While Naive Policy 2 does result in high power of $0.744$, the Square Root policy leads to 7.6\% higher power with \textit{substantially} fewer resource requirements, reducing the capacity required by 1.5\% and the number of subjects required by 90\%. This effect arises because, while the Naive Policy 2 recognizes that there is an operational dosage effect and that scaling capacity with the number of subjects is required to maintain the pilot effect size ($0.057$), it does not recognize that the effect size can be increased through higher capacity and increased operational dosage. The Square Root policy recognizes that the effect size can be larger than the pilot effect size, targets an effect size of $0.182$, close to the upper bound $\theta^{ss}(\infty, N_1)=0.184$, and with this larger effect size, recognizes that fewer subjects are required to reduce the variance. In fact, the capacity required to serve this reduced number of subjects almost instantaneously is less than the capacity required to serve the number of subjects in Naive Policy 2 at a lower quality of service.

Thus, if the goal is to ascertain whether the intervention has an effect, the queueing informed power analysis simultaneously leads to higher statistical power and lower resource requirements.%

\section{Discussion}\label{s:discussion}

It goes without saying that the dosage of a drug in a pharmaceutical trial will influence the measured treatment effect. The argument of this paper, at the highest level, is that the capacity constrained resources in service interventions create an analogous ``operational dosage'' corresponding to how \textit{quickly}, \textit{consistently}, and/or \textit{often} subjects receive the prescribed interventions. As a result, the right question to ask, when evaluating service interventions trial outcomes is not ``did it work?'' but rather ``did it work at this level of operational capacity?''.  

This insight, though simple, leads to contributions to the growing academic literature on experimentation with behavioral interventions (\S\ref{ss.conclusiontheory}), as well as suggestions for best practices for RCT design (\S\ref{ss.conclusionplanning}) and interpretation (\S\ref{ss.conclusioninterpretation}). 

\subsection{Theoretical Contributions}\label{ss.conclusiontheory}

First, we contribute to the growing literature on SUTVA violations in experimentation due to interference, when one subject's treatment affects others' outcomes \cite{rosenbaum2007interference, hudgens2008toward, imbens2010rubin, angrist2014perils}. The existing literature has shown that ignoring spillover effects can lead to large biases in estimation of treatment effects social networks \cite{manski2013identification, ugander2013graph, eckles2016design, aronow2017estimating, saveski2017detecting, athey2018exact, leung2020treatment, yu2022estimating, brennan2022cluster, cortez2022staggered,  harshaw2023design, jia2023clustered, viviano2023causal, ugander2023randomized, bajari2023experimental, boyarsky2023modeling, shirani2024causal} and, more recently, in marketplaces \cite{blake2014marketplace, fradkin2019simulation, holtz2020reducing, wager2021experimenting, munro2021treatment, liu2021trustworthy, li2022interference, johari2022experimental, bright2022reducing, dhaouadi2023price, li2023experimenting, johari2024does, masoero2024multiple, delarue2025pricing, weng2024experimental,  hays2025double, zhang2025debiasing}.
We identify a type of spillover effect that arises not through interactions of participants with each other, but through interactions of individual participants with a capacity constrained system. This type of spillover effect further leads to a violation of ``no hidden variation'' \citep{vanderweele2013causal} of SUTVA: interference leads to hidden variation in wait times through what we term the ``operational dosage'' effect. This type of spillover effect is most similar to the work on spillover effects in marketplaces, where supply and demand constraints create dependencies across subjects.%
Our methods contribute to the line of recent work using Markov chain models to model interference patterns \cite{hu2022switchback,johari2022experimental,farias2022markovian,farias2023correcting,li2023experimenting,hays2025double} and, most closely related, Li et al. \cite{li2023experimenting}, who study queueing systems arising in online platforms. 

In contrast to the existing literature, our work has a distinct focus on the selection of capacity and number of participants as a design choice. Prior work on interference in marketplaces largely focuses on experiments on digital platforms (A/B tests) that run on existing users, removing the need to study issues like trial recruitment or scaling. In contrast, our work focuses on settings inspired by healthcare, education, and social services where the experimenter develops a new intervention and decisions of capacity selection, user recruitment, and scaling are of primary concern.

We are able to systematically capture the operational dosage effects by leveraging techniques from queueing theory, a branch of mathematical modeling from operations research designed to study wait times and congestion in complex systems \cite{shortle2018fundamentals}. While the existing literature on queueing theory largely focuses on staffing decisions and does not address experimentation, our work shows how to adapt these tools \cite{whitt1992asymptotic, whitt2006fluid,vericourt2008dimensioning, vericourt2011nurse} to analyze questions in experiment design.
Framing the experimentation setting as a queueing system allows us to identify a critical threshold $r = (\lambda + \tau + \mu p)/\lambda$ in the participant-to-server ratio that marks the transition between two distinct queueing regimes that dictate the size of the treatment effect and the relationship between sample size and power, explaining the non-linear, but predictable, changes in effect size with added participants.

Second, we contribute to the active discussion on external validity and (the lack of) replicability of behavioral interventions. %
While initial explanations focused on type-I error and questionable research practices, recent work has identified systematic sources of heterogeneity in treatment effects \cite{bryan2021behavioural, vivalt2020much, bryan2021behavioural, vivalt2020much, gelman2018dont, almaatouq2024beyond, almaatouq2023some,krefeld2024exposing}. This growing ``heterogeneity revolution'' shows that some of this variability in experiment outcomes  
can be attributed to contextual factors like socioeconomic status, demographics, and implementation details.
Our work identifies capacity constraints as an additional source of systematic heterogeneity both within and across RCTs. For example, within a trial, operational dosage may lead to heterogeneous treatment effects if one group of subjects routinely becomes eligible for service during times of lower congestion than another group. Across trials, operational dosage can lead to different estimated effect sizes if different trials of the same intervention use different capacity levels.

Further, the variability introduced by operational dosage can contribute to issues of `voltage drops' \cite{list2022voltage, list2024optimally} where promising small-scale interventions become less effective at scale. When scaling an intervention and increasing the number of participants, our work highlights that the capacity must also be appropriately scaled, or else operational dosage and the resulting treatment effect size may drop. These capacity-dependent effect sizes must be considered in any cost-benefit analysis of the intervention. (see \ref{ss.conclusioninterpretation}.)

Finally, our work relates to an existing body of work on interventions when the intervention designer is aware of capacity constraints. 
A body of work studies how to best target resources,  by identifying individuals that are most likely to benefit \cite{azizi2018designing,killian2019learning,dasgupta2024preliminary,liu2025measuring,baek2023policy}. Recent work has also studied methods for learning optimal targeting strategies through experimentation \cite{klasnja2015microrandomized,owen2020optimizing,wilder2025learning}. This literature typically treats capacity as fixed and focuses on optimizing targeting criteria. 
Our work studies the outcomes of RCTs when the experimenter is not aware of the capacity constraints and focuses on the question of how to choose the capacity and number of participants.

\begin{figure}
     \centering
     \begin{tabular}{p{0.48\linewidth}p{0.48\linewidth}}
     \toprule
     \multicolumn{2}{c}{\textbf{SERVICE INTERVENTION TRIAL KPIs}}\\
     \textit{Trial Planning} & \textit{Trial Interpretation} \\
     \midrule
        \textbf{Individual Operational Dosage} \newline 
        - What service level is necessary for the individual protocol being tested to work? \vspace{0.2cm} \newline 
        \textbf{System Workload} \newline 
        - Based on individual behavior, how much workload should be expected? \vspace{0.2cm} \newline
        \textbf{Required Capacity} \newline 
        - How many servers are required to handle the workload and achieve the appropriate operational dosage? &
        \textbf{Subject Count} \newline 
        - How many subjects were being treated at any time of the trial? \vspace{0.2cm} \newline  
        \textbf{Server Count} \newline 
        - How many servers were operating at any time of the trial? 
        \vspace{0.2cm} \newline
        \textbf{Service Wait Times} \newline 
        - How long do subjects wait for service, once they are eligible? \vspace{0.2cm} \newline
        \textbf{Queue Length} \newline 
        - How many subjects are waiting for service at any time? \vspace{0.2cm} \newline
        \textbf{Server Utilization Rate} \newline 
        - What proportion of the time are the servers busy? \vspace{0.2cm} \newline
        \textbf{Completion Rate} \newline 
        - What proportion of subjects receive the required operational dosage? \vspace{0.2cm} \newline
        \textbf{Abandonment Rate} \newline 
        - How frequently do subjects leave the queue (i.e., become ineligible again) without being served?\\
     \bottomrule
     \end{tabular}
     \caption{Key Performance Indicators (KPIs) to consider when planning and interpreting service intervention trials.}
     \label{fig:KPIs}
\end{figure}

\subsection{Implications for Trial Planning}\label{ss.conclusionplanning}
A central implication of our analysis is that trial outcomes—both the estimated treatment effect and the likelihood of achieving statistical significance—depend jointly on the sample size and the service capacity. %
Together, these quantities determine the operational dosage of the intervention, which in turn governs both the magnitude of the observed effect and the power of the trial. This means that trial planners must 
carefully consider the operational system in which the intervention is delivered. To aid in this process, our analysis highlights three guiding questions for experimenters.

\begin{enumerate}
\item \textbf{What is the objective of my trial?}
The role of operational dosage also depends on whether the trial is designed to test efficacy or effectiveness. For efficacy trials, the primary objective is to establish the existence of an effect; thus, planners typically want to maximize operational dosage to ensure that any potential benefit can be detected. In contrast, effectiveness trials seek to assess performance under realistic conditions. %
For effectiveness trials, service levels during the trial should be aligned with those expected in real-world deployment, so that the observed outcomes reflect the likely implementation environment.

\item \textbf{Is my operating system capable?}
Designing an intervention protocol that specifies what service an individual should receive under given circumstances is not sufficient to ensure that the intervention will be consistently implemented in practice. Service delivery depends on the capacity of the operating system that executes the protocol. If that system becomes congested, particularly if the participant-to-server ratio exceeding the critical threshold, the intervention received by participants will fall short of the intended design, altering the operational dosage and thus biasing the estimated treatment effect. Trial planners must therefore distinguish between the idealized intervention design and the operational capacity required to realize it, and ensure that the latter is adequate to deliver the intended dosage. 

\item \textbf{Can I ensure appropriate statistical power?}
Even when the operating system and trial objectives are clear, planners must ensure that the experiment has sufficient power. This requires choosing sample size and service capacity in tandem, recognizing that both jointly determine operational dosage. A practical first step is to conduct analyses of the kind we have presented here, which explicitly link congestion and capacity constraints to expected treatment effects---perhaps using data collected in a pre-trial pilot. Our mathematical modeling and simulation approach illustrate two methods of conducting such analyses.
A complementary step is for trial planners to explicitly document service capacity assumptions in trial registrations. We are not aware of any registered protocols (including those for \textit{Keheala}) that predict the level of capacity required to maintain the intended participant experience. 
Incorporating such planning not only improves the credibility of trial findings but also facilitates more reliable translation to practice.
\end{enumerate}

\subsection{Implications for Trial Interpretation}\label{ss.conclusioninterpretation}

Our analysis also has implications for how the outcomes of service intervention trials should be interpreted. At a minimum, it reminds us that the findings of such RCTs are conditional on the operational dosage experienced by participants. Beyond this general caution, our analysis yields a set of practical recommendations for interpreting trial results. These apply broadly to policymakers and researchers across medicine, behavioral science, education, and other fields where service interventions are evaluated.

\begin{enumerate}
    \item \textbf{Seek evidence of operational dosage.}
    When interpreting the results of a service intervention trial, it is not enough to focus solely on the reported average treatment effect. Equally important is an understanding of the operational dosage---the extent to which participants actually experienced the intervention as designed. %
    In order for a policy-maker to gauge this, experimenters should report operational statistics %
    on how the service was delivered. Relevant measures include (a) the participant-to-server ratio, (b) the utilization rate of servers, (c) the average queue time for participants who were eligible but had to wait, and (d) the completion rate of service delivery. Together, such statistics reveal whether participants received the intervention promptly and consistently. %
    Without this information, estimates of treatment effects risk being misinterpreted, as they reflect not only the intervention itself but also the operational system in which it was embedded. 

    \item \textbf{Distinguish protocol failure from operational failure.} 
    A diminished or null treatment effect does not necessarily signal that the intervention protocol itself is ineffective; it may instead reflect limitations in its operational implementation under capacity constraints. Misattributing such effects to protocol failure risks overlooking interventions that could succeed if delivered with sufficient capacity. 
    To avoid this pitfall, policymakers must consider congestion effects and delays in order to separate true protocol limitations from operational shortcomings.

    \item \textbf{Do not assume larger samples imply `better' estimates.}
    Well-intentioned researchers might endeavor to enroll as many subjects as possible in a service intervention trial, seeking greater statistical power. Yet, as we have illustrated, this can have the unintended consequence that the protocol becomes diluted by congestion and delays, especially if the participant-to-server ratio exceeds the critical threshold, thereby lowering the observed treatment effect. This point has practical implications for policy makers: larger trials are not always more informative. If two studies of the same intervention yield different results, the instinct to trust the larger one is not necessarily the right choice. In fact, a smaller trial may provide a clearer indication of the intervention’s impact, provided that service capacity is sufficient to carry out the intended protocol. %
    Recognizing this distinction is essential to interpreting evidence and making decisions about implementation.

    \item \textbf{Plan for scaling and replication with capacity in mind.}
    Trials of service interventions should be considered an opportunity to learn about operational implementation (at scale) as well as treatment effects. The first thing a policy-maker should realize is that there is no single cost and no single effectiveness estimate per subject for a service intervention; rather, there exists a trade-off curve in which both costs and benefits vary with service capacity and participant numbers. Most importantly, the benefit decreases when the participant-to-server ratio exceeds the critical threshold. A rigorous approach to understand this trade-off curve is to use trial data to estimate the primitive quantities that govern the day-to-day impact of the intervention---such as the effect of an individual interaction---and then simulate how the system would perform at different capacity levels (as we have done in \S \ref{s.keheala}). Such simulations allow decision-makers to predict how the observed benefits are likely to be affected by scaling. In the absence of a full modeling exercise, it remains essential to consider the queueing regime observed in the trial. Explicitly situating trial results within their queueing regime ensures that scaling decisions are guided not just by average treatment effects, but also by %
    expectations about operational performance.
\end{enumerate}

\subsection{Conclusion}
The objective of this work is to help experimenters and policymakers to manage---i.e., plan, execute, interpret---RCTs of service interventions. The main insight is that, for service interventions, capacity constraints impact the treatment effect and statistical power of trials, thereby affecting external validity and replicability. 
We demonstrate these insights in a theoretical model, as well as through simulation experiments based on field data.

Our work is not without limitations. First, the proportion of time spent in a `desired state'---our main outcome---is, in some cases, only a proxy for the ultimate success of an intervention. For healthcare applications, for example, medication adherence is important but only a step towards being cured. Second, for some applications, a specific pattern or history of desired behavior (e.g., sustained stay in the desired state for more than a certain time) is more important than the cumulative sum of time in the desired state. Such considerations are not captured in our models.

Despite this, we believe that the insights we uncover are are critical for both researchers and policymakers across a variety of domains.
For example, an education tutoring intervention that has large treatment effects with high tutor-to-student ratios may fail when implemented at scale with fewer tutors per student. A promising social program piloted with manageable caseloads may show no effect when case workers are overwhelmed at higher case loads. 

We make a range of recommendations for the planning, reporting, and interpretation of service intervention trials. Most importantly, we highlight the need to make operational capacity visible through better measurement and reporting, a necessary step toward improved detection and deployment of promising interventions.

\appendix

\bibliographystyle{plain}
\bibliography{body/bibliography}

\newpage
\appendix

\section*{Appendix for ``Operational Dosage: Implications of Capacity Constraints for the Design and Interpretation of Experiments''}
~
The Appendices are organized into five sections: Section \ref{si:sec:theory} provides additional details for the mathematical framework; Section \ref{si:sec:theoryresults} presents derivations and proofs for the main theoretical results; Section \ref{si:sec:theorynumerics} provides numerical validation of the theoretical results; Section \ref{si:sec:hyp_testing_power_analysis} provides supplementary details on the power analysis approaches; and Section \ref{si:sec:Keheala} describes the development of the Keheala simulation model.

% 1. Additional details for mathematical framework. 
\section{Additional details on mathematical framework}\label{si:sec:theory}

This section provides the mathematical foundation for the theoretical model introduced in the main text. We begin by describing subject behavior and system dynamics, defining the states, transitions, and service processes that generate the congestion effects of interest (Subsection \ref{si:ssec_model}). We then analyze the steady-state behavior of the resulting Markovian system, characterizing its long-run properties (Subsection \ref{si:ssec_steady}). Finally, we embed this system within an experimental framework, leading to Lemma 1, which establishes the connection between the steady-state queueing results and the primary experimental outcome (Subsection \ref{ssec:methods_experiment_setup}).

\subsection{Model}\label{si:ssec_model}
The model described in \S \ref{s:model} induces a continuous time Markov chain $\{X_{M,N}(t)\}$ for $t\geq 0$. This Markov chain is governed by transition rate matrix $\mathcal{Q}$, where $\mathcal{Q}_{M,N}(i,j)$ represents the rate that the system moves from state $X_{M,N}(t)=i$ to state $X_{M,N}(t)=j$. 

First consider the rate of subjects entering the queue. If there are $i$ subjects in the queue, then there are $N-i$ subjects in a state $0$ and so the rate at which subjects enter the queue is $(N-i)\lambda$. 

Now, consider the rate of subjects leaving the queue, which depends on the congestion level. Each server has a service rate $\mu$ and so they can serve the subjects at a rate of $\min\{i\mu, M\mu\}$ (limited by the number of subjects $i$ in the queue and number of servers $M$). 
If the number of subjects in the queue at time $t$ is less than the number of servers (i.e., $X_{M,N}(t)=i\leq M$), then subjects transition out of the queue with a transition rate of $ip\mu + i\tau$. If the number of subjects in the queue at time $t$ exceeds the number of servers (i.e., $X_{M,N}(t)=i>M$) then subjects transition out of the queue with rate $Mp\mu + i\tau$. If the subject remains in the undesired state, despite being served, the subject rejoins the queue. Summarizing:

\begin{equation} \label{eq:queue_dynamics}
\begin{aligned}
    Q_{M,N}(i, i+1) &= (N-i) \lambda, \qquad  \text{for}\ 0\leq i<N, \\
    Q_{M,N}(i, i-1) &= 
    \begin{cases}
    i\mu p + i\tau, &  \text{if}\ i \leq M,\\
    M \mu p + i\tau, & \text{if}\ M < i \leq N.
    \end{cases}
\end{aligned}
\end{equation}
The rate of all other transitions $Q_{M,N}(i, j)$ for $i\neq j$ is $0$. Finally, the quantity $\mathcal{Q}_{M,N}(i,i)$ is defined so that the sum of each row is $0$. 

\subsection{Steady-state analysis}\label{si:ssec_steady}
The Markov chain devined above is ergodic with unique stationary distribution $\pi_{M,N}$, where $\pi_{M,N}(k)$ denotes the steady state probability of $k$ subjects in the undesirable state. $\pi_{M,N}(k)$ satisfies the following equations:

\begin{equation}
    \pi_{M,N}(j)=
    \begin{cases}
         \pi_{M,N}(j-1) \cdot \frac{(N-j+1)\lambda}{j\mu p + j\tau}
         & \text{for } 1\leq j \leq M\\
         \pi_{M,N}(j-1) \cdot \frac{(N-j+1)\lambda}{M\mu p + j\tau}
         & \text{for } j\geq M.
    \end{cases}
\end{equation}
with normalization condition $\sum_{j=0}^{N} \pi_{M,N}(j)=1$. 

The steady state \textit{expected queue length} is  
\begin{equation}
\label{eq:finite_exp_queue_length}
    \overline{K}_{M,N} =  \sum_{j=1}^N j \cdot \pi_{M,N}(j).
\end{equation}

\subsection{Experimental Framework}
\label{ssec:methods_experiment_setup}

We now use the queueing system to describe the system dynamics in an experiment. 

\subsubsection{Treatment and Control Group Behavior} The treatment and control groups form two distinct stochastic systems. Treatment group subjects interact through a queueing system with dynamics governed by the transition rate matrix $\mathbf{Q}_{M_1,N_1}$ and steady-state distribution $\boldsymbol{\pi}_{M_1,N_1}$. Control group subjects evolve independently according to the natural transition rates $\lambda$ and $\tau$ without access to the service intervention, or equivalently,  with transition rate matrix $\mathbf{Q}_{0,N_0}$ and steady-state distribution $\boldsymbol{\pi}_{0,N_0}$.

\subsubsection{Outcome metric} 

The metric of interest is the expected time spent in the undesirable state $\E_{\pi_{M,N}} \left[ x_{M,N}(u,t) \right]$, where the expectation is taken with respect to the stationary distribution. We note that since our model considers homogeneous users, this quantity is the same for all users $u$ in the system; we can extend this model to consider a finite number of heterogeneous user types as well.

Note that for the control group, there there are no servers ($M=0$) and participants transition between states based on the rates $\lambda$ and $\tau$. The expected time spent in the undesirable state simplifies to $\E_{\pi_{0,N}} \left[ x_{0,N_0}(u,t) \right]= \lambda / (\lambda + \tau)$.

The system is more complex in the treatment group, where there are non-zero servers ($M_1\geq 1$), since these shared servers induce dependencies across participants. A key observation in our analysis is that the proportion of time the participant spends in the undesirable state can be expressed as a function of the steady state queue length $\overline{K}_{M, N}$. This equivalence arises because any time that a user spends in the unverified state is spent either in the queue or being served. This observation enables us to leverage existing queueing theory results on functions of queue length in our analysis of these service intervention experiments. 

\begin{lemma}[Relationship between proportion of time verified and queue length]
\label{lemma:prop_verified_to_queue_length}
Let $X_{M,N}(t)$ denote the number of people being served or waiting in the queue. 
Then the proportion of time the user spends in the undesirable state, averaged across users, is
    \begin{equation*}
        \frac{1}{N} \cdot \sum_{u} Y_{M,N}(u,T) 
        = \frac{1}{NT} \cdot \int_{t=0}^T X_{M,N}(t) dt
    \end{equation*}
The long run (steady state) proportion of time the user spends in the undesirable state can be expressed as
    \begin{equation*}
        \E_{\pi_{M,N}}\left[ x_{M,N}(u, t) \right] =  \overline{K}_{M, N}/N.
    \end{equation*}
\end{lemma}

\begin{proof}[Proof of Lemma \ref{lemma:prop_verified_to_queue_length}.]
For the equation for finite $T$,
\begin{align*}
    \frac{1}{N} \cdot \sum_{u} Y_{M,N}(u,T) 
    &= \frac{1}{N} \sum_u \frac{1}{T} \int_{t=0}^T x_{M,N}(u,t)dt\\
    &= \frac{1}{NT}\int_{t=0}^T \sum_u x_{M,N}(u,t) dt\\
    &= \frac{1}{NT} \cdot \int_{t=0}^T X_{M,N}(t) dt
\end{align*}

Note that the left-hand side is the average time that each user spends in the queue and the right-hand side is the average queue length divided by $N$, over the fixed time horizon $T$. 
The equation for the steady state quantity follows from taking the limit of both sides of the above expression as $T\to\infty$. Since the Markov chain is ergodic, both long run averages converge to the steady state quantities.

\begin{gather*}
    \lim_{T\to\infty} \frac{1}{N} \cdot \sum_{u} Y_{M,N}(u,T) 
    = \E_{\pi_{M,N}}\left[ x_{M,N}(u, t) \right] \\
    \lim_{T\to\infty}\frac{1}{NT} \cdot \int_{t=0}^T X_{M,N}(t) dt 
    = \overline{K}_{M, N}/N.
\end{gather*}
\end{proof}

\clearpage

% 2. Derivation and proofs for main results
\section{Derivation and proofs for main results}\label{si:sec:theoryresults}

This section presents the detailed mathematical derivations and proofs supporting our main theoretical results.
The complexity of service intervention systems, with their dependencies between subjects competing for limited capacity, makes exact analysis of treatment effects and variance difficult for systems with finite $N_1, N_0, M_1$, and $T$. Thus we use two asymptotic frameworks to provide tractable approximations of the finite system behavior. 

In subsection \ref{si:ssec:clt}, we provide the proof of the Central Limit Theorem, which establishes the asymptotic distribution of the treatment effect estimator for an RCT with a fixed number of servers $M_1$ and subjects $N_1$ and $N_0$, as the observation time $T \to \infty$. This approach characterizes the sampling distribution of the treatment effect estimator and motivates variance approximations used for calculating the power of the test. 

In subsection \ref{si:ssec:fluid_limit}, we provide fluid model approximation and characterize its equilibrium properties, and the characterization critical threshold that delineates the quality-driven and efficiency-driven regimes in the queueing system. The fluid model approximates  the system behavior of the treatment group with $M_1$ servers and $N_1$ subjects as both $M_1, N_\to \infty$, while maintaining a fixed subject to server ratio. This analysis allows us to approximate the treatment effect $\theta^{ss}(M_1,N_1)$ and the mean of the estimator $\hat{\theta}(M_1, N_1, N_0, T)$ and, further, prove the threshold behavior discussed in \S\ref{ss:treatmenteffect}. Note that this fluid model is deterministic and cannot give approximations for variance, for which we rely on the CLT approximation.

%We first provide the proof of the Central Limit Theorem, which establishes the asymptotic distribution of the treatment effect estimator. We then derive the fluid model approximation and characterize its equilibrium properties, followed by the derivation of the critical threshold that delineates the quality-driven and efficiency-driven regimes in the queueing system.

\subsection{Finite system when $T \to \infty$}\label{si:ssec:clt}
In this subsection, we provide proofs for Theorem \ref{thm:clt} and Corollary \ref{cor:consistency}, both of which are derived for the finite system case, when the time horizon $T \to \infty$.

\subsubsection{Theorem \ref{thm:clt}: Central Limit Theorem for Finite System}
Our main theorem is as follows. 

%\subsubsection{Proof of Central Limit Theorem}
% \label{si:ssec:clt_proof}
\begin{theorem}[Central Limit Theorem]
\label{thm:clt}
Let 
$A(M_1,N_1,T) =  \sum_{u: Z_u=1}1 - Y_{M_1,N_1}(u,T ) / N_1
$
and 
$ B( N_0,T) =
    \sum_{u: Z_u=0} 1- Y_{0, N_0}(u,T)  / N_0 
$
denote the treatment and control terms of Equation \eqref{e:estimator}, so that $\hat{\theta}(M_1, N_1, N_0, T) = A(M_1, N_1,T) - B(N_0,T)$. 

\medskip
Then the following quantities converge in distribution

\begin{align*}
    T^{1/2} \cdot \left( A(M_1,N_1,T) - \mu_1 \right) 
    &\implies N(0, \tilde{\sigma}_1^2) 
    \\
    T^{1/2} \cdot \left( B(N_0,T)  - \mu_0\right) &\implies N(0, \tilde{\sigma}_0^2)
\end{align*}
where 

\begin{gather*}
\mu_1 = \mu_1(M_1,N_1) = 1 - \overline{K}_{M_1,N_1}/N_1,
\quad \mu_0 = \frac{\tau}{\lambda + \tau}\\
\tilde{\sigma}^2_1 = \tilde{\sigma}_1^2(M_1,N_1) = \frac{2}{N_1^2} \sum_{j=0}^{N_1-1} \frac{1}{(N_1-j) \pi_{M_1,N_1}(j)} \left[ \sum_{i=0}^j (i - \bar{K}_{M_1, N_1})\pi_{M_1,N_1}(i) \right]^2\\
 \tilde{\sigma}_0^2 = \tilde{\sigma}_0^2(N_0) = \frac{1}{N_0} \cdot \frac{2 \lambda \tau}{(\lambda + \tau)^3}
\end{gather*}
\medskip

Then the scaled estimator converges

\begin{equation*}
T^{1/2} \cdot \left(\hat{\theta}(M_1, N_1, N_0, T)  - (\mu_1 - \mu_0) \right) \implies N\left(0, \ 
\tilde{\sigma}_1^2 + \tilde{\sigma}_0^2\right). 
\end{equation*}
\end{theorem}

\proof{Proof of Theorem \ref{thm:clt}.}
%\begin{proof}[Proof of Theorem \ref{thm:clt}]
    The proof proceeds in three parts: 1) showing existence of a CLT for the estimator, 2) characterizing the mean of the limiting distribution, and 3) characterizing the variance of the limiting distribution.

    \textit{Step 1: Existence of a CLT.} We utilize Theorem 3.1 in \cite{liu2015central} on the existence of CLTs for continuous time Markov chains. This result shows for a positive recurrent CTMC $X_t$, with stationary distribution $\pi$, and real valued function $f$, a CLT holds for the sample mean $S(t) = 1/t \int_{0}^t f_{X_s}ds, t\geq 0$, a CLT of the form $t^{1/2} \left[S(t) - \pi(f)\right] \implies N(0, \sigma^2(f)$ when certain conditions are satisfied for $X_t$ and $f$. We apply this theorem to our setting, letting $X_t$ be the state of our CTMC and $f$ be the identify function, i.e., $f(x) = x$ for any state $x$.

    We need to check that the conditions $X_t$ is strongly ergodic and $\E_\pi[f(x)^2]<\infty$ hold. First note that the Markov chain is ergodic, since it is possible to transition between any two states in finite time. Further, since the state space is finite, the expected hitting time of any state is finite and so ergodicity implies strong ergodicity \citep{tweedie1981criteria}. The second condition $\E_\pi[f(x)^2]<\infty$ holds since the state space is finite. Thus a CLT exists. 

    \textit{Step 2: Characterizing the mean.}
    We analyze the treatment patients and the control patients separately. We start with the treatment patients. Recall from Lemma \ref{lemma:prop_verified_to_queue_length} that
    \begin{equation*}
        \frac{1}{N} \sum_{i=1}^{N} \frac{Y_{M,N}(u,T)}{T}
        = \frac{1}{NT} \int_{i=0}^T X_{M,N}(t)dt. 
    \end{equation*}
    In other words, we can calculate the proportion of time non-verified $Y_{M,N}(u,T)$, averaged across users, as the time average of the queue length, normalized by the number of patients $N$. By the Ergodic theorem for Markov chains \cite{meyn2012markov}, queue length time average converges to the expected length of the queue under the steady state distribution $\overline{K}_{M,N}$ and so the average proportion of time that users spend verified in the treatment group converges to $1 - \overline{K}_{M, N_1}/N_1$.

    The control group consists of $N_0$ individuals that evolve independently. For any individual $u$, this individual switches between two states $x(u,t)=0$ and $x(u,t)=1$. By the Ergodic theorem, proportion of time that $u$ spends verified on $[0,T]$ converges to the expectation in steady state that the user is in state $x(u,t)=1$. Thus the proportion of time each user $u$ spends verified converges to $\tau/(\lambda+\tau)$, and the average across all $N_0$ users is the same quantity.

    \textit{Step 3: Characterizing the variance.}
    Note that the set of treatment patients is independent from the set of control patients, so we can analyze the behavior of the two systems separately. 
    For both systems, we first characterize the variance of the time average of the queue length and then relate this variance to the variance of the estimator. 

    We start with the treatment patients. For the treatment patients, the queueing system is a birth death chain. We use the following result from \cite{whitt1992asymptotic}, attributed to \cite{burman1980functional}, for the asymptotic variance for a birth death process.

%First, let $Y(t)$ be any real-valued continuous time stochastic process with sample mean $(\overline{Y}(t) = t^{-1} \int_0^t Y(s) ds$ for $s>0$ and the asymptotic variance is $\tilde{\sigma}^2 = \lim_{t\to\infty} t \ \Var(\overline{Y}(t)).$

\begin{quote}
\begin{proposition}[Asymptotic Variance  \cite{whitt1992asymptotic,burman1980functional} ]
Let $X(t)$ be a birth and death process on a subset of integers $\{ 0, \ldots, n\}$ with birth and death rates $\lambda_i$ and $\mu_i$, respectively. Let $Y(t) = f(X(t))$ where $f$ is a real-valued function on the state space of the BD process $X(t)$ and let $\bar{f} = \sum_{i=0}^n \pi_i f(i)$ be the steady state mean. Define the time average (sample mean)  $(\overline{Y}(t) = t^{-1} \int_0^t Y(s) ds$ for $s>0$ and define the asymptotic variance to be
$\tilde{\sigma}^2 = \lim_{t\to\infty} t \ \Var(\overline{Y}(t)).$

Then
\[
\tilde{\sigma}^2 = 2 \sum_{j=0}^{n-1} \frac{1}{\lambda_j \pi_j} \left[ \sum_{i=1=0}^j (f(i) - \bar{f})\pi_i \right]^2.
\]
\end{proposition}
\end{quote}

Note that the average number of jobs in the queue can be written in the above formulation as $f(i) = i$ and then we can directly use the formula in our setting. 

Let $\tilde{\sigma}^2_{M,N}$ be the asymptotic variance of the time average queue length in a system with $M$ servers and $N$ users. Then
\begin{equation}
\label{eq:asymptotic_var_finite_m_n}
\tilde{\sigma}^2_{M,N} = 2 \sum_{j=0}^{N-1} \frac{1}{(N-j) \pi_{M,N}(j)} \left[ \sum_{i=0}^j (i - \bar{K})\pi_{M,N}(i) \right]^2
\end{equation}
where $\overline{K}$ itself is a function of $\pi_{M,N}$. 

Since the time average proportion of time verified for the treatment group is simply the time average of the queue length normalized by $N_1$, we get that the treatment group variance contribution is 
\begin{equation*}
    \tilde{\sigma}^2 = \tilde{\sigma}^2(M,N_1, N_0) = \frac{2}{N_1^2} \sum_{j=0}^{N_1-1} \frac{1}{(N_1-j) \pi_{M,N_1}(j)} \left[ \sum_{i=0}^j (i - \bar{K}_{M, N_1})\pi_{M,N_1}(i) \right]^2.
\end{equation*}

For the control group, patients are independent from each other and so we can begin by analyzing a single patient. A single patient forms a birth death chain with only two states $\{0, 1\}$. Thus we can use Equation \eqref{eq:asymptotic_var_finite_m_n} to calculate the time average of the given patient's verification proportion. For this simple two state case, the resulting expression simplifies to 
\begin{equation*}
    \frac{2 \lambda \tau}{(\lambda + \tau)^3}.
\end{equation*}
All control users are independent and so the average across users has variance 
\begin{equation*}
    \frac{1}{N_0} \cdot \frac{2 \lambda \tau}{(\lambda + \tau)^3}.
\end{equation*}

Since the treatment and control groups are independent from each other, the asymptotic variance of the estimator is the sum of the respective asymptotic variance for the treatment and control groups. Thus completes the proof. 

\endproof
%\end{proof}

\subsubsection{Corollary \ref{cor:consistency}: Consistency}

A corollary of Theorem \ref{thm:clt} is that $\hat{\theta}(M_1, N_1, N_0, T)$ is a consistent estimator for the the long term treatment effect $\theta^{ss}(M, N)$, when $M=M_1$ and $N=N_1$. That is, the estimator recovers the true treatment effect if the experiment is run for long enough. 
This result does not hold for general $M\neq M_1$ and $N\neq N_1$.
\begin{corollary}[Consistency]
\label{cor:consistency}
For any $N_1, N_0>1$, 
    \begin{equation*}
        \lim_{T \to\infty} \hat{\theta}(M_1, N_1, N_0, T) \xrightarrow{p} \theta^{ss}(M_1, N_1)
    \end{equation*}
where $p$ denotes convergence in probability.
\end{corollary}

\begin{proof}
    The result follows directly by observing that $\theta^{ss}(M_1,N_1)$ is defined to be $\mu_1-\mu_0$ and that $\mu_1-\mu_0$ is a constant, and so the convergence in distribution in Theorem \ref{thm:clt} implies convergence in probability. 
\end{proof}

\subsection{Fluid Model}\label{si:ssec:fluid_limit}
We now consider the case of a treatment group with $M_1$ servers and $N_1$ subjects as both $M_1, N_\to \infty$, while maintaining a fixed subject to server ratio. We first characterize the fluid model, then prove Lemma \ref{lemma:ode_ss}, before applying the model to an experimental setting, leading to Theorem \ref{thm:treatment_effect_mean_field} and the corresponding proof. 

\subsubsection{Characterization of Fluid Model}
\label{si:sss:fluid_model_characterization}

To capture the implications of congestion arising from capacity constrained interventions, we must characterize how the size of the treatment effect defined in 
\eqref{eq:treatment_effect} changes with the number of servers and users 
or, equivalently, characterize the steady state queue length $\overline{K}_{M_1,N_1}$ of the treatment group with $M_1$ servers and $N_1$ users. Due to the complexity of the system, instead of directly directly analyzing $\overline{K}_{M_1,N_1}$, we utilize a \textit{fluid model} to approximate this quantity in large systems.

We construct a continuous time fluid approximation \cite{whitt2006fluid} of the system by replacing the (stochastic) evolution in the Markov chain with (deterministic) system of differential equations. This approach to analyzing the system
results in close approximations of the finite, stochastic model in large systems but
leads to simpler analytical expressions for quantities of interest. 
(See \cite{whitt2002stochastic} for the general method and \cite{vericourt2008dimensioning} for an example in a closed queueing system).

% The fluid model is motivated by a scaling of the system where we increase
% both the number of servers $M_1$ and number of users $N_1$ proportionally (such that $N_1/M_1$ approaches a constant) and consider the \textit{average} behavior of the system, which becomes deterministic even for finite $T$. 

%Formally, we consider a sequence of systems indexed by $i$, where the number of servers is $M_1^{(i)}$ and the number of users is $N_1^{(i)}$.%, with both $M_1^{(i)},N_1^{(i)}\to \infty$ as $\i \to \infty$. 
%The fluid model approximates a large scale system where we scale both $M_1^{(i)}, N_1^{(i)}\to\infty$ proportionally such that the limiting ratio of users to servers is \newhl{$\lim_{i\to \infty} N_1^{(i)} / M_1^{(i)} = \kappa$}. 

%We define the fluid model, motivated by this large system limit. 

We consider a unit mass of participants being served by a mass of providers, such that the ratio of participants to providers is $\kappa$. The state of the system $z_{\kappa}(t)$ is defined to be the fraction of users in the undesirable state at time $t$. Then $z_{\kappa}(t)$ obeys the following ordinary differential equation:
\begin{subnumcases}
%\label{eq:ode}
{\frac{dz_{\kappa}(t)}{dt}=}
    \lambda (1-z_{\kappa}) - z_{\kappa} \cdot  \mu p - z_{\kappa} \tau,  & \text{for } $z_{\kappa}< 1/\kappa$, \label{eq:ode_z_small}\\
    \lambda (1-z_{\kappa}) -  \mu p / \kappa - z_{\kappa} \tau, & \text{for } $z_{\kappa} \geq 1/\kappa$ \label{eq:ode_z_large}.
\end{subnumcases}
Importantly, the two equations above show that the evolution of this system depends on whether the fraction of subjects in the undesirable state $z_{\kappa}(t)$ is greater than or less than the the ratio of servers to subjects $1/{\kappa}$. When the mass of subjects in the undesirable state is less than $z_{\kappa}(t)<1/\kappa$, there is ample capacity in the system and the rate at which subjects receive service undesirable state is limited by $z(t)$. Conversely, when $z_{\kappa}(t) > 1/\kappa$, the system is capacity constrained and the rate at which users receive service is limited by $1/\kappa$, e.g., the ratio of providers to participants.

Analogous to the finite system steady state quantity $\overline{K}_{M,N}/N$, we can define the steady state proportion of users in the system (undesirable state) in the fluid model, defined to be the value at which $dz/dt=0$, which we denote $z^*_{ \kappa}$. 
The following result characterizes the steady state behavior and shows that the steady state has a threshold behavior that depends on the relative size of the user-to-server ratio $\kappa$ and the critical threshold $r$. We show in SI \ref{si:sec:theorynumerics} that the fluid steady state gives a good approximation for the average queue length in the finite, stochastic model.

\begin{lemma}[Fluid model steady state]
\label{lemma:ode_ss}
    For a given value of $\kappa<\infty$, the ODE system given by \eqref{eq:ode_z_small} - \eqref{eq:ode_z_large} has the following steady state solution:
    \begin{equation}
    {z^*_{\kappa}
    = }
    \begin{cases}
    \frac{\lambda}{\lambda + \tau + \mu p}, & \text{if } \kappa \leq r  \ \text{(Quality-Driven)}, 
    \\[10pt]
    \frac{\lambda - \mu p / \kappa }{\lambda + \tau}, & \text{if } \kappa > r \  \text{(Efficiency-Driven)}\\
    \end{cases}
    \end{equation}
    
Further, $z^*$ is globally asymptotically stable, i.e., given any initial condition $z(0)$, we have $\lim_{t\to\infty}z(t) = z^*$. 
\end{lemma}

The expression in Lemma \ref{lemma:ode_ss} gives an analytical characterization for the relationship between the user-to-server ratio $\kappa$ and the fraction of people in the undesirable state, which is not immediately apparent from the queue length expression in  \eqref{eq:finite_exp_queue_length} for finite settings. These regions correspond with the well known Quality-Driven and Efficiency-Driven regimes in queueing theory \cite{gans2003telephone}. In particular, note that for small participant-to-server ratio $\kappa<r$, the mass of servers is large enough that increasing the relative mass of users does not change the queue length. This corresponds to the Quality-Driven regime in queueing theory  where there is ample capacity. For large participant-to-server ratio $\kappa>r$, the treatment effect decreases non-linearly in $\kappa$ as the system becomes capacity constrained, corresponding to the Efficiency-Driven regime. In particular, note that the steady state solution approaches $\lambda/(\lambda+\tau)$ as $\kappa\to\infty$, which corresponds to the proportion of users in the unverified state in the \textit{control} group. As the user-to-server ratio increases, users have to wait longer to be served, and, in essence, receive a lower dose of the service, with the limiting scenario being that they receive no service. 

Returning to the behavior of the treatment group for finite $M_1$ and $N_1$,
%For a finite system with large enough values of $M$ and $N$, 
we can use the behavior of the fluid model to approximate the average queue length in the finite system by plugging in $\kappa = N_1/M_1$:
\begin{equation*}
    \overline{K}_{M_1,N_1} \approx N_1 \cdot z^*_{N_1/M_1}.
\end{equation*}
Indeed, we observe that the the steady state distribution of the finite system converges is close to the fluid model steady state for large $M_1$ and $N_1$. See SI \ref{si:sec:theorynumerics} for comparisons of simulations of the finite system and the fluid model.

\proof[Proof of Lemma \ref{lemma:ode_ss}]

We suppress the notation $\kappa$ and let $z(t)$ denote the state at time $t$ and $z^*$ denote the steady state. 

We show that the system has a unique steady state $z^*$ where $dz^*/dt = 0$ and use a Lyapunov argument to show that $z^*$ is globally asymptotically stable.

We begin by dividing the problem into three cases. 

\begin{itemize}
    \item Case 1: $1/r< 1/ \kappa$
    \item Case 2: $1/r = 1/\kappa$
    \item Case 3: $1/r > 1 / \kappa$. 
    % \item Case 1: $\kappa< r$
    % \item Case 2: $\kappa = r$
    % \item Case 3: $\kappa > r$. 
\end{itemize}

In each case, we will define a candidate Lyapunov function $V:\mathbb{R}\to\mathbb{R}$ which is continuously differentiable and verify the following properties
\begin{itemize}
    \item V(z) is positive definite, i.e., $V(z)>0$ for all $z\neq z^*$ and $V(z^*)=0$.
    \item The time derivative satisfies $dV(z)/dt \leq 0$ for all $z \in \mathbb{R}$, and
    \item $dV(z)/dt < 0$ for all $z \neq z^*$.
\end{itemize}
Once we show that $V(z)$ satisfies the above properties, we can conclude that $V(z)$ is a valid Lyapunov function and $z^*$ is globally asymptotically stable (see Theorem 4.1 of \cite{khalil2002nonlinear}).

\medskip

We will use the following observations about $dz/dt$. When $z \leq 1 / \kappa$, Equation $\eqref{eq:ode_z_small}$ holds and 
\begin{align*}
    \frac{dz}{dt}<0 \ & \text{when } \lambda/(\mu p + \lambda + \tau)<z\\
    \frac{dz}{dt}=0 \ & \text{when }\lambda/(\mu p + \lambda + \tau)=z \\ 
    \frac{dz}{dt}>0 \ & \text{when }\lambda/(\mu p + \lambda + \tau)>z.
\end{align*}

Similarly, when $z> 1 / \kappa$, Equation $\eqref{eq:ode_z_large}$ holds and 
\begin{align*}
    \frac{dz}{dt}<0 \ & \text{when } (\lambda - \overline{M}\mu p)/(\lambda+\tau)<z\\
    \frac{dz}{dt}=0 \ & \text{when }(\lambda - \overline{M}\mu p)/(\lambda+\tau)=z \\ 
    \frac{dz}{dt}>0 \ & \text{when }(\lambda - \overline{M}\mu p)/(\lambda+\tau)>z.
\end{align*}

We now proceed with the three cases.

\medskip
%%%%%%%%%%%%%%%%%%%%%%
% {\noindent\textit{Case 1: $1/r < 1/\kappa$ (Small $\kappa$).}}
\noindent\textit{Case 1: $\kappa < r$ (Small $\kappa$).}

Let 
\[V(z) = \frac{1}{2}(z-1/r)^2\]
be the candidate Lyapunov function, where 
$r= (\lambda + \tau + \mu p)/ \lambda$
%$r=\lambda / (\lambda + \tau + \mu p)$ 
denotes the critical threshold. 
$V(z)$ is continuously differentiable on $z\in \mathbb{R}$. 
We have $V(z) > 0$ for all $z\neq r$ and $V(z)=0$ for $z={1/r} = \lambda / (\lambda + \tau + \mu p)$ and so $V(z)$ is positive definite. 

We now compute the time derivative 
\begin{equation*}
    \frac{dV(z)}{dt} 
    = \frac{dV(z)}{dz} \cdot \frac{dz}{dt}
    = (z -r) \cdot \frac{dz}{dt}
\end{equation*}
and evaluate $\frac{dV(z)}{dt}$ under four subcases: $z<r$, $z=r$, $r<z\leq 1/\kappa$, and $1/\kappa < z$.

When $\kappa < r$, we have
{$1/r < 1/\kappa$}.
%$r< 1/ \kappa$. 
When %\newhl{$z < 1/\kappa$}, 
{$z<1/r < 1/\kappa$}, 
\eqref{eq:ode_z_small} holds and $dz/dt > 0$, and so $\frac{dV(z)}{dt} = (z -r) \cdot \frac{dz}{dt} < 0$.

When $z={1/r}<1/\kappa$, \eqref{eq:ode_z_small} holds and $dz/dt = 0$. Thus $\frac{dV(z)}{dt}=0$. 

When ${1/r}<z\leq 1/\kappa$, \eqref{eq:ode_z_small} holds and $dz/dt < 0$. Thus $\frac{dV(z)}{dt}<0$.

When $1/\kappa < z$, \eqref{eq:ode_z_large} holds. Since ${1/r} = \lambda / (\lambda + \tau + \mu p)< 1/\kappa$, we have $\lambda < (\lambda + \tau + \mu p)/\kappa $ and so
\begin{equation}
\label{eq:ode_ss_proof_large_ss_lessthan_one_over_kappa}
    \frac{\lambda-\mu p/\kappa}{\lambda + \tau} 
    <
    \frac{(\lambda + \tau + \mu p)/\kappa-\mu p/\kappa}{\lambda + \tau} 
    = 
    \frac{1}{\kappa}.
\end{equation}
Then $z > 1/\kappa \implies z > \frac{\lambda-\mu p/\kappa}{\lambda + \tau}$. Thus $dz/dt<0$ and $\frac{dV(z)}{dt}<0$.

Combining these four subcases, we see that $dV(z)/dt\leq 0$ for all $z$ and $dV(z)/dt < 0$ for all $z \neq {1/r}$. We conclude that $z^*={1/r} = \lambda / (\lambda + \tau + \mu p)$ is a globally asymptotically stable equilibrium when {$\kappa < r$}.

\medskip
%%%%%%%%%%%%%%%%%%%%%%
{\noindent\textit{Case 2: $\kappa = r$ (Medium $\kappa$).} }
% \noindent\textit{Case 2: $r = 1/\kappa$.} 
We define the same candidate Lyapunov function 
\[V(z) = \frac{1}{2}(z-{1/r})^2,\] which is continuously differentiable and we have shown that $V(z)$ is positive definite. We now consider the time derivative $\frac{dV(z)}{dt} = (z -{1/r}) \cdot \frac{dz}{dt}$ and consider three subcases: $z < {1/r} = 1/\kappa$, $z>{1/r}=1/\kappa$, and $z={1/r}=1/\kappa$.

When $z < {1/r} = 1/\kappa$, \eqref{eq:ode_z_small} holds and $dz/dt>0$, thus $dV(z)/dt < 0$. When $z={1/r}$, \eqref{eq:ode_z_small} holds and $dz/dt=0$, thus $dV(z)/dt = 0$. When $z>{1/r}=1/\kappa$, \eqref{eq:ode_z_large} holds. Since ${1/r} = \lambda/(\lambda + \tau + \mu p) = 1 / \kappa$, we have $\lambda = (\lambda + \tau + \mu p) / \kappa$ and so 
\begin{equation*}
    \frac{\lambda - \mu p / \kappa}{\lambda + \tau} 
    = 
    \frac{\frac{\lambda + \tau + \mu p}{\kappa} - \frac{\mu p}{\kappa}}{\lambda + \tau}
    = 
    \frac{1}{\kappa}.
\end{equation*}
Thus $z > \frac{\lambda - \mu p / \kappa}{\lambda + \tau}$ and, from \eqref{eq:ode_z_large}, $dz/dt<0$. 

Combining these three subcases, we conclude that $dV(z)/dt\leq 0$ for all $z$ and that $dV(z)/dt < 0$ for all $z \neq {1/r}$. Thus $z^*=1/r = \lambda / (\lambda+\tau+\mu p)$ is a globally asymptotically stable equilibrium when $\kappa = r$.

\medskip
%%%%%%%%%%%%%%%%%%%%%%
{\noindent\textit{Case 3: $\kappa>r$ (Large $\kappa$).}}
% \noindent\textit{Case 3: $r > 1 / \kappa$.}

In the setting where $\kappa > r$
%$r > 1 / \kappa$
, we define a different candidate Lyapunov function $W:\mathbb{R}\to\mathbb{R}$ as follows 
\[
W(z) = \frac{1}{2} \left( z - \frac{\lambda - \mu p / \kappa}{\lambda + \tau} \right)^2
\]
where we will show that the equilibrium is $z^* = \frac{\lambda - \mu p / \kappa}{\lambda + \tau}$. 
Observe that $W(z)$ is continuously differentiable and positive definite, i.e., $W(z)>0$ for all $z \neq \frac{\lambda - \mu p / \kappa}{\lambda + \tau}$ and $W(\frac{\lambda - \mu p / \kappa}{\lambda + \tau}) = 0$.

The time derivative is
\[
\frac{dW(z)}{dt} = \frac{d W(z)}{dz} \cdot \frac{dz}{dt} = (z - \frac{\lambda - \mu p / \kappa}{\lambda + \tau}) \cdot \frac{dz}{dt}.
\]
We consider five subcases: 
$z\leq 1 / \kappa$, 
$1/\kappa < z \leq {1/r}$, 
$ {1/r} < z < \frac{\lambda - \mu p / \kappa}{\lambda + \tau}$, 
$ {1/r} < z = \frac{\lambda - \mu p / \kappa}{\lambda + \tau}$ and 
$\frac{\lambda - \mu p / \kappa}{\lambda + \tau}<z$.

When $z\leq 1 / \kappa$, \eqref{eq:ode_z_small} holds and $dz/dt>0$.
Note that when $1/\kappa <  {1/r}$, we can lower bound $\lambda$ since  $\lambda/(\lambda + \tau + \mu p) > 1/\kappa \implies \lambda > (\lambda+\tau + \mu p) / \kappa$. Then 
\[
\frac{\lambda - \mu p / \kappa}{\lambda + \tau} 
>
\frac{(\lambda+\tau + \mu p) / \kappa - \mu p / \kappa}{\lambda + \tau}
= 
\frac{1}{\kappa} 
\geq
z.
\]
Thus  $(z - \frac{\lambda - \mu p / \kappa}{\lambda + \tau})<0$, $dz/dt>0$, and $dW(z)/dt<0$.

When $1/\kappa < z \leq  {1/r}$,
\eqref{eq:ode_z_large} holds. Since $ {1/r} = \lambda / (\lambda + \tau + \mu p) > 1/\kappa$, 

\[
\frac{\lambda - \mu p / \kappa}{\lambda + \tau} 
>
\frac{\lambda - \mu p \cdot \lambda / (\lambda + \tau + \mu p)}{\lambda + \tau} 
= \frac{\lambda}{\lambda + \tau + \mu p} =  {1/r}.
\]
Then $z \leq  {1/r} < \frac{\lambda - \mu p / \kappa}{\lambda + \tau} $ and so $dz/dt>0$. Combining this with $z - \frac{\lambda - \mu p / \kappa}{\lambda + \tau} < 0$ we get $dW(z)/dt<0$.

When $ {1/r} < z < \frac{\lambda - \mu p / \kappa}{\lambda + \tau}$, then $z>1/\kappa$ and \eqref{eq:ode_z_large} holds. $dz/dt>0$, $z - \frac{\lambda - \mu p / \kappa}{\lambda + \tau} < 0$, and so $dW(z)/dt<0$.

When $z = \frac{\lambda - \mu p / \kappa}{\lambda + \tau}$, $z>1/\kappa$ and so \eqref{eq:ode_z_large} holds and $dz/dt=0$. Thus $dW(z)/dt=0$.

When $\frac{\lambda - \mu p / \kappa}{\lambda + \tau}<z$, then \eqref{eq:ode_z_large} holds and $dz/dt<0$. Combining this with $z - \frac{\lambda - \mu p / \kappa}{\lambda + \tau} > 0$, we get $dW(z)/dt < 0$.

Combining the five subcases above, we can see that $dW(z)/dt <0$ for all $z \neq \frac{\lambda - \mu p / \kappa}{\lambda + \tau}$ and $dW(z)/dt =0$ for $z = \frac{\lambda - \mu p / \kappa}{\lambda + \tau}$. Thus we conclude that $W(z)$ is a valid Lyapunov function and $z* = \frac{\lambda - \mu p / \kappa}{\lambda + \tau}$ is a globally asymptotically stable point.

\endproof

% \newhl{For "straightforward algebra" part, this comes from
% \begin{align*}
%     \frac{\lambda}{\lambda+\tau+\mu p} &< \frac{1}{\kappa} \\
%     \lambda &< \frac{\lambda+\tau+\mu p}{\kappa}
% \end{align*}
% which implies that
% \begin{align*}
%     \frac{\lambda - \mu p / \kappa}{\lambda + \tau} 
%     &< \frac{\frac{\lambda + \tau + \mu p}{\kappa} - \frac{\mu p}{\kappa}}{\lambda + \tau}\\
%     & = \frac{1}{\kappa}
% \end{align*}
% }

\subsubsection{Applying the RCT Setup to the Fluid Model}

For a system with finite $M$ and $N$, we can approximate the finite system behavior using a fluid model with server-to-user ratio $M/N$. We apply this approximation to the experiment context
with $M_1$ servers, $N_1$ users in the treatment group, and $N_0$ users in the control group, 

We define the \textit{fluid model treatment effect} $\theta^*(\kappa)$, to be the difference between the mass of users in the verified state, between a treatment group with user-to-service ratio $\kappa$ and a control group. 
This quantity is analogous to the finite system steady state treatment effect $\theta^{ss}(M_1,N_1)$
as
\begin{equation}
    \label{eq:mean_field_treatment_effect}
    \theta^{*}(\kappa)
    := 
    \left[1 - z^*_{\kappa} \right] - 
    \left[1 - \frac{\lambda}{\lambda+\tau} \right].
\end{equation}
We approximate the finite system steady state treatment effect $\theta^{ss}(M_1,N_1)$ using the fluid model treatment effect:
\begin{equation*}
    \theta^{ss}(M_1,N_1) \approx \theta^{*}(N_1/M_1\\).
\end{equation*}

We can exactly characterize the fluid model treatment effect, as a function of $N_1/M_1$. 
We use this approximation in \S\ref{ss:treatmenteffect} to highlight how the steady state treatment effect $\theta^{ss}(M_1,N_1)$ changes with $M_1$ and $N_1$. These fluid model approximations $\theta^*(N_1/M_1)$, along with the (finite system) steady state $\theta^{ss}(M_1,N_1)$ quantities are shown in Fig. \ref{fig:numerics}. We observe that the the steady state distribution of the finite system is close to the fluid model steady state for large $M_1$ and $N_1$. See SI \ref{si:sec:theorynumerics} for comparisons of simulations of the finite system and the fluid model.
However, we note that this fluid model is deterministic and thus, when we study the variance of the estimator and the statistical power of the experiment, we use the finite system quantities in Theorem \ref{thm:clt}.

% \subsubsection{Implications of Fluid Model}

%We can exactly characterize the fluid model treatment effect, as a function of $\rho_1$. This provides an approximation for the finite system $\theta^{ss}(M_1, N_1) \approx \theta^*(N_1/M_1)$.
\begin{theorem}[Fluid model treatment effect]
\label{thm:treatment_effect_mean_field}
Fix a participant-to-server ratio $\kappa$ in the treatment group.
The fluid model treatment effect is
\begin{equation*}
\theta^*(\kappa) = 
    \begin{cases}
    \frac{\lambda\mu p}{(\lambda+\tau)(\lambda+\tau+\mu p)} & \text{for } \kappa \leq  {r} \ \text{(QD Regime)}\\
    \frac{\mu p}{\kappa(\lambda+\tau)} & \text{for } \kappa >  {r} \ \text{(ED Regime)}
    \end{cases}
\end{equation*}
In particular, 
as the participant-to-server ratio $\kappa$ increases, 
\[
\lim_{\kappa\to\infty} \theta^{*}(\kappa) 
%= \lim_{\overline{M} \to 0} \hat{\theta}^{*}(\overline{M})
= 0, 
\]
\end{theorem}
\begin{proof}
    The result follows immediately from applying Lemma \ref{lemma:ode_ss} to \eqref{eq:mean_field_treatment_effect}. 
\end{proof}
This approximation demonstrates that treatment effects remain constant when increasing the participant-to-server ratio $\kappa$ in the small $\kappa$, Quality-Driven regime but decrease with $\kappa$ in the Efficiency-Driven regime. Intuitively, in the Quality-Driven regime, there is an excess capacity of servers, so that all subjects can be served immediately. 
In the Efficiency-Driven regime the service rates are constrained by capacity and the addition of more subjects means that existing subjects wait longer to be served. The proof follows immediately by applying the fluid model steady state characterization in Proposition \ref{lemma:ode_ss} to the treatment group system.

Theorem \ref{thm:treatment_effect_mean_field} also implies that when the treatment group is run with a different $\kappa'$ than the intended level $\kappa$ when scaled up, then it is possible that the estimator from the experiment does not recover the treatment effect at scale-up.

The proof follows immediately from the fluid model steady state expression in Lemma \ref{lemma:ode_ss}.

\clearpage

% 3. Numerical validation of theory
\section{Numerical validation of theory}\label{si:sec:theorynumerics}

This section presents numerical validation of the theoretical model developed in the main text (as distinct from the simulation analysis of the real-world system described in Section \ref{si:sec:Keheala}). We first outline the numerical setup and parameter choices used to generate Fig. \ref{fig:numerics_power} in the main text. We then conduct simulations to assess the accuracy of the theoretical approximations, including the Central Limit Theorem and fluid model results. 

\subsection{Parameters for Fig. \ref{fig:numerics_power}}
Fig. \ref{fig:numerics_power} presents numerics of the continuous time Markov chain (CTMC) model using both CLT and fluid model approximations. 
The CTMC parameters used to generate these figures are: 
    $\lambda=0.63, \tau=0.34, \mu=5.5, p=0.07$.

These parameters are calibrated to achieve similar levels of treatment effect, proportion of time spent in the desirable state, and number of calls made as the Keheala RCT described in \S \ref{s.keheala} of the main text. Under these parameters, a control group participant with no access to intervention spends $0.353$ proportion of time in the desirable state. The capacity-unconstrained effect is $\theta^{ss}(\infty, N_1)=0.184$. The critical ratio is $r=2.16$.

We note that while the results of this calibrated CTMC model are qualitatively similar to the full Keheala simulation model (described in detail in Section \ref{si:sec:Keheala} of this document), there are slight deviations due to the relaxation of some of our modeling assumptions (necessary for the theoretical development) in the full simulation model--namely allowing for the non-Markovian and heterogeneous behavior of subjects. 

\subsection{Validation of theoretical approximations}
In \S\ref{s:mainresults}, we use the asymptotic approximations for the mean and variance of the estimator $\hat{\theta}(M_1, N_1, N_0, T)$ suggested by Theorem \ref{thm:clt}. We refer to these approximations as the CLT approximations. These approximations fix finite $M_1, N_1, N_0$ but consider the asymptotic behavior as $T\to\infty$. Additionally, in \S \ref{ss:treatmenteffect} \eqref{eq:fluid_limit_effect_size} we use an additional large system fluid model approximation in order to characterize how the effect size and estimator change with $M_1$ and $N_1$. %

These approximations are convenient, as the the finite $M, N_1, N_0$ approximation motivated by Theorem \ref{thm:clt} allows us to characterize the behavior at different $M, N_1, N_0$ without running computationally heavy simulations. The fluid model approximation further allows us to prove the non-linearity in the decrease in the effect size when increasing the ratio of users to providers. 

We now verify that these methods give us reasonable approximations of a system with finite $M_1, N_1, N_0$ and finite time horizon $T$. This boils down to confirming that the average queue length and the variance in the average queue length in simulations is similar to the quantities given by the approximations. 

\paragraph{Numerical setup}

We run numerical simulations with varying system parameters, capacity levels, and number of subjects and compare the simulation mean and variance with the quantities given by the mean field and CLT approximations. We vary arrival rates $\lambda \in [0.1, 0.3]$, service rates $\mu \in [2, 4]$, abandon rates $\tau \in [0.1, 0.2]$, and call effect probabilities $p \in [0.1, 0.3]$. We vary the number of subjects $N \in 25, 300]$ and the number of servers (capacity) $M \in [5,10]$. We also fix two time cutoffs $T_0 = 50$ and $T_{max}=200$. 

For each instance with server and user pair $(M,N)$ and parameter set $(\lambda, \mu, \tau, p)$, we run 500 simulation runs. For each run $i$, we analyze the system at different time horizons $T \in [T_0, T_{max}]$ and for the interval $[T_0, T]$ we calculate the estimated effect size $\hat{\theta}(M_1, N_1, N_0, T)$. For each $T$, we then plot the sample mean and variance of the $\hat{\theta}(M_1, N_1, N_0, T)$ across simulation runs $i$. We include bootstrapped 95\% confidence intervals, and compare with the mean given by the mean field limit approximation and the mean and variance given by the CLT approximation. 

\subsubsection{Validation of CLT approximation}
The CLT approximation is used to approximate both the mean and variance of the treatment effect estimate $\hat{\theta}(M_1,N_1,N_0,T)$ for \textit{finite} values of $M_1, N_1, N_0,$ and $T$. We will show that the sample mean of the estimates $\hat{\theta}(M_1,N_1,N_0,T)$ is closely approximated by the limiting mean $\theta^{ss}(M_1,N_1)$ and that the  variance of the estimates is closely approximated by the scaled limiting variance $(\sigma_1^2+\sigma_0^2)/T$.

Fig.  \ref{fig:simulation_comparison_mean_by_N} compares the sample average of the effect size estimates $\hat{\theta}(M_1,N_1,N_0,T)$ in simulations with the CLT limiting effect size $\theta^{ss}(M_1,N_1)$, for a fixed $T$ and for varying levels of providers $M_1$, number of treatment participants $N_1$, arrival rates $\lambda$, service rates $\mu$, abandon rates $\tau$, and call effect probabilities $p$.  Across all instances, the sample average of the effect size estimates $\hat{\theta}(M_1,N_1,N_0,T)$ is close to the limiting mean. 

\begin{figure}
    \includegraphics[width=\textwidth]{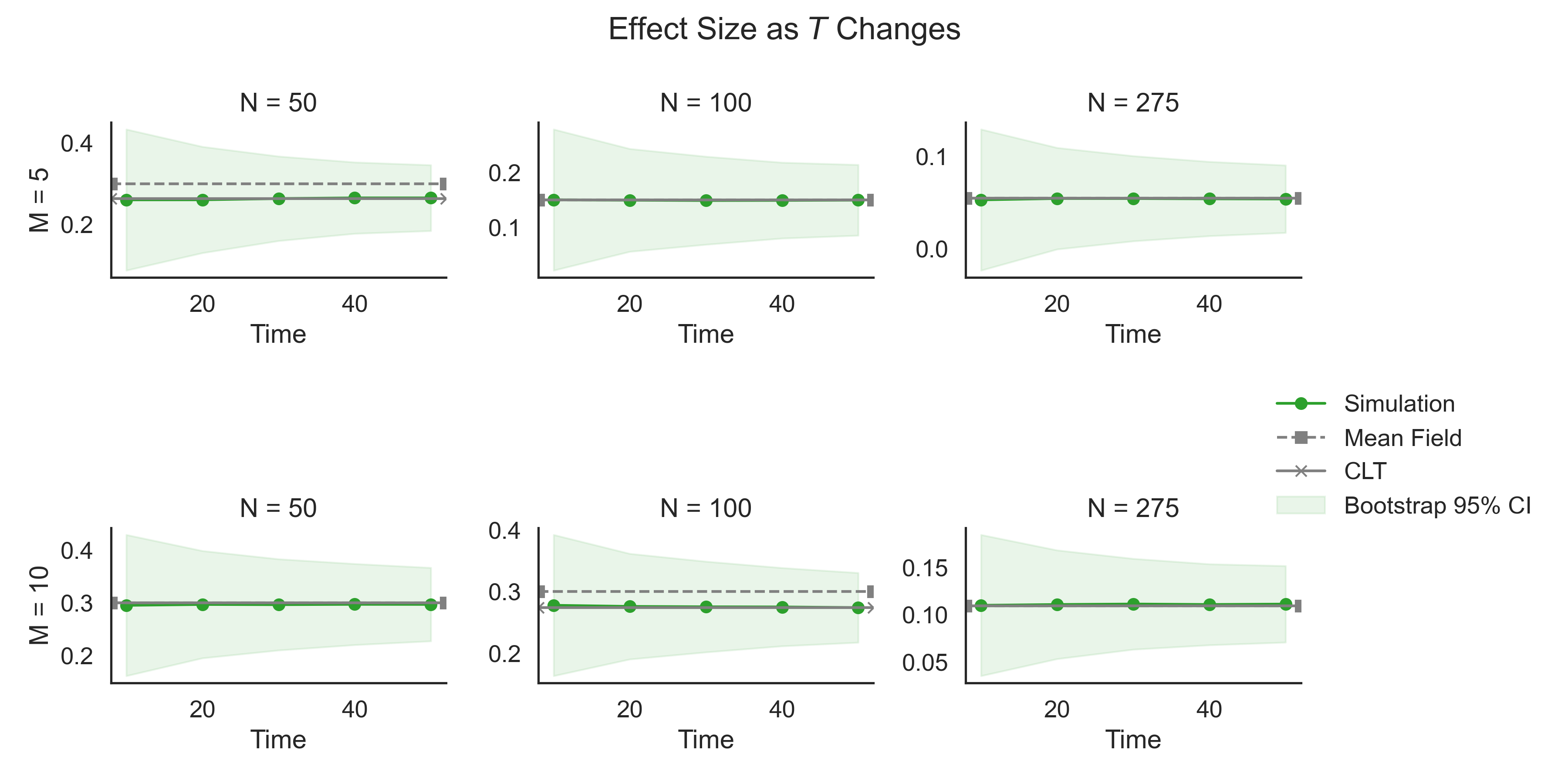}
    \caption{Comparison of sample mean of estimated effect size in simulation runs with CLT approximation and mean field approximation, as $T$ varies.}
    \label{fig:simulation_comparison_mean_by_T}
\end{figure}

Fig. \ref{fig:simulation_comparison_mean_by_T} compares the sample average of the effect size estimates with the limiting mean, as $T$ varies. The distribution of effect size estimates is centered around the limiting mean and the variance of this distribution decreases in $T$.

\begin{figure}
    \includegraphics[width=\textwidth]{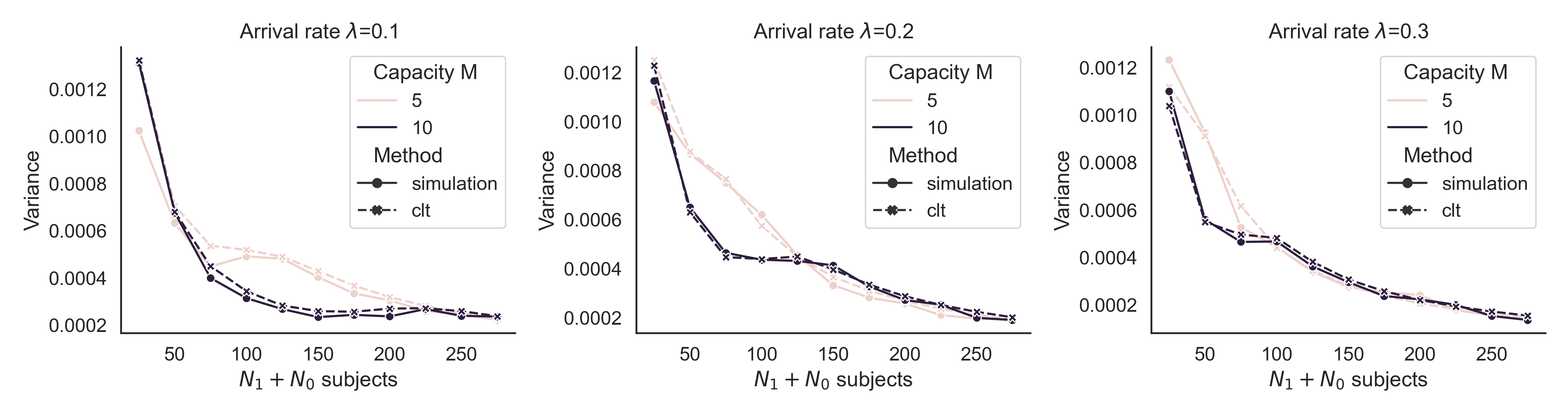}
    \includegraphics[width=\textwidth]{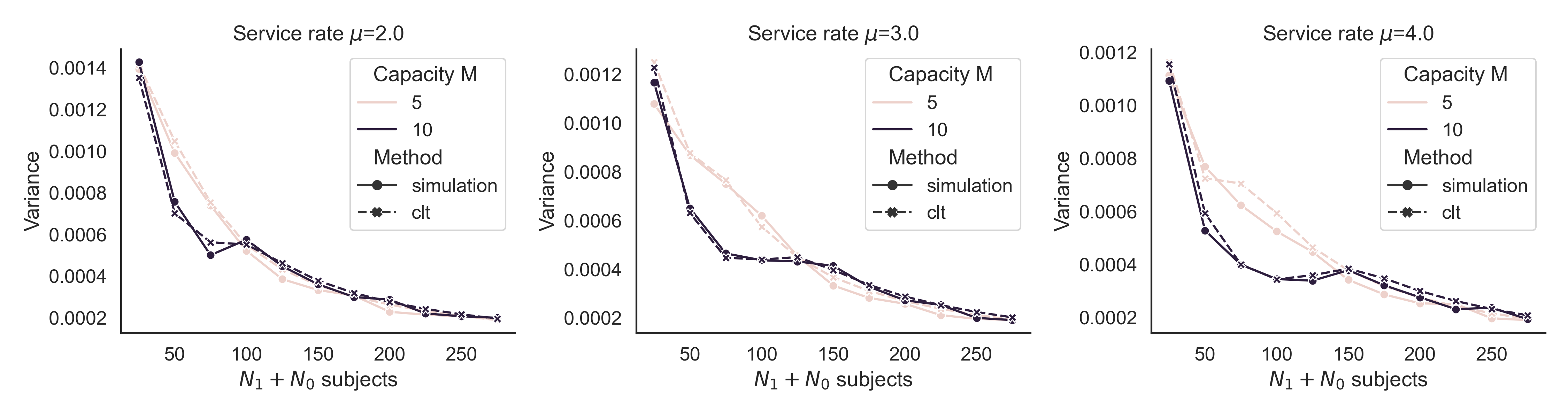}
    \includegraphics[width=\textwidth]{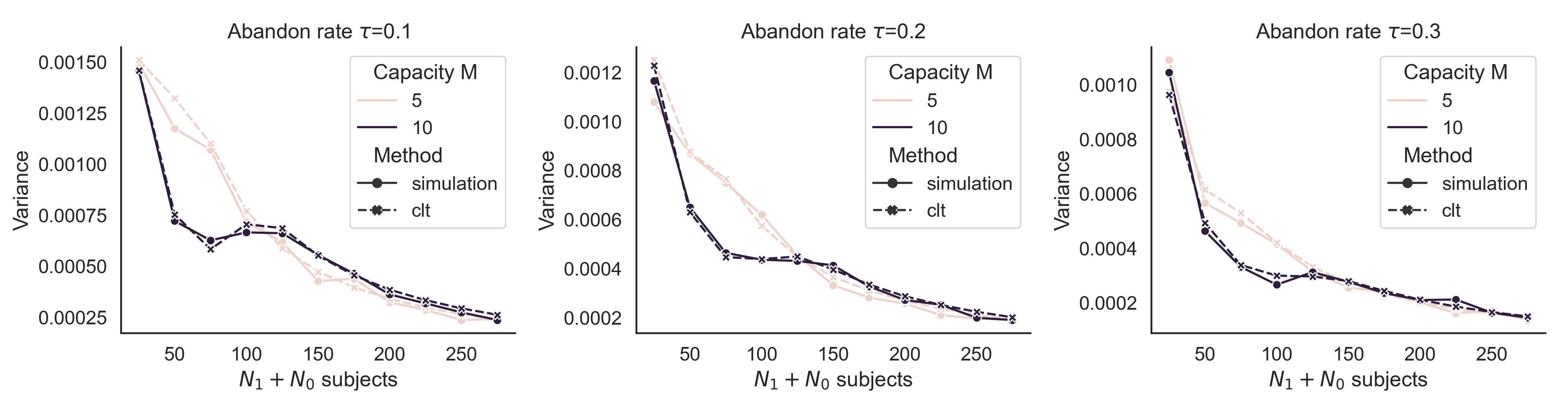}
    \includegraphics[width=\textwidth]{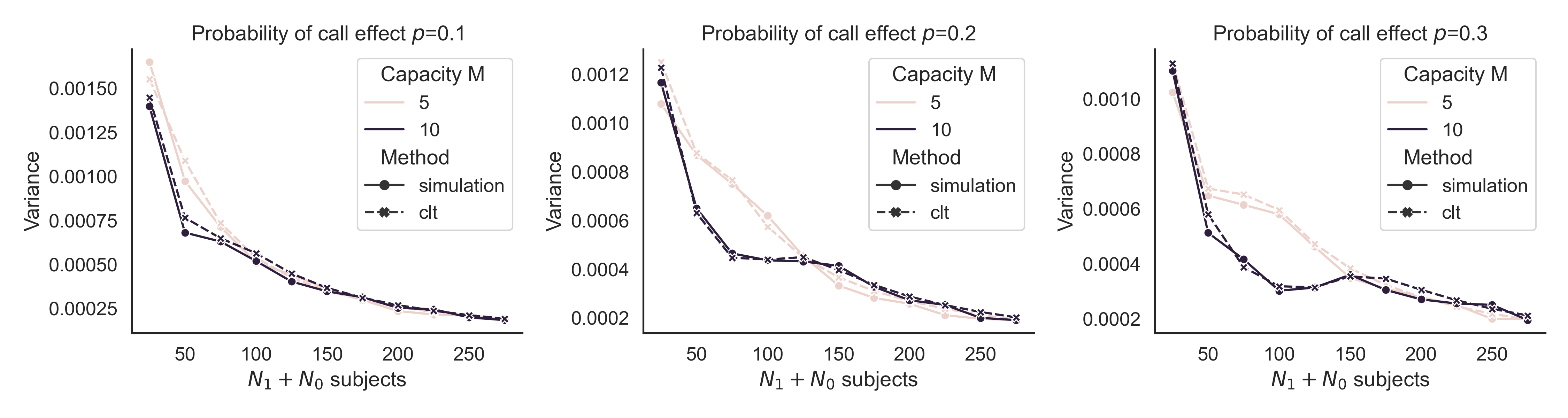}
    \caption{Comparison of sample variance of estimated effect sizes in simulation runs with CLT approximation, as $N_1+N_0$ varies.}
    \label{fig:simulation_comparison_variance_by_N}
\end{figure}

Fig.  \ref{fig:simulation_comparison_variance_by_N} compares the sample variance of the effect size estimates $\hat{\theta}(M_1,N_1,N_0,T)$ with the scaled limiting variance $(\sigma_1^2+\sigma_0^2)/T$, for a fixed $T$ and for varying levels of providers $M_1$, number of treatment participants $N_1$, arrival rates $\lambda$, service rates $\mu$, abandon rates $\tau$, and call effect probabilities $p$. The variance of the effect size estimates is closely approximated by the scaled limiting variance across all parameter instances. 

\begin{figure}
    \includegraphics[width=\textwidth]{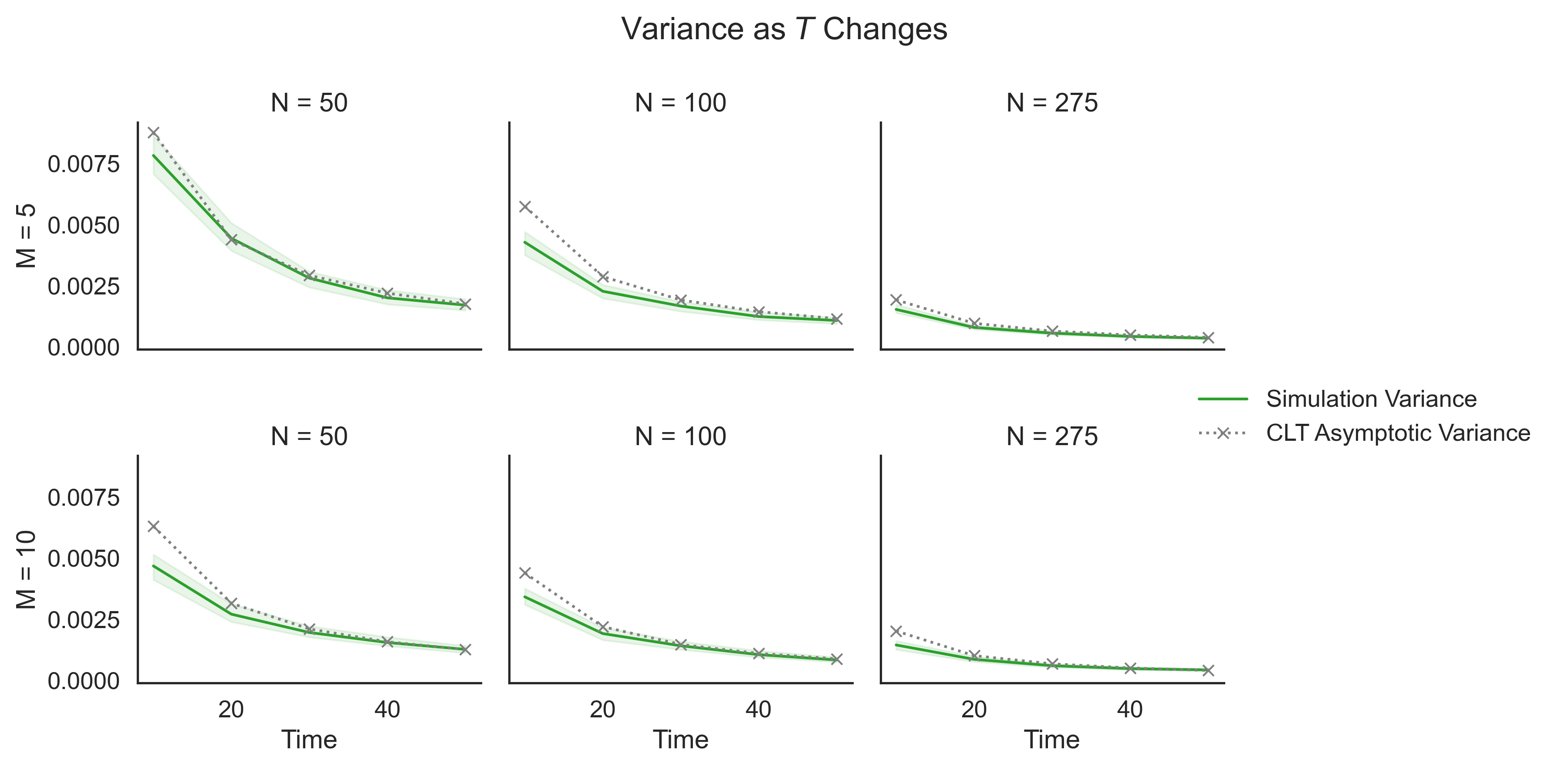}
    \caption{Comparison of sample variance of estimated effect size in simulation runs with CLT approximation, as $T$ varies.}
    \label{fig:simulation_comparison_variance_by_T}
\end{figure}

Fig. \ref{fig:simulation_comparison_variance_by_T} compares the sample variance of the effect size estimates with the scaled limiting variance, as $T$ varies. The simulations show that the variance of the effect size estimate is closely approximated by the scaled limiting variance
for large enough $T$ (approximately $T\approx 20$ in the settings shown).

\subsubsection{Validation of fluid approximation}
The fluid model is used to approximate the mean of the treatment effect $\theta^{ss}(M_1,N_1)$ and the treatment effect estimates $\hat{\theta}(M_1,N_1,N_0,T)$ for large enough $M_1,N_1$.

\begin{figure}
    \includegraphics[width=\textwidth]{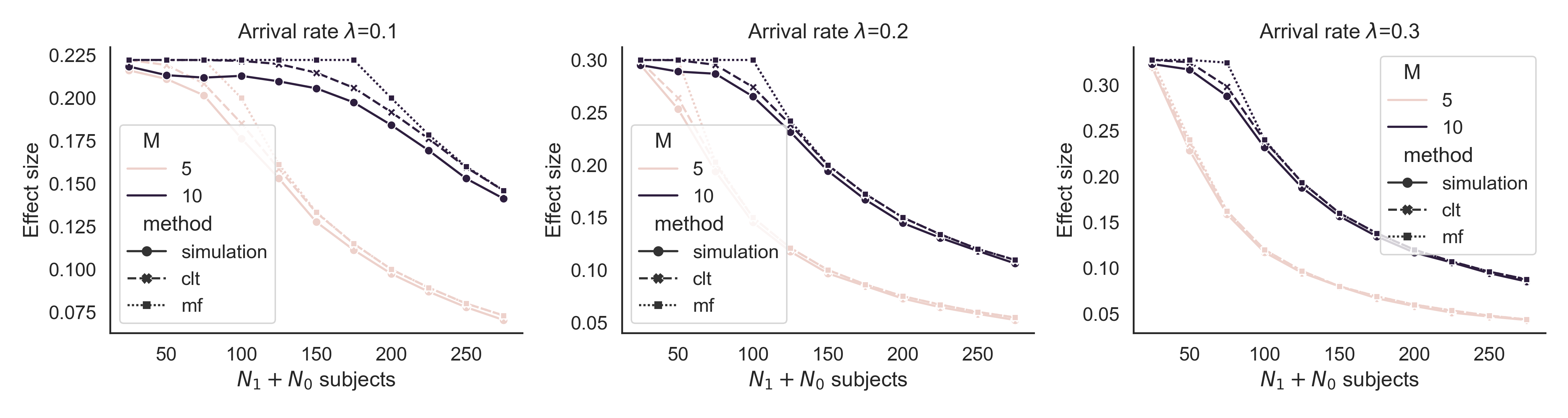}
    \includegraphics[width=\textwidth]{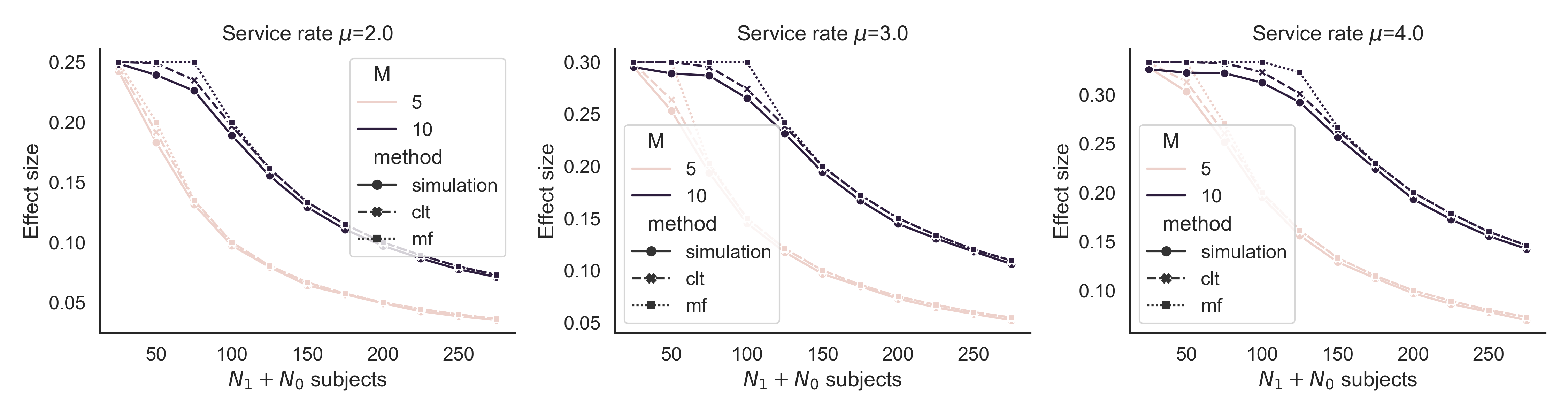}
    \includegraphics[width=\textwidth]{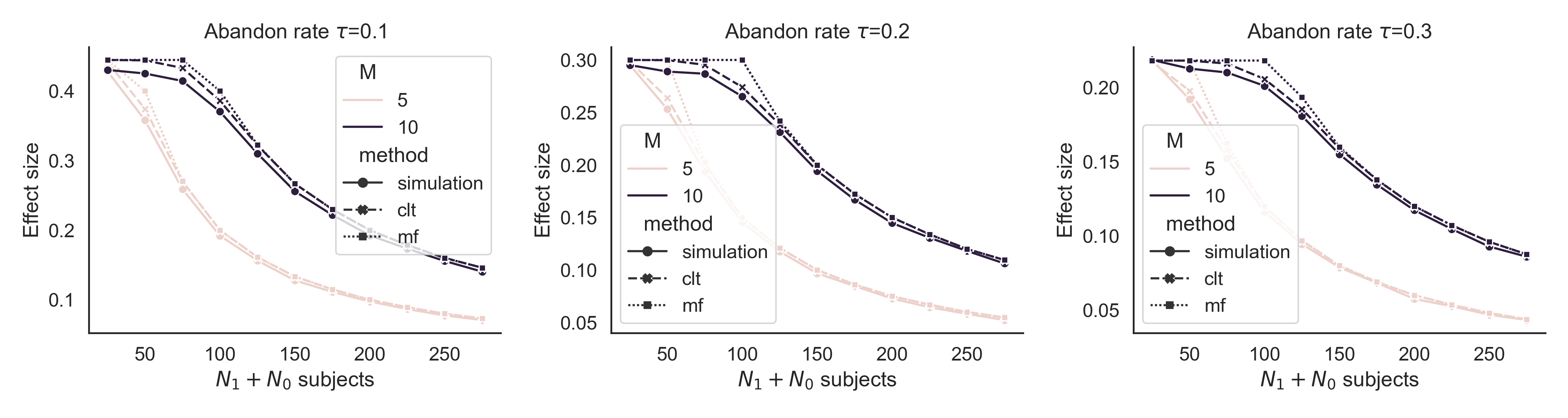}
    \includegraphics[width=\textwidth]{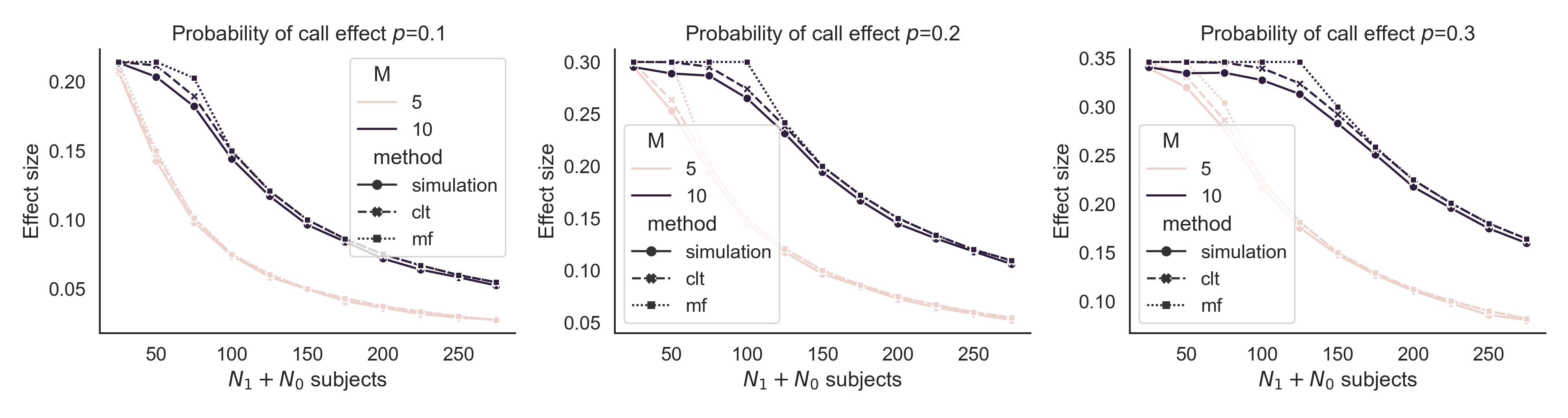}
    \caption{Comparison of effect sizes in simulation versus effect size given by CLT and mean field approximations, as $N_1+N_0$ varies.}
    \label{fig:simulation_comparison_mean_by_N}
\end{figure}

Fig \ref{fig:simulation_comparison_mean_by_N}
compares the sample average of the effect size estimates $\hat{\theta}(M_1,N_1,N_0,T)$ in simulations with the fluid model effect size and the CLT limiting effect size, for varying parameters (providers $M_1$, number of treatment participants $N_1$, arrival rates $\lambda$, service rates $\mu$, abandon rates $\tau$, and call effect probabilities $p$). For both values of $M_1$ shown, the sample average of the effect size estimates is close to the mean field approximation for most $N_1$, except for the $N_1$ near the phase transition in the fluid model. This phase transition, where the effect size changes from constant in $N_1$ to decreasing in $N_1$, corresponds to the value of $N_1$ such that $N_1/M_1$ is close to the critical ratio $r$. Near this threshold, the fluid model approximation predicts a sharper change in behavior (see the threshold behavior in \ref{thm:treatment_effect_mean_field}) than the CLT or simulation outcomes. The fluid approximation provides a close approximation in the Quality-Driven and Efficiency-Driven regimes to the left and the right of this threshold, respectively.

\clearpage

% 4. Additional details for hypothesis testing and power analysis approaches
\section{Statistical hypothesis testing and power analysis approaches}
\label{si:sec:hyp_testing_power_analysis}
This section elaborates on the power analysis methods introduced in the main text. We begin by discussing the general framework for hypothesis testing in the context of service intervention experiments, highlighting how capacity constraints alter standard assumptions about independence and effect size. We then present the detailed formulations of the power analysis approaches, including both the naïve and queueing-informed methods, and discuss their implications for experimental design and statistical efficiency.

\subsection{Statistical Power for Capacity Constrained Interventions}
The experimenter fixes significance level $\alpha$ 
and tests the null hypothesis $H_0: \theta^{ss}(M_1,N_1)=0$ with the alternative hypothesis $H_a: \theta^{ss}(M_1,N_1)\neq 0$.

Recall that, in the setting where the treatment effect is not constant but depends on the number of participants and number of servers, we define statistical power as the probability of detecting an effect at least as large as the true effect size at the given $N_1$ and $M_1$. Mathematically, statistical power is defined to be:
\begin{equation*}
Pr(\text{reject} H_0 | \text{true effect} = \theta^{ss}(M_1,N_1)).
\end{equation*}
The asymptotic normality of the estimator $\hat{\theta}(M_1,N_1,N_0,T)$ (see Theorem \ref{thm:clt}) motivates the following approximation for power for a two-sided test with $H_0: \theta^{ss}(M_1,N_1)=0$ and alternative hypothesis $H_a: \theta^{ss}(M_1,N_1) \neq 0$.

\begin{equation}
\label{eq:power_approximate_mde_true_effect_two_sided}
     {
    \Phi\left( \frac{\theta^{ss}(M_1, N_1)}{\sqrt{\tilde{\sigma}_1^2  (M_1, N_1)/T
    +
    \tilde{\sigma}_0^2 (N_0)/T}} 
    -
    \Phi^{-1}(1 - \alpha / 2)
    \right)
    + %
    \Phi\left(-  \frac{\theta^{ss}(M_1, N_1)}{\sqrt{\tilde{\sigma}_1^2  (M_1, N_1)/T
    - 
    \tilde{\sigma}_0^2 (N_0)/T}} 
    -
    \Phi^{-1}(1 - \alpha / 2)
    \right),
    }
\end{equation}
where $\Phi$ is the CDF of the standard normal $N(0,1)$ distribution.

The non-monotonic relationship between sample size and power is due to the fact that both the numerator and the denominator of the normalized effect size
\[
 \frac{\theta^{ss}(M_1, N_1)}{\sqrt{\tilde{\sigma}_1^2  (M_1, N_1)/T
    +
    \tilde{\sigma}_0^2 (N_0)/T}} 
\]
decrease in $N_1$. As Fig. \ref{fig:numerics} illustrates, the numerator decreases faster than the denominator in sections of the Efficiency-Driven regime. 

\subsection{Power Analysis Policies}
\label{si:sssec:power_analysis_definitions}

We consider a standard procedure in which the experiment designer first runs a small scale \textit{pilot study}, conducts a \textit{power analysis} using the outcomes from the pilot study, and then uses the results to decide how many users to recruit in the full study. We consider two naive policies for designing the large scale experiment as well as a third, queueing-informed, policy. We assume a fixed experiment length $T$ where all participants are active on the platform during the experiment time horizon.

Let $M_1^p$, $N^p_1$, and $N^p_0$ denote the number of servers, treatment subjects, and control subjects in the pilot, respectively.
After running the pilot study, the experimenter obtains an estimate of the treatment effect $\hat{\theta}(M_1^p, N^p_1, N^p_0, T)$ and an estimate for the variance $\widehat{\Var}\left(\hat{\theta} (M_1^p, N^p_1, N^p_1, T)\right)$. Based on the estimates from the pilot study (and, in the case of the naive pilot policies, a specified Minimum Detectable Effect $MDE$), the experimenter decides how to set the number of users and servers in the full experiment. 

A power analysis policy thus specifies an $(M_1, N_1, N_0)$ pair. 
We focus on policies with an equal number of users in treatment and control $N_1 = N_0$, although the observations can be extended for any constant ratio of treatment to control participants. 

\paragraph{Naive Policy 1: No $M_1$ Scale-up.}
Recall that the long-term treatment effect is $\theta^{ss}(M_1, N_1, N_0)$ and the variance of the estimator is ${\Var}\left(\hat{\theta} (M_1, N_1, N_0,T) \right) \approx 1/T \cdot \left(\tilde{\sigma}^2_1(M_1, N_1,T) + \tilde{\sigma}^2_0( N_0,T)\right)$.

For this first naive policy, the experimenter is not aware of the dependencies of both the treatment effect and the variance on $M_1, N_1, N_0$ and makes two (incorrect) assumptions about the experiment:
\begin{enumerate}
    \item They wrongly assume that the treatment effect $\theta^{ss}(M_1, N_1)$
    $\approx \hat{\theta}(M_1^p, N_1^p, N_0^p, T)$
    for all $M_1$ and $N_1$.
    \item %
    They wrongly assume that there exist constants $\tilde{\sigma}_1^2$ and $\tilde{\sigma}_0^2$ such that $\tilde{\sigma}_1^2  (M_1, N_1) = \tilde{\sigma}_1^2/N_1$ and $\tilde{\sigma}_0^2 (N_0) = \tilde{\sigma}_0^2/N_0$ for all $M_1, N_1, N_0$.
\end{enumerate}

In other words, assumption (1) states that the naive experiment designer believes that there is a constant  treatment effect for any $M_1$, $N_1$, and $N_0$ and that all experiment designs will lead to a similar treatment effect as the pilot study. Thus, if the experimenter knew the treatment effect of the pilot study, they would set $\MDE={\theta}^{ss}(M_1^p, N^p_1)$. Further, an experimenter who is not aware of the impact of congestion does not increase the number of servers beyond $M^p_1$. 

We make several simplifying assumptions that make the naive experimenter's task easier. First, the experimenter does not know ${\theta}^{ss}(M^p, N^p_1)$ and only obtains an estimate $\hat{\theta}(M_1^p, N^p_1, N^p_0, T)$. Further, the question of how to estimate the variance of the estimator $\hat{\theta}(M^p, N^p_1, N^p_0, T)$ is an open question, as the estimator involves outcomes that are correlated through time and through capacity constraints. To simplify matters, we assume that the experimenter knows three additional pieces of information after the pilot study: they have oracle access to $\theta^{ss}(M^p_1, N^p_1, N^p_0), \ \tilde{\sigma}^2_1(M^p_1, N^p_1)/T$, and $\tilde{\sigma}^2_0( N^p_0)/T$. We note that oracle access to these quantities makes the naive experimenter's task easier, and we will still show that the naive policy underperforms with respect to the queueing-informed policy. In other words, the experimenter has perfect knowledge of the sampling distribution of the estimator in the pilot study, but is not given additional information about the sampling distribution at other $M_1$, $N_1$, or $N_0$.

Under this oracle knowledge and Assumption (2), the naive experimenter (incorrectly) assumes that the treatment group variance term scales with $N_1$ and so the variance at $N_1 \neq N^p_1$ is $\tilde{\sigma}^2_1(M^p_1, N^p_1)/T \cdot N^p_1 / N_1$. %
Likewise, the experimenter (correctly) assumes that the control group variance term scales with $N_0$ and so the variance at $N_0 \neq N^p_0$ is $\tilde{\sigma}^2_0( N^p_0)/T \cdot N^p_0 / N_0$.

Given these assumptions, the naive experimenter chooses $N_1$ as follows.

\begin{subequations}
\begin{align}
\nonumber
    {\mathrm{minimize}} \:  \quad & N_1 \\
    \nonumber
    \mathrm{subject\,to}\quad
    &  {\Phi\left( \frac{\MDE}{\sqrt{f(N_1) + g(N_0)}} 
    -
    \Phi^{-1}(1 - \alpha / 2)
    \right)} \\
    &  {\quad + %
    \Phi\left(-  \frac{\MDE}{\sqrt{f(N_1) + g(N_0)}} 
    -
    \Phi^{-1}(1 - \alpha / 2)
    \right) \geq \beta} \\
    \label{eq:naive_opt_mde}
    & \MDE = {\theta}^{ss}(M_1^p, N^p_1)\\
    \label{eq:naive_opt_m}
    & M_1 = M^p_1\\
    \label{eq:naive_opt_var_treat}
    & f(N_1) = \frac{ N^p_1}{N_1 \ T}\cdot \tilde{\sigma}^2_1(M^p_1, N^p_1) \\ \label{eq:naive_opt_var_control}
    & g(N_0) = \frac{N^p_0}{ N_0 \ T} \cdot \tilde{\sigma}^2_1( N^p_0)  \\
    \nonumber
    & N_1 = N_0 \\
    \nonumber
    & N_1 \in \mathbb{Z}^n_+
\end{align}
\end{subequations}
Note that \eqref{eq:naive_opt_mde}-\eqref{eq:naive_opt_m} reflect Assumption (1) and \eqref{eq:naive_opt_var_treat}-\eqref{eq:naive_opt_var_control}  reflect Assumption (2). 
Let $N_1^{naive}$ denote the solution to the naive optimization problem. Then the first Naive Policy of not scaling $M_1$ would recommend
$(M_1^p, N_1^{Naive}, N_1^{Naive})$ for the full experiment.

\paragraph{Naive Policy 2: Proportional $M_1$ Scale-up.}
We consider a (less naive) second policy where the experimenter makes Assumption (2) but not Assumption (1). This experimenter knows that $\theta^{ss}(M_1,N_1)$ depends on both $M_1$ and $N_1$, but does not necessarily know how. In this setting, experimenter similarly sets $MDE = \theta(M_1^p, N_1^p)$ and prescribes the same $N^{naive}$ as above, but takes the natural step of scaling $M_1$ proportionally with $N_1$, so that 
    $M_1 = \left\lceil M_1^p \cdot \frac{N^{naive}}{N_1^p} \right\rceil.$
The Naive Policy (Proportion $M_1$ Scale-up) recommends $(\lceil M^p_1 \cdot \frac{N^{naive}}{ N^p}\rceil, N_1^{Naive}, N_1^{Naive})$ for the full experiment.

\paragraph{Queueing Informed Square Root Staffing Policy.}

We now consider a policy informed by queueing theory, which utilizes the square root staffing rule, a common heuristic to balance efficiency and quality of service. This experimenter intelligently corrects for both Assumption (1) and Assumption (2).

The queueing informed experimenter solves the following optimization problem to solve for $M_1, N_1, N_0$.

\begin{subequations}
\begin{align}
    \nonumber
    {\mathrm{minimize}} \:  \quad & N_1 \\
    \nonumber
    &  {\Phi\left( \frac{\MDE}{\sqrt{f(N_1) + g(N_0)}} 
    -
    \Phi^{-1}(1 - \alpha / 2)
    \right)} \\
    &  {\quad + %
    \Phi\left(-  \frac{\MDE}{\sqrt{f(N_1) + g(N_0)}} 
    -
    \Phi^{-1}(1 - \alpha / 2)
    \right) \geq \beta} \\
    \label{eq:queueing_opt_mde}
    & \MDE = {\theta}^{ss}(M_1, N_1)\\
    \label{eq:queueing_opt_m}
    & M_1 = \left\lceil r N_1 + \gamma \sqrt{N_1}\right\rceil\\
    \label{eq:queueing_opt_var_treat}
    & f(N_1) = \tilde{\sigma}^2_1(M_1, N_1)/T \\ 
    \label{eq:queueing_opt_var_control}
    & g(N_0) = \tilde{\sigma}^2_1(N_0)/T  \\
    \nonumber
    & N_1 = N_0 \\
    \nonumber
    & N_1 \in \mathbb{Z}^n_+
\end{align}
\end{subequations}
Let $N^{sqrt}_1$ denote the solution $N_1$ to the above problem. 
Compared to the naive optimization problem, the queueing informed optimization problem corrects for Assumption (1) in \eqref{eq:queueing_opt_mde},  implements the square root staffing rule in \eqref{eq:queueing_opt_m}, and uses the variance approximation from Theorem \ref{thm:clt} in \eqref{eq:queueing_opt_var_treat}-\eqref{eq:queueing_opt_var_control}.
The Square Root Staffing Policy then proposes $(\left\lceil r N_1 + \gamma \sqrt{N_1}\right\rceil, N_1^{sqrt}, N_1^{sqrt})$ for the full experiment. 

Note, however, that both calculating the expressions in \eqref{eq:queueing_opt_mde}-\eqref{eq:queueing_opt_var_control} requires knowing $\lambda$, $\tau$, $\mu$, and $p$. In practice, these parameters are unknown and experimenter would need to obtain estimates $\hat{\lambda}, \hat{\tau}, \hat{\mu}$, and $\hat{p}$ from the pilot study, and then use these estimates to compute \eqref{eq:queueing_opt_mde}-\eqref{eq:queueing_opt_var_control}. Similar to the discussion of variance estimation for the naive policy, in order to isolate the issue of scaling up $M_1$ and $N_1$, we assume that the experimenter has oracle access to the true quantities and leave the issue of estimation of queueing quantities to future work.

\subsection{Choice of $\gamma$ value}
\label{ssec:sqrt_gamma}
The theoretical properties of the Square Root rule are established for queueing systems with exponential inter-arrival and service times ($M/M/N$ queues) in \cite{halfin1981heavy}. In this case (as in many other systems) the constant $\gamma$ is chosen to be positive. This staffing rule positions the system in the QED regime. The formula for the number of servers has two components; $M_1$ to match the offered load on the system $r N_1$ (the amount of traffic the queue is expected to handle) and a safety buffer proportional to $\sqrt{N_1}$. In other words, to maintain a fixed level of service when increasing $N_1$, the chosen $M_1$ should not be linear in $N_1$ but rather needs to account for variability in the system and incorporate this $\sqrt{N_1}$ term. For a mathematical justification of this rule, see \cite{halfin1981heavy, shortle2018fundamentals}. 

Importantly, for our purposes, \cite{vericourt2008dimensioning} point out that in closed queueing systems, optimal staffing levels are sometimes obtained using a negative $\gamma$. For the case of service interventions, the intuition for this is clear for two reasons. First, since the sample size is fixed (for a given trial) the maximum number of subjects waiting for service (i.e., the longest possible queue) is fixed. This is not the case for open queueing systems. Second, the subjects of service interventions may transition from the undesired state (in which they are eligible for service) to the desired state (in which they are not eligible for service) in the absens of an intervention. This means that the service provider may not have to serve every single subject that becomes eligible for service at some point. 

As noted before, the parameter $\gamma$ controls the specific position around the boundary between the Quality-driven regime and the Efficiency-driven regime. Larger positive $\gamma$ values push the system towards the Quality-Driven regime with higher service quality but increased resource requirements. Smaller (or negative) values move toward the Efficiency-Driven regime with lower resource requirements but diminished treatment effects. As part of our numerical experiments, we will highlight the range of capacity decisions---and the resulting statistical power---that are compatible with the square root staffing rule.

\subsection{Numerics comparing power analysis policies}
\label{si:ssec:numerics_power_analysis_policies}
Using numerics from the CTMC model, we compare the naive power analysis policies with the Queueing Informed Square Root Staffing Policy. We note that the results in the CTMC model mirror the results in the Keheala-calibrated simulations in the main text \S\ref{s.keheala}.

\paragraph{Numerical experiment setup}

We consider a setting with $\lambda = 0.4$, $\tau = 0.35$, $\mu = 3$, $p=0.1$, and $T=10$.  We focus on a setting with $\lambda<0.5$ and $\tau<0.5$ to capture the effect where states are ``sticky'', i.e., without intervention, the user is more likely to remain in the state they were in yesterday. 

We assume a standard target of $\beta=0.80$ for statistical power, a significance level of $\alpha=0.05$, and consider two scenarios for the pilot: 
\begin{itemize}
    \item \textbf{Scenario 1 -- Small participant-to-server ratio (QD) pilot:} Initial pilot study with $M_1^p = 5$ and $N_1^p = N_0^p = 10$ subjects. 
    \item \textbf{Scenario B --  Large participant-to-server ratio (ED) pilot:}  Initial pilot study with $M_1^p=5$ but more $N_1^p = N_0^p = 25$ subjects. 
\end{itemize}

We compare the power at each of the three power analysis policies: Naive Policy 1 (No $M_1$ Scale-up), Naive Policy 2 (Proportional $M_1$ Scale-up), and the Queueing Informed Square Root Staffing Policy. Using Equation \eqref{eq:power_approximate_mde_true_effect_two_sided}, we evaluate the statistical power of the experiment with $(M_1, N_1, N_0)$ to detect a treatment effect at least as large as the true treatment effect $\theta^{ss}(M_1, N_1)$ of the experiment, where we approximate estimator mean and variance using Equations \eqref{eq:asymptotic_mean_approximation} and \eqref{eq:asymptotic_var_approximation}.

\paragraph{Comparison of power analysis policies}

\begin{figure}
    \begin{subfigure}[b]{0.5\linewidth}
        \centering
        \includegraphics[width=.8\linewidth]{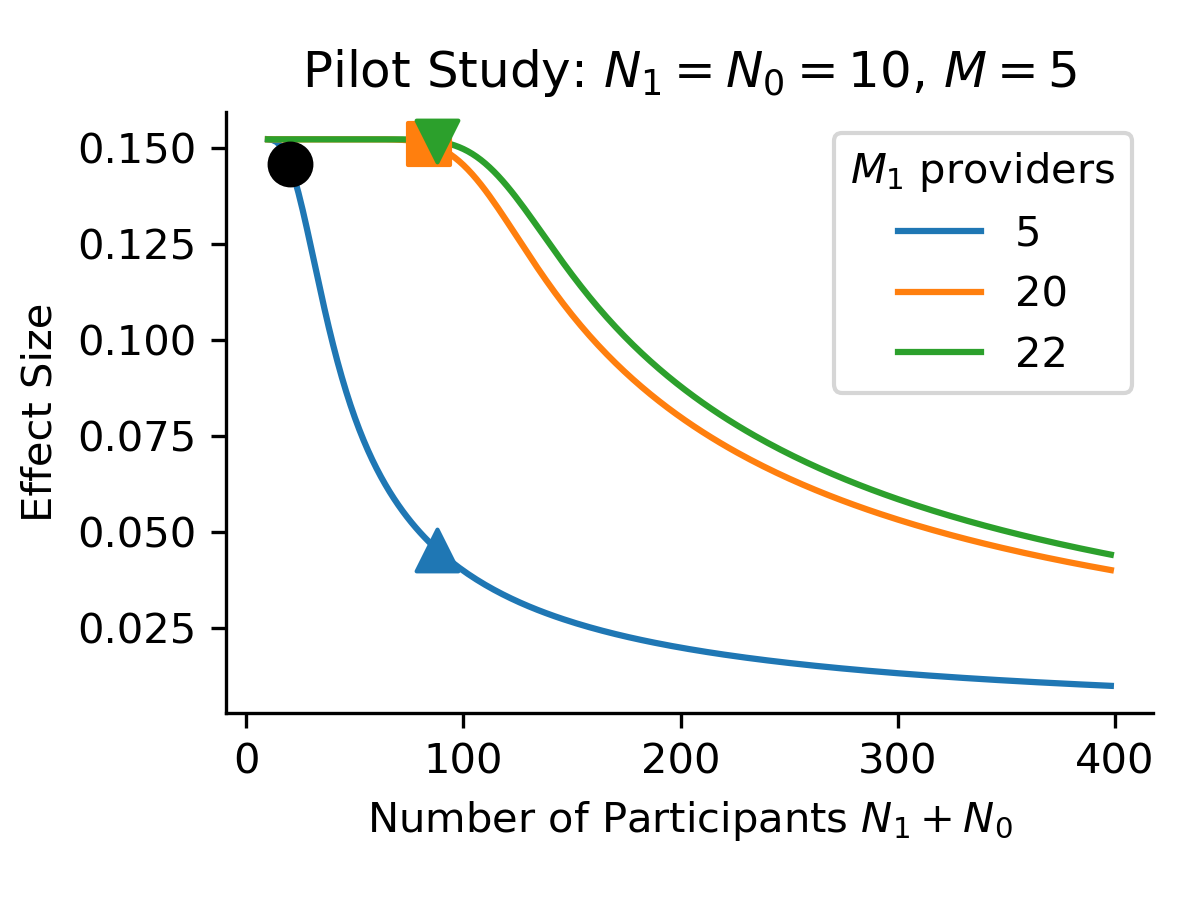} 
        \caption{QD pilot, Treatment effect}
    \end{subfigure}
    \hfill
    \begin{subfigure}[b]{0.5\linewidth}
        \centering
        \includegraphics[width=.8\linewidth]{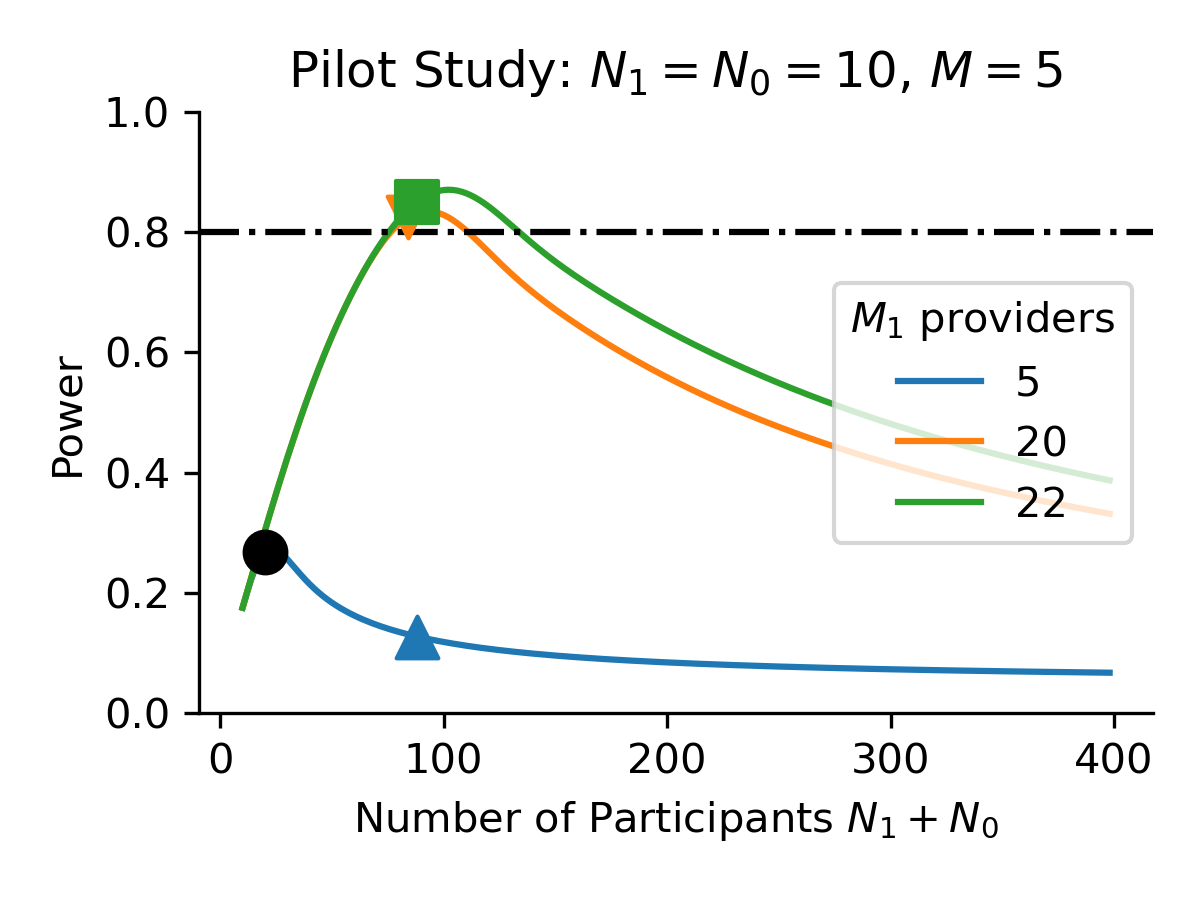}
        \caption{QD pilot, power}
    \end{subfigure}

    \begin{subfigure}[b]{0.5\linewidth}
        \centering
        \includegraphics[width=.8\linewidth]{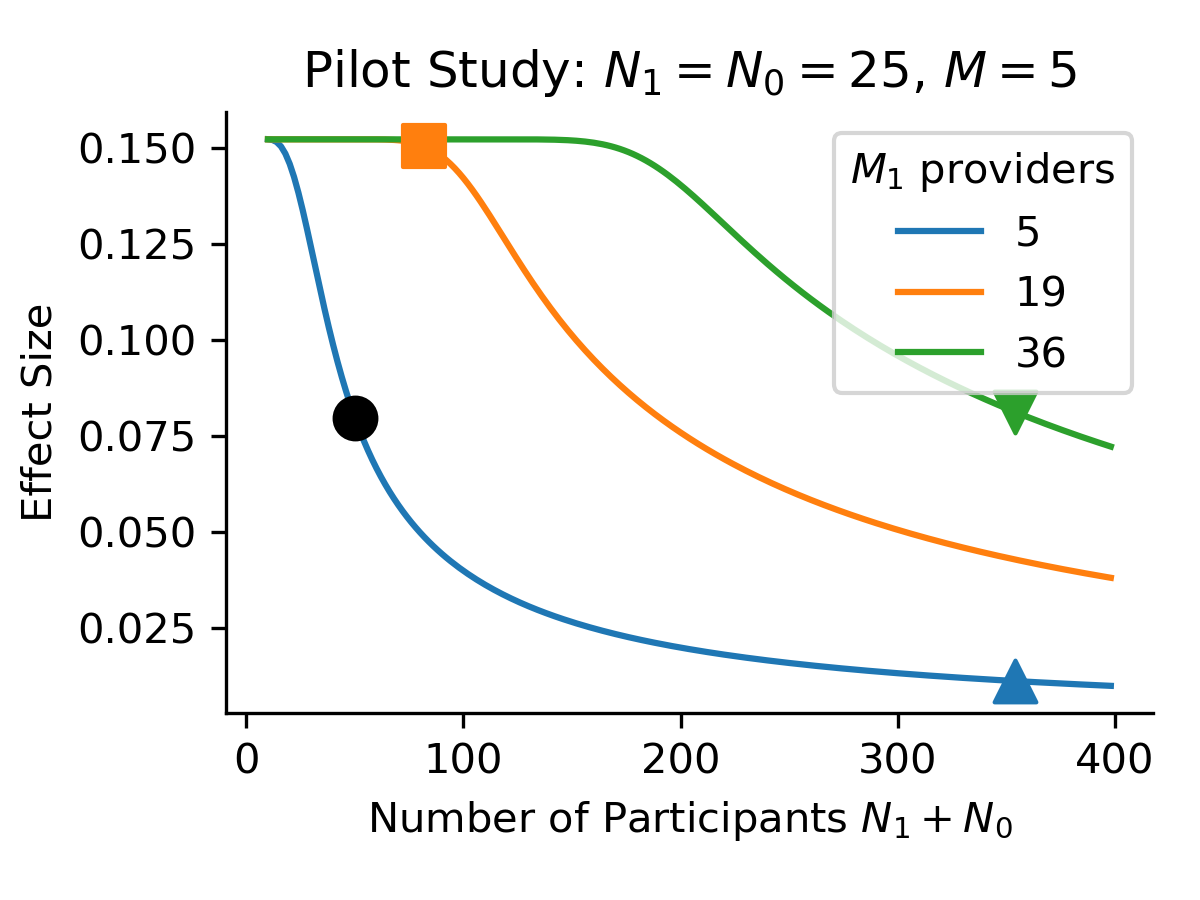}
        \caption{ED pilot, treatment effect}
    \end{subfigure}
    \hfill
    \begin{subfigure}[b]{0.5\linewidth}
        \centering
        \includegraphics[width=.8\linewidth]{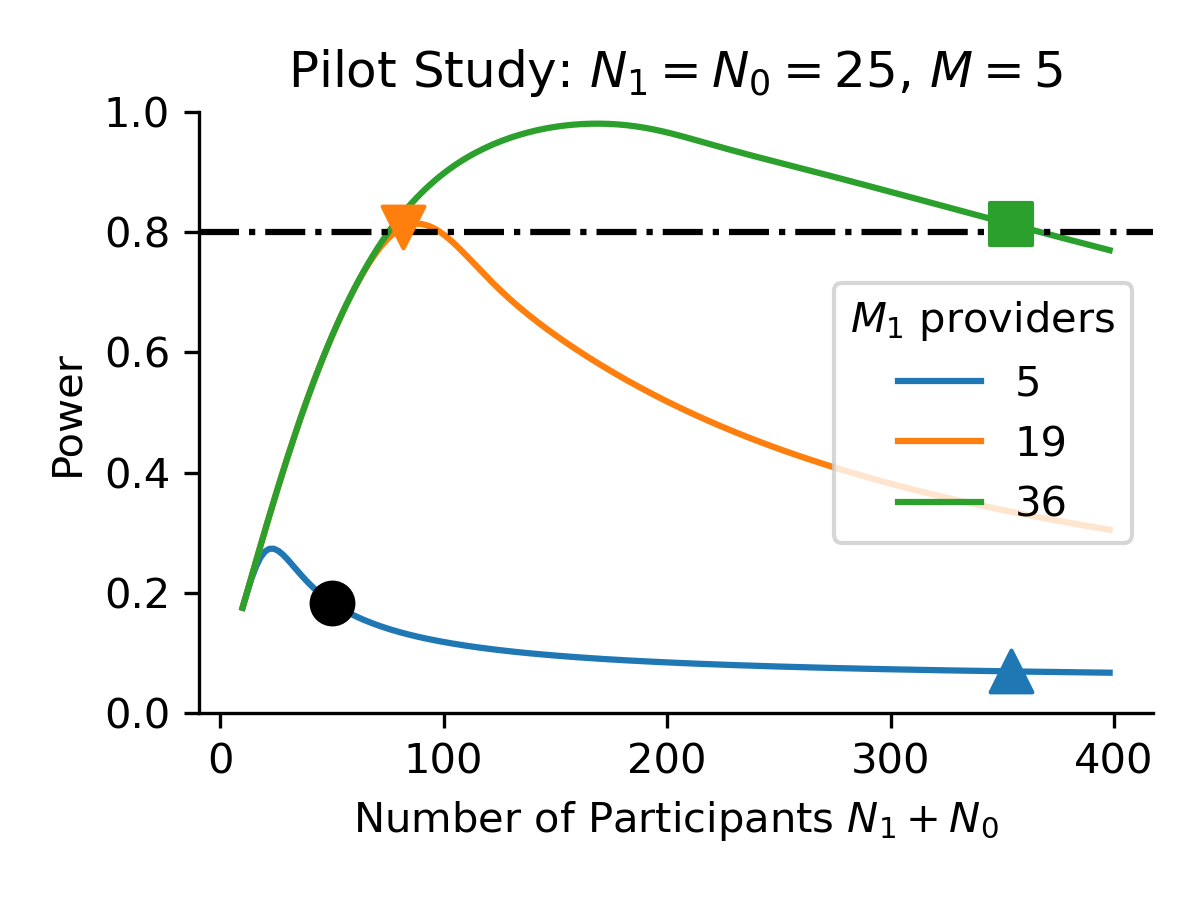}
        \caption{ED pilot, power}
    \end{subfigure}

    \caption{Numerics depicting  pilot studies in the Quality-Driven (QD) and Efficiency-Driven (ED) regimes, the power analyses methods, and the results in the full experiment. The figure shows the treatment effect (panels (a) and (c)) and statistical power (panels (b) and (d)) for various combinations of patient enrollment ($N_1 + N_0$) and service capacity ($M_1$ providers). The top row (panels (a) and (b)) depicts the results of a pilot study with $N_1 = N_0 = 10$ and $M = 5$, placing it in the Quality-Driven regime. The bottom row (panels (c) and (d)) depicts a pilot study with $N_1 = N_0 = 25$ and $M = 5$, placing it in the Efficiency-Driven regime. In each panel, different colored lines represent different provider capacities. The black circles indicate outcomes under the pilot study parameters, the triangle represents the Naive No Scale-Up policy, the upside down triangle represents the Naive Proportional policies, and the square represents the Square Root Policy. The horizontal dashed lines in panels (b) and (d) mark the 80\% power threshold.
     }
    \label{si:fig_numerics_power_analysis_policies}
\end{figure}

Figure \ref{si:fig_numerics_power_analysis_policies} shows the performance of the three policies, for two difference pilot study configurations of $M^1_p$ and $N_1^p$. Since these initial pilot study parameters $M^1_p$ and $N_1^p$ are typically chosen for feasibility reasons and not optimized, a power analysis policy would ideally perform well given any instance of $M^1_p$ and $N_1^p$. While the queueing informed Square Root Staffing Policy can give the desired power $\beta$ for both scenarios, Naive Policy 1 (No $M_1$ Scale-up) does not achieve the desired power in either scenario, and Naive Policy 2 (Proportional $M_1$ Scale-up) achieves the desired policy but, depending on the initial pilot study conditions, may require significantly more resources than the Square Root Staffing Policy.

Under these parameters, the critical ratio is $r=0.38$ and Scenario 1 ($M_1^p/ N_1^p = 0.5$) corresponds to the small $N_1$ (Quality-Driven) regime whereas Scenario 2 ($M_1^p/ N_1^p = 0.2$) corresponds to  the large $N_1$ (Efficiency-Driven) regime. Importantly, as Fig.  \ref{si:fig_numerics_power_analysis_policies} shows, the treatment effect is larger for $M_1^p = 5$ and $N_1^p = 20$ than it is for $M_1^p=5$ and more users $N_1^p = 50$, which drives the differences in the power analysis policies, as explained below.

In Scenario 1 (small $N^p_1$ pilot), the treatment effect in the pilot study is $\theta^{ss}(5, 10) = 0.146$. Naive Policy 1 (No-$M_1$ Scale-up) chooses $(M_1,N_1, N_0)=(5, 44, 44)$, Naive Policy 2 (Proportional $M_1$ Scale-up) chooses $(M_1,N_1, N_0)=(22, 44, 44)$, and the Square Root Staffing Policy chooses $(M_1,N_1, N_0)=(20, 42, 42)$. Naive Policy 1 (No $M_1$ Scale-up) does not achieve the desired power, and actually has less power than the initial pilot experiment, due to the decreased effect size when increasing $N_1$ but not increasing $M_1$. Both the Naive Policy 2 (Proportional $M_1$ Scale-up) and the Square Root Staffing Policy achieve $80\%$ power. However, the Square Root Staffing Policy is able to achieve this power with slightly fewer users \textit{and} servers, which reduces costs for the experiment.

In Scenario 2 (large $N^p_1$ pilot), the treatment effect in the pilot study is $\theta^{ss}(5, 25) = 0.080$, which is lower than the treatment effect in Scenario 1, due to the higher user load. We observe a similar pattern where the Naive Policy 1 (No $M_1$ Scale-up) is underpowered but both Naive Policy 2 (Proportional $M_1$ Scale-up) and Square Root Staffing policies achieve 80\% power.  Naive Policy 1 (No-$M_1$ Scale-up) chooses $(M_1,N_1,N_0)=(5, 177, 177)$, Naive Policy 2 (Proportional $M_1$ Scale-up) chooses $(M_1,N_1, N_0)=(36, 177, 177)$, and the Square Root Staffing Policy chooses $(M_1,N_1,N_0)=(19, 41, 41)$. In this setting, the Square Root Policy is able to achieve the same power as Naive Policy 2, but with 47\% fewer servers and 77\% fewer users, requiring far fewer resources for the same power. Both naive policies choose a large number of users because they fail to realize that the small treatment effect size in the pilot study is due to a high user load on the system. Both naive policies attempt to increase the number of users in order to detect an effect as small as the pilot study observed effect size. The Queueing Informed Square Root Staffing rule recognizes that the treatment effect can be larger if there are more servers and thus requires fewer users in the experiment to detect this effect.

The Queueing Informed Square Root Staffing Policy is able to achieve similar statistical power with fewer resources because it is able to leverage information about the queueing structure. In the example given, the Naive Policy 2 requires a large number of resources because the initial pilot resulted in a small treatment effect $\theta^{ss}(M_1,N_1)=0.080$, leading the experimenter to assume that a large number of subjects is required to detect an effect. An experimenter using the Square Root Staffing Policy is able to discern that the treatment effect $\theta^{ss}(M_1,N_1)$ would be larger if there were fewer subjects, relative to providers, and so the number of subjects and providers needed is smaller than what the Naive Policy 2 suggests.

We note that we primarily see the benefit in the Queueing Informed Square Root Staffing Policy over Naive Policy 2 when the initial pilot study happens to be in the Efficiency-Driven (large $N^p_1$) regime. If the initial pilot study is in the Quality-Driven regime, both policies lead to similar recommendations and results.

\clearpage

% 5. Details on Keheala simulation calibration

\section{Additional details on simulations of Keheala RCTs}\label{si:sec:Keheala}
We calibrate our simulation model to the Keheala trial data using a double machine learning (double ML) approach \citep{chernozhukov2018double} to predict whether a patient will self-verify treatment adherence on day $t+1$ as a function of their demographic features, cumulative and recent behavior, as well as platform outreach on day $t$.\footnote{This model is distinct from but inspired by (and based on the same data as) the simulation model used for the numerical analysis in \cite{baek2023policy}. However, that analysis was focused on the impact of targeting (i.e., which patients to prioritize for outreach) while this analysis is focused on the implications of operational capacity on statistical power.} The double ML framework allows for estimating causal effects in the presence of high-dimensional covariates by flexibly controlling for confounding through machine learning-based nuisance models. In our context, it isolates the causal effect of a single outreach on next-day self-verification while accounting for individual heterogeneity in adherence behavior. We use this model to simulate counterfactual trial outcomes under varying levels of service capacity and patient enrollment.

\paragraph{Data.}
The dataset comprises 5,433 individuals assigned to the Keheala intervention arm of a large, four-arm randomized controlled trial conducted across 902 clinics in Kenya between April 2018 and December 2019 \citep{Yoeli_2019}. All individuals were undergoing TB treatment, had at least two months of therapy remaining, and met eligibility criteria including access to a mobile phone and ability to communicate in Kiswahili or English. This results in a total of 416,071 patient-days, which we use to train our models. The dataset includes baseline socio-demographic and clinical characteristics as well as operational data for patient usage of Keheala's services, their interactions with Keheala, and their self-verification behavior (see more detail, below). 

\paragraph{Features.}
Our machine-learning model uses 8 static socio-demographic and clinical features collected at enrollment, and 5 dynamic behavioral features describing recent and cumulative engagement with the \textit{Keheala} platform. Below is a more detailed description of each set of features. 

\begin{itemize}
    \item \textbf{Static features}
    \begin{itemize}
        \item \textit{Demographics:} Age, Weight, Height, Gender, Primary language (all features either numeric or categorical indicators). 
        \item \textit{Clinical history:} HIV status unknown, HIV co-infection, TB type (categorical indicators). 
        \item \textit{Geographical location:} Administrative zone (categorical indicators). 
    \end{itemize}
    \item \textbf{Dynamic features}
    \begin{itemize}
        \item \textit{Enrollment duration:} Number of days since enrollment on the Keheala platform. 
        \item \textit{Verification behavior.} Two features: (a) binary indicators for verification on each of the last 7 days, (b) total number of verifications since enrollment. 
        \item \textit{Sponsor outreach.} Two features: (a) binary indicators for receiving outreach on each of the last 7 days, (b) total number of outreach received since enrollment.
    \end{itemize}
\end{itemize}

\paragraph{Outcomes and treatment.}
Let $Y_{i,t}\in\{0,1\}$ denote whether patient $i$ self-verified adherence on day $t$, the outcome of interest. Let $D_{i,t}$ denote whether the treatment, whether a support sponsor performed outreach to the patient on day $t$. 

\paragraph{Model specification.}

We fit the following models (following \cite{chernozhukov2018double}):
\begin{itemize}
    \item Outcome models: $g_0 = E[Y_{i, t+1|X_{i,t}, D=0]}$ trained on the set of patients who do not receive outreach and $g_1 = E[Y_{i, t+1|X_{i,t}, D=1]}$ trained on the set of patients who do receive outreach, and
    \item Propensity model: $m (X_{i,t})=P(D_{i,t}=1|X_{i,t})$.
\end{itemize}
The two models $g_1$ and $g_0$ above predict the next-day verification probability with and without platform outreach, conditional on the patient's past behavior.  The functions $g_0, g_1$, and $m$ are estimated using random forests with five-fold cross-fitting.

\paragraph{Model performance.}
We report the model performance on held out folds. For $g_1$, the AU-ROC is 0.79, computed on held-out observations with $D_{i,t}=1$. For $g_0$, the AU-ROC is 0.92, computed on held-out observations with $D_{i,t}=0$.

\paragraph{Mapping models to simulation.}
We use the outcome models to simulate next-day verification outcomes of patients. 
For a patient-day with covariates $X_{i,t}$, let $p_1(X_{i,t})$ and $p_0(X_{i,t})$ denote their probability of verifying the next day, with and without outreach, respectively.  

For a patient that does not receive outreach, they verify the next day with probability $p_0(X_{i,t}) = g_0(X_{i,t})$. We simulate their next day outcome as a random variable $Bernoulli(p_0(X_{i,t}))$.

For a patient that does receive outreach, we calculate $p_1(X_{i,t})$ by sampling the effect of the outreach as follows. 
On the subset of patient-day observations $X_{i,t}$ where the patient did not verify on day $t$, we can interpret $g_1(X_{i,t}) - g_0(X_{i,t})$ as 
the impact of a call on next day verification rate. 
Across this subset, the $5$th and $95$th percentile predicted increases are  $0.071086$ and $0.078015$, respectively, with a mean of 0.07455. In the simulations, when a patient is called, we draw a $\Delta \sim Unif[0.071086, 0.078015]$ and let $p_1 = \min\{1, p_0(X_{i,t})+\Delta\}$. We simulate their next day outcome as a random variable $Bernoulli(p_1(X_{i,t}))$

\paragraph{Simulation of RCTs.}
We use the models fit above to simulate out sample paths of counterfactual RCTs. The steps below outline the procedure to simulate a single trial over a fixed time horizon of $T=180$ days, 
\begin{itemize}
    \item Initialization: We sample patients for the control arm and treatment arm from the empirical patient pool. For each sampled patient, we initialize their behavioral trajectory using the first 10 days of data from the real trial, to determine $X_{i, 1}, \ldots, X_{i,10}$. Once this ten day history is established, we use the double ML models to simulate counterfactual sample paths for the rest of trial duration, as explained below. 
    \item Control arm: For patients in the control arm, no outreach is delivered. Next-day verification is simulated using $p_0(X_{i,t})=g_0(X_{i,t})$, described above. 
    \item Treatment arm: Each day, patients who did not self-verify are eligible for outreach and join a first-come-first-served queue. On each day, the number of outreach instances available  is drawn from a Poisson distribution with mean $C$, the capacity of the system, which we vary across simulations. A patient with features $X_{i,t}$ has next-day verification probability of $p_1(X_{i,t})$, described above, if they are called and $p_0(X_{i,t})$ otherwise. If a patient is called and does not verify on the next day, they re-enter at the end of the queue; if they verify, they leave the queue and only rejoin if they miss again in the future. Unserved patients remain in order and roll over to the next day.
    \item Outcome generation: For each patient-day, we compute $p_0(X_{i,t})$ and $p_1(X_{i,t})$. If the patient is called, we draw the next-day verification $Y_{i, t+1}\sim Ber(p_1(X_{i,t}))$ and otherwise we draw $Y_{i, t+1}\sim Ber(p_0(X_{i,t}))$.
\end{itemize}

\paragraph{Simulated trial outcomes and metrics.}
For each patient, we observe the proportion of days that they self-verify adherence. We then compute the average verification rate between the treatment and control group and estimate the average treatment effect using the difference-in-means estimator \eqref{e:estimator}. Within each simulated trial, we compute:
\begin{itemize}
    \item Estimated standard error, computed according to Welch's formula and allowing for unequal variances across groups.
    \item Welch's $t$-statistic and the associated p-value for a two-tailed $t$-test, using the estimated standard error and the Welch-Satterthwait degrees of freedom. 
    \item Cohen's $d$ standardized effect size.
    \item Analytical power (at significance level $\alpha=0.05$) of detecting an effect at least as large as the observed effect size, based on the noncentral $t$ distribution for a two-tailed $t$-test. 
\end{itemize}

\paragraph{Simulation parameters.}

We simulate RCT trajectories for varying numbers of enrolled patients ($N_0$ and $N_1$, varying from 20 to 650) and daily call capacities ($C$, varying from 1 to 80). For each combination of enrolled patients and call capacities, we run 25 simulated trials.

\paragraph{Summary statistics and critical threshold calculation.}

Under this simulation method, the average verification probability across control group patients is $0.375$, averaged across simulation runs. The average verification probability of the treatment group depends on the service capacity and the number of enrolled patients (the treatment group sample size).

We must also calculate the critical threshold (defined for the CTMC in \S\ref{s:model}), which is used in the Queueing-Informed Square Root Staffing policy, for the discrete time setting. 
For each of the parameters in the continuous time Markov chain model described in \S\ref{s:model}, we calculate the analogous quantity in the discrete time simulations.
\begin{itemize}
    \item $\lambda$: probability $P(Y_{i,t+1}=0 | Y_{i,t}=1, D_{i,t}=0)$ that a patient verifying at time $t$ does not verify at time $t+1$. 
    \item $\tau$: probability $P(Y_{i,t+1}=1 | Y_{i,t}=0, D_{i,t}=1)$ that a patient not verifying at time $t$ starts verifying at time $t+1$, in the absence of outreach.
    \item $\mu$: average number of calls made per unit capacity, defined in our model to be 1.  
    \item $p$: outreach-induced lift $P(Y_{i,t+1}=1 | Y_{i,t}=0, D_{i,t}=1)-P(Y_{i,t+1}=1 | Y_{i,t}=0, D_{i,t}=0)$ in next-day verification probability. 
\end{itemize}

We estimate the first two probabilities from the control group outcomes in a single simulation run, averaged across all participants in the control group, resulting in the quantities $\lambda \to 0.465$ and $\tau\to 0.289$, respectively. 
We set the average number of calls per unit capacity to be $1$. 
We set the outreach induced lift to be the mean lift $0.075$, calculated according to the of the effect of an outreach described above (``Mapping models to simulation'').
Plugging these values into the critical threshold definition \eqref{eq:critical_ratio} gives a critical threshold of $1.77$.

\end{document}